\PassOptionsToPackage{pdfpagelabels=false}{hyperref} 
\documentclass[useAMS,usenatbib]{mnras}
\pdfoutput=1
\usepackage[pdftex]{graphicx}
\usepackage{amsmath}
\usepackage{natbib}
\usepackage{txfonts}
\usepackage{times}
\usepackage{xspace}
\usepackage{bm}
\renewcommand{\vec}[1]{\mbox{\boldmath$#1$}}

\newcommand{\fracspace}{\vphantom{\int}}

\newcommand{\msun}{\,M$_{\odot}$\xspace}
\newcommand{\msunyr}{M$_{\odot}$\,yr$^{-1}$\xspace}
\newcommand{\etaM}{$\eta_{\rm M}$\xspace}
\newcommand{\vrad}{$v_{\rm rad}$\xspace}

\usepackage{etoolbox}
\makeatletter
\patchcmd\@combinedblfloats{\box\@outputbox}{\unvbox\@outputbox}{}{\errmessage{\noexpand patch failed}}
\makeatother

\volume{{\rm submitted}}

\title[TNG50: Outflows]{First Results from the TNG50 Simulation: \\Galactic outflows driven by supernovae and black hole feedback}
\author[D. Nelson et al.]{Dylan Nelson$^{1}$\thanks{E-mail: dnelson@mpa-garching.mpg.de},
Annalisa Pillepich$^{2}$, 
Volker Springel$^{1,3,4}$, 
R{\"u}diger Pakmor$^{1,3}$, \newauthor 
Rainer Weinberger$^{5}$,
Shy Genel$^{6}$,
Paul Torrey$^{7}$, 
Mark Vogelsberger$^{8}$, \newauthor
Federico Marinacci$^{9}$, 
Lars Hernquist$^{5}$
\\\\
$^{1}$Max-Planck-Institut f\"{u}r Astrophysik, Karl-Schwarzschild-Str. 1, 85741 Garching, Germany\\
$^{2}$Max-Planck-Institut f\"{u}r Astronomie, K\"{o}nigstuhl 17, 69117 Heidelberg, Germany\\
$^{3}$Heidelberg Institute for Theoretical Studies, Schloss-Wolfsbrunnenweg 35, 69118 Heidelberg, Germany\\
$^{4}$Zentrum f\"{u}r Astronomie der Universit\"{a}t Heidelberg, ARI, M\"{o}nchhofstr. 12-14, 69120 Heidelberg, Germany\\
$^{5}$Harvard-Smithsonian Center for Astrophysics, 60 Garden Street, Cambridge, MA, 02138, USA\\
$^{6}$Center for Computational Astrophysics, Flatiron Institute, 162 Fifth Avenue, New York, NY 10010, USA\\
$^{7}$University of Florida, Department of Astronomy, 211 Bryant Space Sciences Center, Gainesville, FL 32611, USA\\
$^{8}$Kavli Institute for Astrophysics and Space Research, Department of Physics, MIT, Cambridge, MA, 02139, USA\\
$^{9}$Department of Physics and Astronomy, University of Bologna, Viale Berti Pichat 6/2, I-40127 Bologna, Italy\\
}

\begin{document}

\maketitle

\begin{abstract}
We present the new TNG50 cosmological, magnetohydrodynamical simulation -- the third and final volume of the IllustrisTNG project. This simulation occupies a unique combination of large volume and high resolution, with a 50 Mpc box sampled by $2160^3$ gas cells (baryon mass of $8 \times 10^4$\msun). The \textit{median} spatial resolution of star-forming ISM gas is $\sim$100$-$140 parsecs. This resolution approaches or exceeds that of modern `zoom' simulations of individual massive galaxies, while the volume contains $\sim$ 20,000 resolved galaxies with $M_\star \ga 10^7$\msun. Herein we show first results from TNG50, focusing on galactic outflows driven by supernovae as well as supermassive black hole feedback.
We find that the outflow mass loading is a non-monotonic function of galaxy stellar mass, turning over and rising rapidly above $10^{10.5}$\msun due to the action of the central black hole. Outflow velocity increases with stellar mass, and at fixed mass is faster at higher redshift. The TNG model can produce high velocity, multi-phase outflows which include cool, dense components. These outflows reach speeds in excess of 3000 km/s out to 20 kpc with an ejective, BH-driven origin. Critically, we show how the relative simplicity of model inputs (and scalings) at the injection scale produces complex behavior at galactic and halo scales. For example, despite isotropic wind launching, outflows exhibit natural collimation and an emergent bipolarity. Furthermore, galaxies above the star-forming main sequence drive faster outflows, although this correlation inverts at high mass with the onset of quenching, whereby low luminosity, slowly accreting, massive black holes drive the strongest outflows.
\end{abstract}



\section{Introduction}

Energetic outflows driven out of galaxies signpost the resistance of astrophysical feedback processes against the inevitability of gravitational collapse. By itself, the accretion of gas into dark matter halos and its subsequent cooling onto a centrally forming galaxy \citep{silk77,wr78} leads to an excess of baryons across the mass scale \citep{thoul95}. This is evidenced most clearly by the differential shape of the galaxy stellar mass function with respect to the underlying dark matter halo mass function from which it must emerge \citep{kauffmann93,cole94}.

To successfully regulate the stellar mass content of galaxies gas must be prevented from joining the star-forming phase, efficiently ejected out of this phase, or both. Preventive feedback alone is not thought to be sufficient \citep{benson03}, while galactic outflows are both theoretically expected \citep{tomisaka88} and observationally detected \citep{heckman90}. Physically, they are expected to be driven by the energy released from stars and supernovae \citep{cc85}, as well as from supermassive black holes \citep{begelman91}.

No model for galaxy formation currently succeeds without incorporating both mechanisms \citep{spr03,croton06}. Outflows and feedback are therefore a fundamental aspect of galaxy formation. However, the underlying physical mechanisms are complex \citep{murray05} and operate at extremely small scales -- a supermassive black hole accretion disk (10$^{-5}$ pc) is roughly ten orders of magnitude smaller than its host galaxy or the intracluster medium (10$^{5}$ pc) whose thermodynamics it can dominate \citep{tabor93,mcnamara00,fabian03}. As feedback-driven outflows are both difficult to simulate and difficult to observe, the topic invites study \citep[reviewed in][]{veilleux05,king15,heckman17}.

\subsection{Observations of galactic outflows}

Pioneering studies have demonstrated the existence of outflows in the local universe \citep{heckman90,lehnert96} and at high redshifts $z>2$ \citep{pettini01}. Galactic outflows are understood to be `ubiquitous' in star-forming galaxies \citep{shapley03,martin05,weiner09,zhu15}, implying that normal main-sequence galaxies continuously drive winds. Observational probes for the incidence and properties of outflows extend from $z \sim 0$, including analyses of SDSS \citep{chen10} through $z \sim 0.5-1$ \citep{rubin10,rubin11} to the peak epoch of cosmic star formation at $z \sim 2-3$ in both absorption \citep{steidel10} and emission, focusing on star formation (SF)-driven outflows \citep{newman12,genzel14} and blackhole/active galactic nuclei (AGN) sources \citep{harrison16,leung17,circosta18} even up to $z \gtrsim 5$ quasars \citep[e.g.][]{bischetti18}.

Gas which makes up such a galactic-scale wind can be identified in a number of different phase tracers \citep{cicone18}: cold gas (e.g. $\lesssim 10^4$ K) including molecules such as CO \citep{fluetsch18} as well as neutral HI and metals such as NaD \citep{concas17b,bae18}. Cool gas (e.g. $\sim 10^4$ K) including metal ion tracers such as FeII, MgII and OIII \citep{martin09,rubin14}, as well as hydrogen in the form of H$\alpha$ \citep{shapiro09}. Warm gas (e.g. $\sim 10^5$ K - $10^6$ K) can also be traced by ionized metals such as OVI \citep{kacprzak15,nielsen17}. Finally, hot gas (e.g. $\gtrsim 10^6$ K) as observed at x-ray wavelengths \citep{lehnert99,strickland09}.

Except in the most local examples, observations of more than one of these outflow phases in the same system are rare. When available, such data imply that outflows are multi-phase, having several co-spatial, possibly kinematically coherent components with a large range in density and temperature \citep{heckman17}. Different phases can have significantly different inferred outflow velocities \citep{grimes09}. Typically, information is only available on one phase, so an inferred outflow rate of mass, energy, or momentum may only be a fraction of the total \citep{forsterschreiber18b}. This is particularly problematic when comparing data and theoretical models, for example of mass loading factors $\eta = \dot{M}_{\rm out} / \dot{M}_\star$ or outflow velocities, as we explore below.

Only in the past few years have we begun to spatially resolve outflows with integral field units (IFU) surveys \citep{rupke17}. Recent explorations with MUSE have demonstrated the power of blind, deep fields out to high-z \citep{finley17b,feltre18} as well as resolved detail at low-z \citep{venturi18}. 
Data from the KCWI instrument has revealed E+A galaxies hosting strong conical outflows, and shown how MgII-traced outflows can be mapped in spatially resolved emission \citep[][\textcolor{blue}{Burchett et al. in prep}]{baron18}. Studies using SINFONI have shown how AO-assisted ground-based IFU data can connect galaxy and outflow properties on resolved, kpc scales at $z \sim 2$ \citep{forsterschreiber18a,davies18,circosta18}. 

From the observational point of view, even identifying a given data feature as an outflow can be challenging. For direct `down the barrel' spectral observations of a host galaxy, two orthogonal techniques are at play: absorption and emission.
A blue-shifted absorption line can be interpreted as an intervening mass of the given gas tracer, flowing towards the observer and so away from the galaxy, blocking some of its background continuum light. While the line center offset from systemic gives an indication of the velocity of the bulk of the material, the shape (equivalent width) simultaneously encodes the full velocity distribution of the outflowing material, the covering factor of the observed phase, and the optical depth \citep{chen10}. 
Optical emission-line profiles are often asymmetric in shape, plausibly consisting of superimposed narrow and broad components, the latter preferentially blueshifted. These can be understood as a galactic-scale outflow, where obscuration by the interstellar medium (ISM) of the host galaxy hides the receding, redshifted gas \citep{armus89}. 


Studies in absorption as well as emission must make a large number of assumptions to convert observed quantities to physical values. There are enough caveats that final uncertainties are large \citep{chisholm17}. Given the breadth of techniques, definitions, galaxy selection functions, mass and redshift coverage, and physical assumptions, even fundamental scaling relations between outflow properties (e.g. velocity, mass/energy loading) and host galaxy properties (e.g. stellar mass, star formation rate, black hole luminosity) are often revised \citep{rupke05c,perna17}.

As a result, and despite the abundance of evidence for the existence of outflowing gas around galaxies, questions of its origin (launch mechanism, energy source) as well as its eventual fate (ability to escape the galaxy, dark matter halo, or to return as a recycled fountain flow) remain largely open topics. Of particular interest is our ability to discriminate winds with stellar versus black hole feedback origins. In particular, the correlations of outflow properties with the star formation and black hole activity of the galaxy itself can shed light on this dichotomy \citep{forsterschreiber18b}, ultimately providing a crucial view of, and constraint on, astrophysical feedback processes.

\subsection{Cosmological hydrodynamical simulations}

One promising avenue to study this interplay of feedback, outflows, and the subsequent cycle of baryons in and around galaxies is through cosmological hydrodynamical simulations \citep[see review in][]{somerville15}. Recent large-volume efforts have clarified the need for efficient outflows in order to make realistic galaxy populations whose characteristics and integral properties are roughly in agreement with observational constraints \citep{vog14a,genel14,crain15,schaye15,dubois16}. These simulations have started to disentangle the dominant physics driving galaxy formation, understanding not only mean relations and `typical' galaxies \citep{genel18}, but also outliers and unique examples \citep{zhu18} -- providing, as a result, interpretation for a diverse array of observations.

The IllustrisTNG project \citep{pillepich18b,naiman18,nelson18a,marinacci18,springel18} is a recent addition to this class of simulation. The first two simulations of this effort, TNG100 and TNG300, realize volumes of $\sim 100^3$ and $\sim 300^3$ comoving Mpc$^3$, respectively. This allows them to sample galaxy populations which include hundreds of thousands of objects with $M_\star \geq 10^9$\msun by $z=0$. However, the unavoidable trade off in such population-level statistical power is -- by construction -- limited resolution.

With baryon mass resolutions of $\sim$\,10$^6$\msun (corresponding roughly to the canonical `1 kpc' spatial resolution) of the current generation of cosmological volumes, we can realistically study the structure of galaxies with $M_\star \gtrsim 10^9$\msun. This hinders investigation of many interesting scientific questions, particularly those related to the structural properties of the stellar and gaseous components of galaxies, the internal structure of disks and their ISM, the co-evolution of black holes and galactic nuclei, feedback, and the resulting baryon cycle of inflows and outflows.

These small-scale phenomena are frequently investigated with `zoom' simulations of individual galaxies \citep{hop14fire}, with project campaigns of order 10 to 100 galaxies in total \citep{wang15}, often oriented around a focus such as Milky Way mass halos \citep{grand17}, Local Group analogs \citep{sawala16} or the challenging regime of galaxies in high-density cluster environments \citep{bahe17b,tremmel18}. However, zoom simulations immediately sacrifice one of the greatest advantages of cosmological simulations -- namely, the ability to make statistically robust statements about large, unbiased populations and thereby comment on galaxy formation and evolution in full generality. 

Together with the companion paper (\textcolor{blue}{Pillepich et al. 2019}) we here present the new TNG50 simulation, the third and final volume of the IllustrisTNG project. TNG50 has been designed to overcome the resolution/volume limitation by simulating a large, cosmological volume at a resolution which approaches or even exceeds that of modern `zoom' simulations of individual galaxies. It enables an insightful view into the structure, chemodynamical evolution, and small-scale properties of galaxies and their halos.

\subsection{Simulation work on outflows}

At small scales there are many idealized studies aimed at understanding the details of supernova-driven wind launching mechanisms \citep{girichidis16,fielding17b,kim17b} or detailed phase-interaction physics \citep{richings18,schneider18a}. General relativistic (GR)MHD simulations explore black hole outflow phenomena including both winds and jets \citep{blandford99,mckinney14}, recently reviewed in \cite{yuan14}.

One of the long-term goals of such efforts is to capture and parameterize the emergent behavior at larger scales in order to develop effective models which can be used in full galaxy or cosmological simulations. The sophistication of such effective models is presently limited by several factors, most critically the finite numerical resolution available. As it will be impossible to ever resolve the energy injection scales of critical astrophysical processes, sub-resolution (or sub-grid) approximations are unavoidable.

As a result, cosmological simulations have rarely if ever been compared directly to outflow observations (i.e. the baryon cycle), but rather to indirect properties such as the amount of metals left in the ISM \citep{torrey18}, the ionized metal content of the circumgalactic medium \citep{nelson18b}, or the stellar formation efficiency of the galaxy itself \citep{pillepich18b}.

Notably, \cite{oppenheimer10} studied SN wind-driving models in early cosmological simulations, emphasizing the importance of recycling and the dependence of recycled accretion on the balance between mass loading and outflow velocity. In the context of the more realistic FIRE model, \cite{muratov15} presented measurements of the scaling of mass loading (and outflow velocity) which decrease (increase) strongly with $M_\star$, such that lower-mass progenitors at early times would have had much higher mass loadings than their present day descendants. 
Similarly, \cite{christensen16} evaluated outflow scalings \citep[and metal content;][]{christensen18} over a similar mass range (halos $< 10^{12}$\msun) and zoom sample size ($\sim$10 - 20) using a delayed cooling blastwave SN feedback model, emphasizing the importance of ejective feedback (i.e. high mass loadings) towards low $M_\star$.

Non-cosmological, idealized simulations of dwarf galaxy disks enable the possibility of higher resolution and sophisticated physical modeling, and have started to explore the phase-structure and feedback dependencies of galactic outflows. \cite{emerick18} highlight how different treatments of the radiative feedback from young stars can modify the mass loading and temperature of outflows. \cite{hu18} demonstrate how different SN energy injection schemes similarly lead to different outflow properties -- in particular, that a terminal momentum injection approach does not produce the same pressure-driven wind as when residual thermal energy is also accounted for \citep[see also][]{smith18}.

None of these works have emphasized any direct observational signatures of their outflows, i.e. forwarding modeling. Recently, \cite{tescari18} made a first comparison between EAGLE `warm gas' and spatially resolved SAMI/IFU H$\alpha$ emission to relate gas kinematics to wind signatures. The notable exception is \cite{ceverino16}, who generated synthetic H$\alpha$ emission line profiles of $z \sim 2$ zoom galaxies and correlated their information content with intrinsic outflow properties. Any robust comparison between outflow observations and hydrodynamical simulations will require such progress in the future, as we discuss below.

All of these investigations have focused \textit{exclusively on SN-driven outflows}, mostly in models specifically neglecting BH feedback, which is however thought to play a dominant role in outflow production in massive galaxies. For example, the \cite{muratov15} measurement of $\eta_{\rm M}$ scalings in FIRE fails to capture the onset of the BH mechanism at $M_\star \gtrsim 10^{10.5}$\msun and therefore misses the inversion of this relation towards higher masses, as we demonstrate below. A notable exception is \cite{brennan18}, which explores a kinetic BH feedback model in a suite of 24 zooms (halos $10^{12} - 10^{13.5}$\msun), showing how black holes can produce faster outflows (up to 1000 km/s) which travel farther and are thus less likely to be subsequently recycled \citep[see also][]{nelson15a}.

The detailed properties (as opposed to the consequences) of SN and BH-driven outflows, as well as the interplay between these two origin mechanisms, have not yet been studied with cosmological simulations, nor across a full sampling of the galaxy mass range, from $M_\star = 10^7$\msun dwarfs to $M_\star = 10^{11.5}$\msun BCGs.

\subsection{The present work}

We first present the TNG50 simulation, a new class of cosmological volume simulation realized at a resolution of modern `zoom' simulations of individual galaxies. Leveraging this new tool, we describe first results focusing on the properties of galactic outflows in the cosmological setting and with respect to the galaxies from which they arise. In this context we are not trying to understand the launching mechanism or the relevant microphysics, but rather the way in which outflows shape the galaxy population as a whole, modulate galaxy evolution, and generate associated observational signatures. As emphasized above, the feedback models in simulations such as TNG are effective in nature, meaning that the outflow properties of the model \textit{at injection} are an \textit{input} rather than an output of the model. We therefore contrast, throughout this work, our results on the emergent properties of galactic outflows with the assumptions of the underlying physical models. 

TNG50 incorporates unchanged a robust scheme for cosmological magnetohydrodynamics coupled to a comprehensive and well-validated model for galaxy formation physics \citep{weinberger17,pillepich18a}. This motivates our study of the resulting outflows, as they reflect the net outcome of a model which produces a realistic galaxy population matching many key properties and scaling relations of observed galaxies (see \textcolor{blue}{Pillepich et al. (2019)} in the context of TNG50 galactic structural properties). The TNG model was designed via calibration against a number of observations, mainly at $z=0$ \citep[see][for a detailed discussion of our calibration/tuning]{pillepich18a}, and parameters related to feedback processes at the smallest scales are arguably the most important ingredient of such a model. We focus here on intermediate and high redshift, namely $z=1$, $z=2$, and above, as particularly interesting and observationally rich epochs of galaxy assembly and galactic-scale outflows.

The paper is organized as follows: in Section \ref{sec_sims} we introduce the TNG50 simulation and the TNG model for galaxy formation, while Section \ref{sec_methods_outflows} details the methodology and analysis details for our study of outflows. Section \ref{sec_results} presents our results: the high redshift galaxy population of TNG50 (Section \ref{sec_results_tng50}); measurement of mass outflow rates and mass loading factors (Section \ref{sec_results_rates}); outflow velocities and the ability of BHs to generate \mbox{$\gtrsim$\,3000 km/s} flows (Section \ref{sec_results_vel}); the multi-phase character of our outflows (Section \ref{sec_results_multiphase}); the angular dependence of winds and their hydrodynamic collimation (Section \ref{sec_results_angle}); and correlations between outflows and the properties of their host galaxy and central black hole, in comparison to observed relations (Section \ref{sec_results_vsgal}). Section \ref{sec_discussion} discusses future directions and comparison with observations, while Section \ref{sec_conclusions} summarizes our conclusions. Finally, Appendix \ref{sec:appendix} presents halo-normalized outflow velocity measurements for comparison.


\section{Methods} \label{sec_methods}

\subsection{The TNG Simulations} \label{sec_sims}

The IllustrisTNG project\footnote{\url{http://www.tng-project.org}} \citep{nelson18a, naiman18, pillepich18b, marinacci18, springel18} is a series of three large cosmological volumes, simulated with gravo-magnetohydrodynamics (MHD) and incorporating a comprehensive model for galaxy formation physics \citep{weinberger17,pillepich18a}. All aspects of the model, including parameter values and the simulation code, are described in these two methods papers and remain in practice \textit{unchanged} for our production simulations, and we give here only a brief overview.

The TNG project includes three distinct simulation volumes: TNG50, TNG100, and TNG300. The larger two have previously been presented: these are \textbf{TNG100}, which includes 2$\times$1820$^3$ resolution elements in a $\sim$\,100 Mpc (comoving) box, and \textbf{TNG300}, which includes 2$\times$2500$^3$ resolution elements in a $\sim$\,300 Mpc box. Here we present the third, final, and by far most computationally demanding simulation of the IllustrisTNG project: \textbf{TNG50}, our high-resolution effort. This run includes 2$\times$2160$^3$ resolution elements in a $\sim$\,50 Mpc (comoving) box. In TNG50 the baryon mass resolution is $8.5 \times 10^4$\msun, the gravitational softening lengths are $290$ parsecs at $z$\,=\,0 for the stars and dark matter, while the gas softening is adaptive with a minimum of $74$ comoving parsecs. TNG50 has roughly fifteen times better mass resolution, and two and a half times better spatial resolution, than TNG100. This represents an unprecedented combination of resolution and volume for a cosmological hydrodynamical simulation, which approaches or exceeds that of modern `zoom' simulations of individual galaxies. The main parameters of the three TNG volumes are compared in Table \ref{simTable}. 

TNG uses the \textsc{Arepo} code \citep{spr10} which solves for the coupled evolution under (self-)gravity and ideal, continuum MHD \citep{pakmor11,pakmor13}. Gravity employs the Tree-PM approach, and the fluid dynamics use a Godunov type finite-volume scheme where an unstructured, moving, Voronoi tessellation provides the spatial discretization. The simulations include a physical model for the most important processes relevant for the formation and evolution of galaxies. Specifically: (i) gas radiative processes, including primordial/metal-line cooling and heating from the background radiation field, (ii) star formation in the dense ISM, (iii) stellar population evolution and chemical enrichment following supernovae Ia, II, as well as AGB stars, with individual accounting for the nine elements H, He, C, N, O, Ne, Mg, Si, and Fe, plus NS-NS byproducts, (iv) supernova driven galactic-scale outflows or winds, (v) the formation, coalescence, and growth of supermassive black holes, (vi) and dual-mode BH feedback operating in a thermal `quasar' state at high accretion rates and a kinetic `wind' state at low accretion rates. Note that TNG does not yet include an explicit treatment of radiation nor cosmic rays, both of which could plausibly affect the phase structure of outflows.

Black holes are seeded in massive halos and then accrete nearby gas at the Eddington limited Bondi rate. Based on this accretion rate, their feedback mode is determined. When the Eddington ratio exceeds a threshold of $\chi = \rm{min} [ 0.002 (M_{\rm BH}/10^8 \rm{M}_\odot)^2, 0.1]$ thermal energy is injected continuously into the surrounding gas. The rate is $\Delta E_{\rm high} = \epsilon_{\rm f,high} \epsilon_{\rm r} \dot{M}_{\rm BH} c^2$ where $\epsilon_{\rm f,high}$ is the high-state coupling efficiency, $\epsilon_{\rm r}$ is the radiative accretion efficiency, and $\epsilon_{\rm f,high} \epsilon_{\rm r} = 0.02$. Below this threshold, kinetic energy is injected as a time-pulsed, oriented `wind' with a random direction which reorients for each event. The rate is $\Delta E_{\rm low} = \epsilon_{\rm f,low} \dot{M}_{\rm BH} c^2$ with $\epsilon_{\rm f,low} \le 0.2$ (its typical value at onset), the efficiency decreasing at low environmental density \citep[see][]{weinberger17}.

{\renewcommand{\arraystretch}{1.2}
\begin{table}
  \caption{Key details of the TNG50 simulation in comparison to its larger volume siblings. These are: the simulated 
           volume and box side-length (both comoving), the number of initial gas cells, dark matter particles, and Monte 
           Carlo tracers. The mean baryon and dark matter particle mass resolutions, in solar masses. The minimum allowed 
           adaptive gravitational softening length for gas cells (comoving Plummer equivalent), the redshift zero softening 
           of the collisionless components in physical parsecs, and CPU time.}
  \label{simTable}
  \begin{center}
    \begin{tabular}{lclll}
     \hline\hline
     
 Run Name                   &                 & \textbf{TNG50}         & TNG100            & TNG300 \\ \hline
 Volume                     & [\,Mpc$^3$\,]   & $\bm{51.7^3}$          & $110.7^3$         & $302.6^3$ \\
 $L_{\rm box}$              & [\,Mpc/$h$\,]   & $\bm{35}$              & 75                & 205 \\
 $N_{\rm GAS}$              & -               & $\bm{2160^3}$          & $1820^3$          & $2500^3$ \\
 $N_{\rm DM}$               & -               & $\bm{2160^3}$          & $1820^3$          & $2500^3$ \\
 $N_{\rm TR}$           & -               & $\bm{2160^3}$          & $2 \times 1820^3$ & $2500^3$ \\
 $m_{\rm baryon}$           & [\,M$_\odot$\,] & $\bm{8.5 \times 10^4}$ & $1.4 \times 10^6$ & $1.1 \times 10^7$ \\
 $m_{\rm DM}$               & [\,M$_\odot$\,] & $\bm{4.5 \times 10^5}$ & $7.5 \times 10^6$ & $5.9 \times 10^7$ \\
 $\epsilon_{\rm gas,min}$   & [\,pc\,]        & $\bm{74}$              & 185               & 370 \\
 $\epsilon_{\rm DM,stars}$  & [\,pc\,]        & $\bm{288}$             & 740               & 1480 \\
 CPU Time                   & [\,Mh\,]        & $\bm{130}$             & 18                & 35 \\
 \hline
 
    \end{tabular}
  \end{center}
\end{table}}

Galactic-scale outflows generated by stellar feedback are modeled using a kinetic wind approach, whereby the available energy from SNII is used to stochastically eject star-forming gas cells from galaxies. The injection velocity of such wind particles is $v_{\rm w} \propto \sigma_{\rm DM}$ where $\sigma_{\rm DM}$ is the local dark matter velocity dispersion, subject to a minimum $v_{\rm w,min} = 350$ km/s. The mass loading of the winds is then $\eta = \dot{M}_{\rm w} / \dot{M}_{\rm SFR} = 2 (1 - \tau_{\rm w}) e_{\rm w} / v_{\rm w}^2$ where $\tau_{\rm w} = 0.1$ is the thermal energy fraction and $e_{\rm w}$ is a metallicity dependent modulation of the canonical $10^{51}$ erg available per SNII. Wind particles are hydrodynamically decoupled from surrounding gas until they exit the dense, star-forming environment. This occurs when the wind reaches a background density lower than 0.05 times the star-formation threshold, or after 0.025 of the current Hubble time elapses, and typically within a few kiloparsecs. The total energy available to drive winds from a gas cell therefore depends on its instantaneous star formation rate as $\dot{E}_{\rm w} = e_{\rm w} \dot{M}_{\rm SFR}$ which is roughly $\simeq 10^{41} - 10^{42}$ erg/s $\dot{M}_{\rm SFR} / (M_\odot / \rm{yr})$ depending on the local gas metallicity \citep[see][]{pillepich18a}.

\begin{figure}
\centering
\includegraphics[angle=0,width=3.3in]{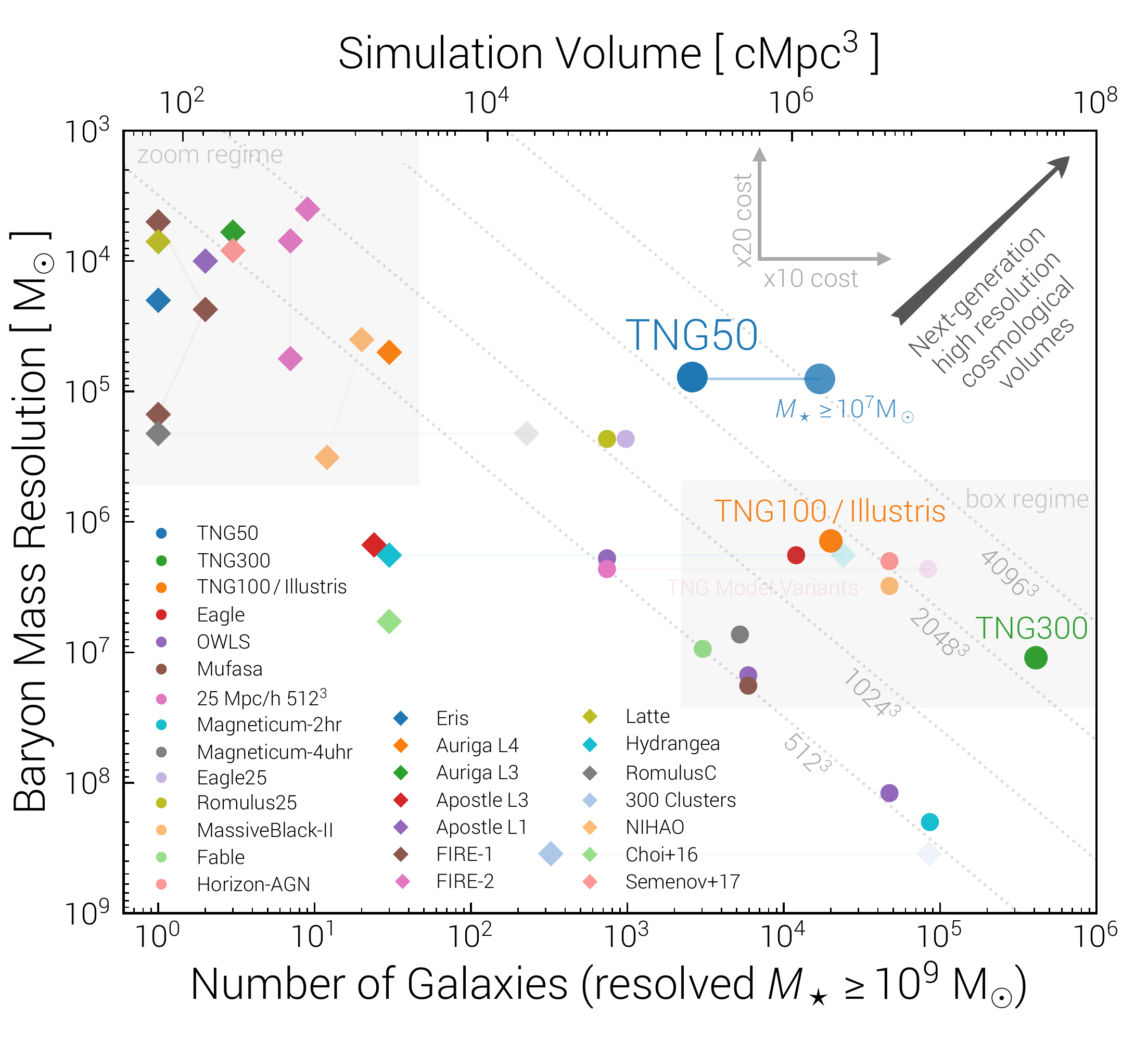}
\caption{ The TNG50 simulation occupies a unique region of parameter space for typical cosmological hydrodynamical simulations. TNG50 includes 2$\times$2160$^3$ resolution elements, giving a baryon mass resolution of \mbox{$8.5 \times 10^4$\msun} with adaptive gas softening down to $74$ comoving parsecs. This approaches or exceeds that of modern `zoom' simulations of individual galaxies, while maintaining the statistical power and unbiased sampling of the full $\sim$\,50 cMpc cosmological volume. Here we show TNG50 (dark blue) in comparison to other cosmological volumes (circles) and zoom simulation suites (diamonds) at $z \sim 0$, based on the total number of resolved galaxies (i.e. at least 100 star/gas particles) with $M_\star \geq 10^9$\msun. For TNG50 we also indicate $N_{\rm gal}$ above $10^7$\msun (rightmost blue circle), which are still resolved at this level -- as in many zooms, but in contrast to other large-volume simulations. The computational difficulty of pushing towards the upper right represents the frontier for next-generation galaxy formation simulations.
 \label{fig_sim_comparison}}
\end{figure}

In Figure \ref{fig_sim_comparison} we offer a comparison between recent cosmological hydrodynamical simulations of galaxy formation, putting TNG50 in context given its combination of volume and resolution, which uniquely positions it in this space. TNG50 is by far the most computationally demanding run of the TNG suite, and one of the most ambitious cosmological simulations to date.

We adopt a cosmology consistent with the \cite{planck2015_xiii} results, namely $\Omega_{\Lambda,0}=0.6911$, $\Omega_{m,0}=0.3089$, $\Omega_{b,0}=0.0486$, $\sigma_8=0.8159$, $n_s=0.9667$ and $h=0.6774$. 

\subsection{Model update for TNG50} \label{sec_sims_tng50}

To execute the TNG50-1 (highest resolution level of the TNG50 volume) simulation a minor change has been made to the fiducial TNG model. With the inclusion of MHD, we found that extremely dense gas cells often have $P_{\rm B} \gg P_{\rm gas}$. The resulting Courant-like constraint on the hydrodynamical timestep, restricted by the magnetic signal velocity instead of the sound speed, requires physical timesteps as small as 10 \textit{years} at the resolution of TNG50-1. Such high-density gas should naturally convert rapidly into stars, but in the fiducial configuration the numerical timestep decreases faster than the star formation probability increases (with increasing gas density), leading to an unfavorable race condition. We instead want that gas cells with extremely short star formation timescales convert into collisionless star particles in a reasonable number of numerical timesteps. We have therefore modified the base \cite{spr03} star formation model, wherein the dependence of the star formation timescale on density $t_\star(\rho)$ is given by (Eqn. 21)

\begin{equation}
t_\star(\rho) = t_0^\star \left( \frac{\rho}{\rho_{\rm th}} \right)^{-\alpha}
\end{equation}

\noindent where $t_0^\star$ is a model parameter which modulates the global gas consumption time-scale, and $\rho_{\rm th}$ is a model parameter setting the threshold density for star formation. In the fiducial configuration, used for all simulations employing this model, the choice of $\alpha = 1/2$ has been made, corresponding to the canonical assumption that $t_\star$ is proportional to the local dynamical time of the gas.

Instead, for TNG50-1, we postulate a somewhat steeper dependence of the star formation rate on gas density \textit{only} for the densest gas. We take $\alpha = 1$ for gas above the runaway threshold, which is defined as the density where the slope of the effective pressure versus density curve $n_{\rm eff} < 4/3$. At this point the effective pressure can no longer support gas against dynamical instabilities, and this occurs at \mbox{24.4 cm$^{-3}$}, which is $\sim$ 230 $\rho_{\rm th}$. Since the star formation rate goes as $\dot{M}_\star \propto \rho^\alpha$, this implies the physical ansatz of more rapid star formation in the densest environments -- the nuclear starburst regime of galactic disks, for instance.

In practice, this change was taken for numerical reasons, and we have verified with test simulations at full resolution ($1024^3$ particles in 25 Mpc/h volumes) that this change affects a negligible amount of gas, by mass or by number, and has no impact on galactic structure nor galaxy population statistics, at any redshift.

\subsection{Measuring Outflow Rates and Velocities} \label{sec_methods_outflows}

To compute mass outflow rates two orthogonal techniques exist: (i) deriving instantaneous fluxes based on gas kinematics at one point in time \citep[e.g.][]{ocvirk08}, and (ii) deriving mass fluxes by tracking the Lagrangian evolution of gas mass across two distinct points in time \citep[e.g.][]{nelson13}. As only the former has the possibility for direct comparisons to observables, we here focus on these instantaneous rates. We compute the radial mass flux of all the gas cells in a thin shell centered on a galaxy as

\begin{equation}
\dot{M} = \left.\frac{ \partial M}{ \partial t }\right\rvert_{\rm \,rad} = \frac{1}{\Delta r} \sum_{\overset{i=0}{|r_i-r_0| < \Delta r/2}}^{N} \left( \frac{ \vec{v_i} \cdot \vec{r_i} }{ |r_i| } m_{i} \right)
\end{equation} 

\noindent where the subscript $i$ enumerates all the gas cells with mass $m_i$ in a particular volume of space, taken as a spherical shell with some thickness $\Delta r$ and mean distance $r_0$ from a central galaxy. We calculate $\dot{M}$ for a range of $r_0$ and corresponding $\Delta r$ values, 5 kpc thick in the inner halo coarsening outwards. The cell position $\vec{r_i}$ is relative to the subhalo center, taken as the position of the particle/cell with the minimum gravitational potential, while the velocity $\vec{v_i}$ is relative to the subhalo center of mass motion, accounting for local Hubble expansion. $v_{\rm rad} > 0$ denotes outflow, and $v_{\rm rad} < 0$ is inflow.


Herein we frequently decompose the outflowing gas by its instantaneous properties, whereby $\dot{M}$ has parameter dependencies:

\begin{equation}
\dot{M} = \dot{M} \left( z, M_\star; r, v_{\rm rad}, \rho, T, Z \right).
\end{equation}

\noindent To compute this distribution for each galaxy (with a given redshift and stellar mass) we calculate the radial mass flux of each gas cell and take the 5D histogram of this mass flux, or mass outflow rate, binning in the gas properties of interest above: radius, radial velocity, density, temperature, and metallicity. A total mass outflow rate is derived by marginalizing over the state of the gas (i.e. $\rho, T, Z$) and satisfying a threshold in outflow velocity of $v_{\rm rad} > v_{\rm thresh}$. We frequently require only $v_{\rm rad} > 0$ km/s to define outflowing material, though higher values are also considered as conservative limits and/or to mimic observational sensitivity limits.

In measuring total mass outflow rates, mass loading factors, or outflow velocities, we consider actual gas cells as well as wind-phase particles, the latter of which are generated by the TNG wind model \citep[see][]{pillepich18a} and are hydrodynamically (though not gravitationally) decoupled from their surrounding gas for the duration of their existence, which is typically restricted to small distances away from galaxies. At $r \gtrsim 10$ kpc, where we generally focus our analysis, wind-phase particles are a negligible contribution and outflows are hydrodynamically resolved. Excluding wind particles has only a minor effect on all our measurements, and we include them as a relevant component of our simulations, one currently modeled at a particularly simple level (i.e. with strong sub-grid assumptions about phase structure and interaction). 

To explore trends of outflow velocity, we must reduce the complex distribution of outflow velocities around a galaxy down to a single number $v_{\rm out}$. We do so by taking mass outflow rate weighted percentiles of $v_{\rm rad}$. In particular, we define the quantity $v_{\rm out,N}$ as the radial velocity above which $(1-N)$\% of the outflow is moving. For example, \mbox{$v_{\rm out,90} = 500$ km/s} implies that ten percent of outflowing mass flux is moving 500 km/s or faster. The `median' value $v_{\rm out,50}$ therefore gives the speed achieved by at least half of the total outflow mass flux. This quantity can be measured at a particular radial distance from the galaxy, $v_{\rm out,50,r=10 kpc}$ for example.

We note that these theoretical velocity percentiles may be similar, but not directly comparable to, observational measurements of outflow velocity (e.g. `$v_{\rm 05}$' or `$v_{\rm 90}$' in the literature), because we consider 3D $v_{\rm rad}$ instead of 1D projected $v_{\rm LOS}$, our velocities are mass flux (not optical depth or emission) weighted, and we do not yet attempt any modeling of synthetic spectral features.

We define and measure a mass loading factor with respect to the star formation rate of the galaxy in the usual way, 

\begin{equation} \label{eqn_eta_M_SN}
\eta_{\rm M}^{\rm SN} = \frac{ \fracspace \dot{M}_{\rm out} }{ \fracspace \dot{M}_\star } .
\end{equation}

\noindent This measure is typically used in the context of supernovae-driven winds, to diagnose the relative efficiency of stellar feedback produced galactic-scale outflows. It has the same dependencies as the mass outflow rate, namely $\eta_{\rm M} = \eta_{\rm M} \left( z, M_\star; r, v_{\rm rad}, \rho, T, Z \right)$.

We typically drop the superscript `SN', which emphasizes that the normalization (denominator) assumes that stellar feedback is the relevant energetic source. As BH energy injection is typically coexistent and always present, we always interpret its possible role in setting $\eta_{\rm M}$ or contributing to any measured $\dot{M}_{\rm out}$.

\subsection{Galaxy Identification and General Properties}

Subhalos are identified with the \textsc{Subfind} algorithm \citep{spr01}, and we exclusively consider central (non-satellite) galaxies with $7 < \log(M_\star / \rm{M}_\odot) < 11$ at redshifts $z \geq 1$.

Throughout this work we take 30 physical kpc aperture values for the stellar mass $M_\star$ of a galaxy. We compute star formation rates $\dot{M}_\star$ within the same aperture and with a temporal smoothing of 100 Myr, as appropriate for observationally derived SFR indicators based on the infrared or 24$\mu m$ measurements \citep{calzetti08}, using an accounting of recently formed stars, rather than the instantaneous star formation rate of dense gas.

We compute the bolometric luminosities of accreting black holes under the assumption of a $\dot{M}$-dependent radiative efficiency \citep[following][]{churazov05,weinberger18} as

\begin{equation}
L_{\rm bol} \, = \,
\left\{\!\begin{aligned}
  \, &\epsilon_r\, \dot{M} \, c^2 \quad &; \quad \lambda_{\rm edd} \ge 0.1\\[1ex]
  \, &10\, \lambda_{\rm edd}^2\, L_{\rm Edd} \quad &; \quad \lambda_{\rm edd} < 0.1
\end{aligned}\right\} ,
\end{equation}

\noindent where $\lambda_{\rm edd} = \dot{M} / \dot{M}_{\rm edd}$ is the Eddington ratio, \mbox{$L_{\rm edd} = \dot{M}_{\rm edd} c^2$} is the Eddington luminosity, and \mbox{$\dot{M}_{\rm edd} = 4 \pi G M_\text{BH} m_p / (\epsilon_r \sigma_{\rm T} c)$} the Eddington accretion rate, with the usual fundamental constants. $L_{\rm bol}$ is an intrinsic luminosity -- obscuration is not included.


\section{Results} \label{sec_results}


\subsection{TNG50: Resolving outflows from resolved galaxies} \label{sec_results_tng50}

\begin{figure*}
\centering
\includegraphics[angle=0,width=6.8in]{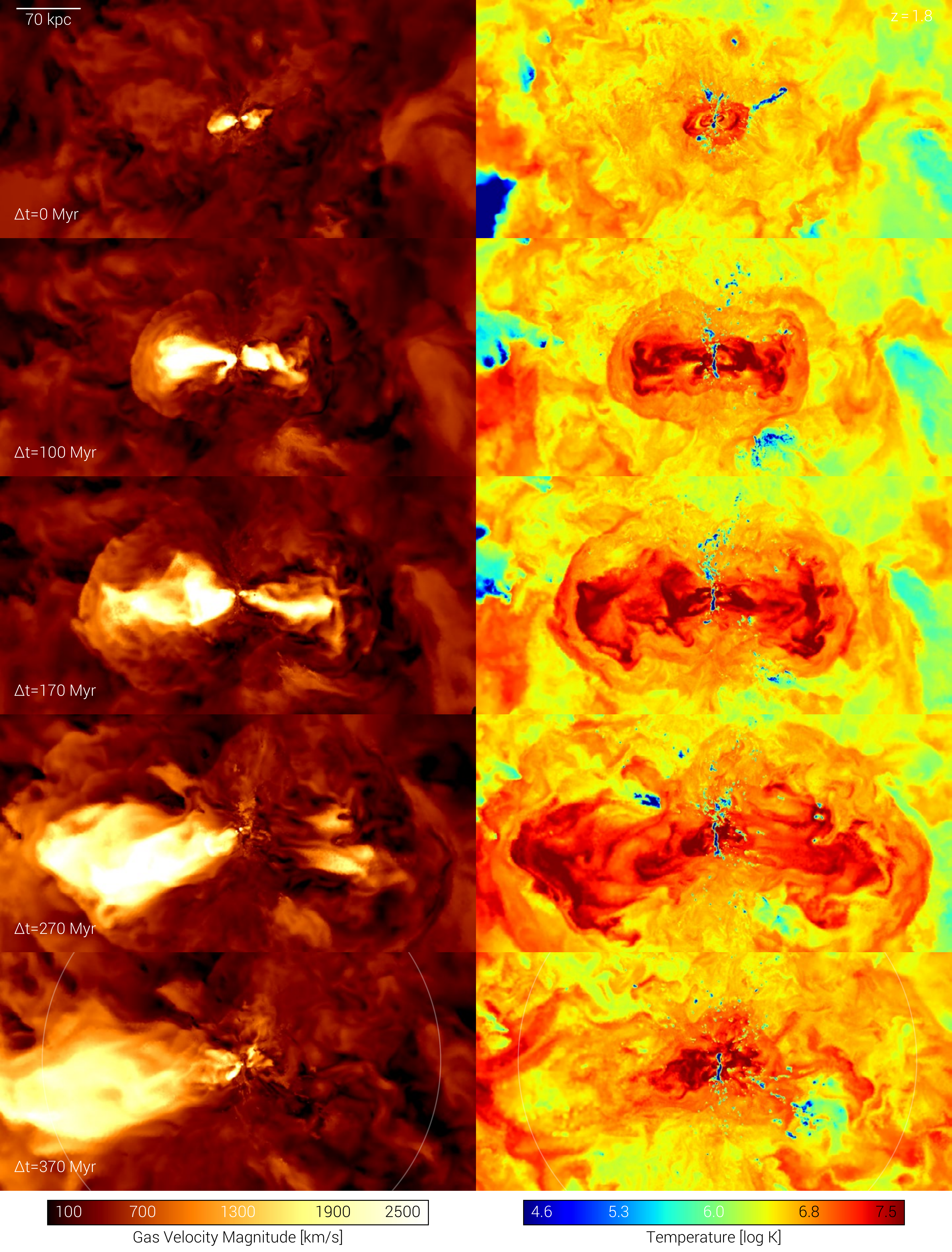}
\caption{ Visualization of black hole feedback driving a large-scale galactic outflow in TNG50. We show the time evolution with five snapshots spanning 370 Myr starting from $z=1.8$ (rows, from top to bottom), tracking a single massive galaxy with $M_\star \simeq 10^{11.4}$\msun which is currently in the process of quenching. The left column shows the gas velocity field, while the right shows gas temperature, on the scale of the virial radius (white circles, final row). Each panel is 550 kpc x 275 kpc, with a thin projection depth of 10 kpc. The dense ISM of the galaxy itself is oriented vertically, edge-on, visible as the blue disk in temperature. The central black hole with a mass of $10^{8.7}$\msun is in the low-accretion state and its kinetic feedback drives a large-scale collimated outflow.
 \label{fig_timeevo1}}
\end{figure*}

\begin{figure*}
\centering
\includegraphics[angle=0,width=6.8in]{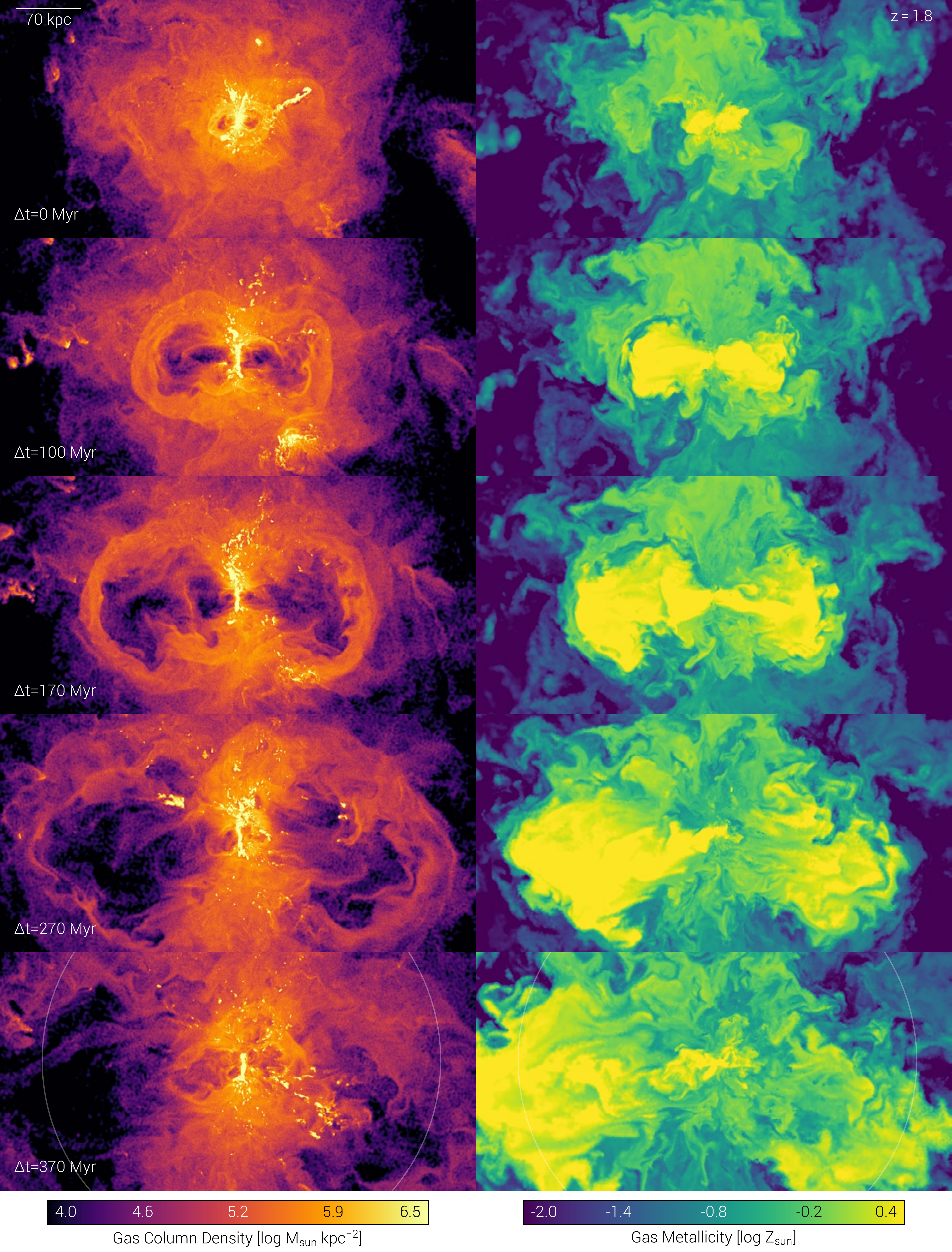}
\caption{ The same time evolution series of a $z \sim 2$ black hole driven outflow as in Figure \ref{fig_timeevo1}, where we show gas column density (left) and gas metallicity (right). The highly directional, jet-like flow drives shells of gas which pile up along an expanding front, producing under-dense cavities in their wake. These cavities are hot, over-pressurized, and expand coherently to the scale of $r_{\rm vir}$; they are able to launch high metallicity ISM gas entirely out of the halo.
 \label{fig_timeevo2}}
\end{figure*}

The fundamental means by which feedback regulates galaxy formation and evolution is by generating non-gravitational motions in the gas. Energy injection gives rise to \textit{outflows}, the subject of this paper, which are particularly evident in the velocity structure of the cosmic gas surrounding galaxies. To set the stage, Figure \ref{fig_timeevo1} therefore visualizes the time evolution of a strong BH-driven outflow in TNG50 originating from a massive $M_\star \simeq 10^{11.4}$\msun galaxy at $z \sim 2$. We show gas velocity (left column) and gas temperature (right column), with time progressing downwards, each row roughly 100 Myr apart. Energy injection from the BH produces a high-velocity, large-scale, and highly collimated (directional) outflow. The gas reaches velocities $\gtrsim$ 2500 km/s even as it crosses the halo virial radius (shown as the circle in the last row).

This halo-scale outflow results from numerous, high time cadence energy inputs within a spatial region of $\lesssim 150$ pc: in the center of the upper right panel, five distinct events are already visible as loop-like features in temperature which have begun to flow outwards. The most recent (smallest) has already reached a distance of $\sim 4$ kpc from the SMBH, which has a mass of $10^{8.7}$\msun and is accreting gas at a rate of \mbox{$\dot{M}_{\rm BH} \simeq 10^{-2}$\msun\,yr$^{-1}$}, such that it is in the `low accretion state' of the TNG model. It has an Eddington ratio of $\lambda_{\rm edd} \simeq 0.002$ and a bolometric luminosity of only $L_{\rm bol} \simeq 10^{42.3}$ erg/s \citep[cf. the system of][]{may18}. This initial `burst' of energy inputs is followed by continual, though less frequent, feedback from the BH over the next $\sim$500 Myr. The central galaxy is the eighth most massive at this time in TNG50, with a SFR of $\simeq$ 50\,\msun\,yr$^{-1}$, log(sSFR) $\simeq$ -9.7 yr$^{-1}$, and is in the process of quenching. 

We see that the outflow heats two centrally offset lobes to temperatures $\gtrsim 10^{7.5} - 10^8$ K. A succession of cocoon like shells pile up as they encounter resistance propagating through the halo gas. Although not shown here, energy dissipation due to hydrodynamical shocks traces the outer envelopes (i.e. density edges) of individual bubbles. The outflow geometry is more jet-like in its collimation than bipolar, being quite linear along the horizontal direction. The overall morphology is reminiscent of early simulations by \cite{suchkov94} of superwind structure, as well as radio observations of \cite{cecil01} who studied the coherence of the magnetic field inside similarly trailing lobes. That work suggested vortex-like dynamics at the edge of the bubble could loft material outwards, while inversion of the observed rotation measure across the boundary implied a separation of outflow and infall regions -- signatures in the magnetic field topology which could be explored in TNG50.

The occasional tilt with respect to the host rotation axis (as seen in the first and last rows) could, if the outflow origin was stellar in nature, reflect asymmetries in the distribution of rapidly star-forming gas and young stars. Here it is instead due to a large momentum injection by the BH in a random direction \citep[e.g.][]{schmitt02,liska18} which then propagates along a preferred direction. We discuss this natural collimation in Section \ref{sec_results_angle}.

The structure is not bi-symmetric as is the case in idealized jet/wind models, and asymmetries of the outflow at larger, halo scales are expected in this more realistic setting \citep{nelson16}. In particular, the left lobe is more pronounced, sustaining higher velocities, than the right lobe in the last two time snapshots. This is plausibly due to density fluctuations in the halo gas of the host galaxy. In particular, we see how an intruding satellite at $\Delta t = 270$ Myr is shredded as it passes through pericenter, populating the inner halo to the right of the central disk with dense gas structures which likely help inhibit outflow propagation in that direction.

Figure \ref{fig_timeevo2} shows the same time sequence, this time in projected gas column density (left) and gas-phase metallicity (right). At the final $\Delta t = 370$ Myr the remnants of this merging satellite are clearly visible as a number of dense gas fragments flowing outwards -- a feature which could easily be mistaken as a feedback-driven outflow. The most striking feature in the density projections is the bipolar cavity structure, which inflates from a few kpc to the scale of the virial radius over the span of a few hundred Myr. Along the edges of each cavity accumulated shells of swept up and ejected mass are clearly visible as strong density contrasts relative to the under dense, post-shock regions in their wakes. Even without further energy injection from the central engine these highly over pressurized bubbles will expand outwards as they dissipate their energy into the halo gas over time.

The evolving gas metallicity distribution over this time span directly shows how galactic-scale outflows enrich the circumgalactic and intergalactic medium with heavy elements \citep{tegmark93}. Although a considerable fraction of the virial volume is already enriched to $0.1 Z_\odot$ by $z \sim 2$, prior to the onset of strong BH activity, the distribution of metals is inhomogeneous and regions of the halo and its immediate environment remain at $\lesssim 0.01 Z_\odot$. Sitting already on the mass-metallicity relation, the central galaxy is, however, highly enriched, and this $\gtrsim Z_\odot$ gas is directly ejected out to scales of 100s of kpc by the BH-driven wind. This high metallicity material is preferentially cospatial with hot gas in the cavities, particularly at early times. A rich turbulent mixing structure develops, producing metallicity sub-structure within the bubbles and possibly across their boundaries. We discuss the multiphase nature and metal content of the outflows in Section \ref{sec_results_multiphase}.

Because the outflow is not isotropic, it allows predominantly metal-poor gas to remain along the major axes of the disk (i.e. in the vertical direction of this projection). Any ongoing accretion will lead to metal dilution over time. Evidence for this fueling of the gas reservoir of the central galaxy, particularly near the end of this strong feedback episode, is also apparent. Cold, overdense gas fragments begin to populate the inner halo at $r \lesssim 0.25 r_{\rm vir}$. They originate predominantly along the major axis of the galaxy, though also appear to form directly in the wake of the outflow. These gas overdensities are clearly infalling, with mixing tails extending radially outward, and often not obviously associated with any substructure in the halo at earlier times. We speculate that they may arise from an in-situ, condensation-like process in the thermally unstable hot halo gas \citep{sharma12,mccourt12,voit15}. Given the ability of TNG50 to resolve detailed structure in the lower density halo gas, a detailed study of precipitation and its ability to feed the central BH and/or regulate the feedback cycle will be an interesting topic for future work.

\begin{figure*}
\centering
\includegraphics[angle=0,width=6.8in]{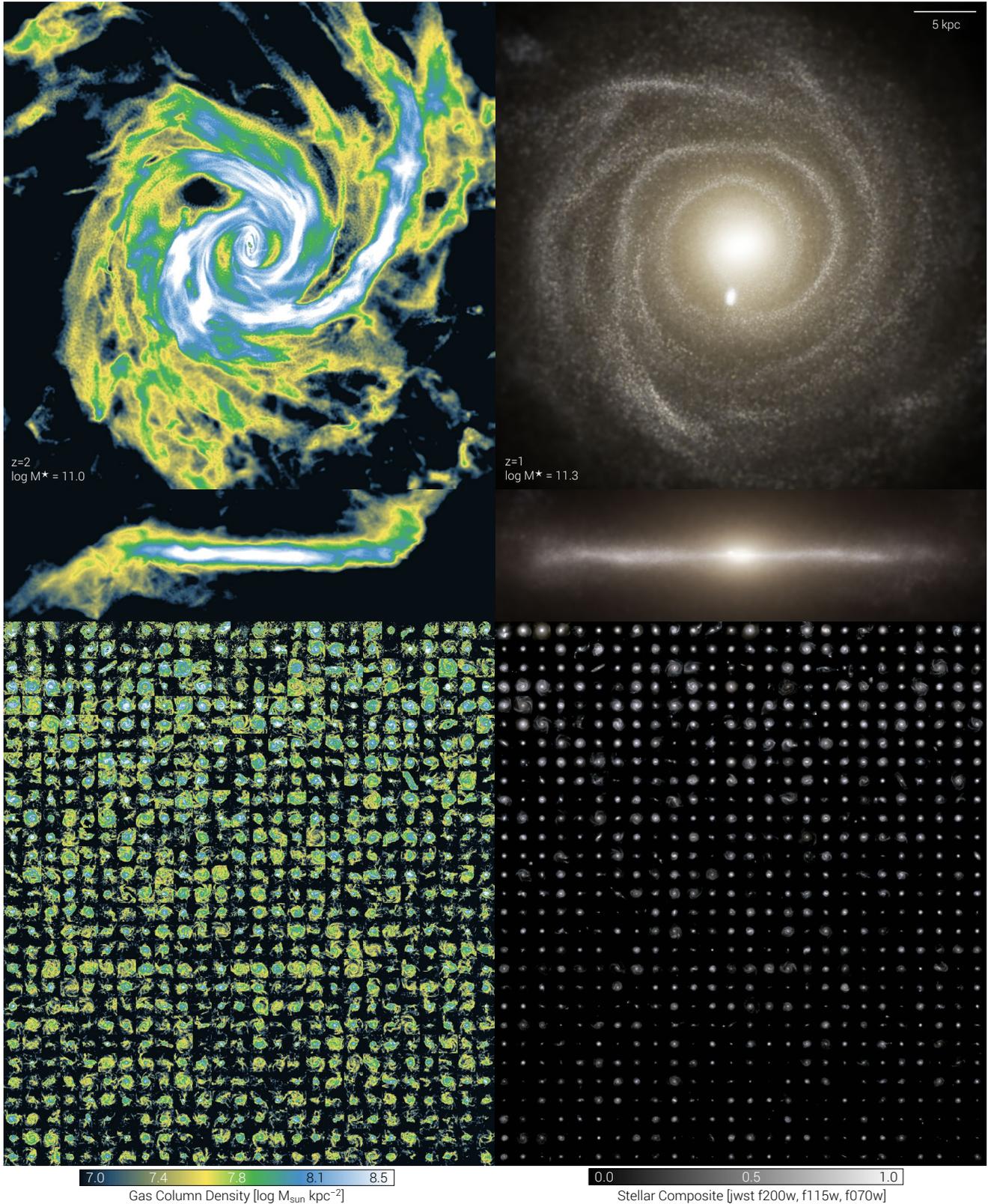}
\caption{ The versatility and power of TNG50 in combining high-resolution (top) with the unbiased statistics of a representative, cosmological galaxy population (bottom). Here we show gas density projections of galaxy disks (left), and mock stellar light images of the same galaxies in JWST \{F200W, F115W, F070W\} rest-frame bands (right). \textbf{(Top)} Two examples of large, well-resolved disks in face-on and edge-on projections. \textbf{(Bottom)} The most massive 754 central galaxies at $z=2$, in descending stellar mass (from left to right, top to bottom), encompassing \mbox{$10^{10.9} < M_{\rm halo}/$\msun$ < 10^{13.4}$}, and \mbox{$10^{8.0} < M_{\star}/$\msun$ < 10^{11.7}$}. All are shown face-on, and each stamp is 40 (30, 20) physical kpc across for $\log(M_{\rm halo})/$\msun$ \le$ 13.5 (12.5, 12.1). At this mass scale and redshift, most galaxies host a rotationally supported, gas-rich, rapidly star-forming disk -- with a black hole at the center -- and will drive extended, gaseous outflows. 
 \label{fig_mosaic}}
\end{figure*}

Sitting at the center of this outflow, nearly invisible at this scale, is the stellar body and gaseous reservoir of the central galaxy. Figure \ref{fig_mosaic} shows one such massive galaxy at $z \sim 2$ in a face-on gas density projection (upper left), and at later times an example of the stellar structure of a $z \sim 1$ descendant of such a system, traced through a mock of its optical stellar light (upper right). In both cases the edge-on view is shown immediately below. The redshift two system hosts a large, flocculent, clumpy spiral structure, roughly $10 - 20$ kpc in radial extent. This disk is relatively thin in its vertical scale-height at the center, but warped at the outskirts, with gas features coupling the outer disk to ongoing accretion from its larger scale environment. The line-of-sight velocity dispersion of the star-forming (i.e. H$\alpha$-traced) component varies across the disk; between $1-2 r_{\rm 1/2,\star}$ the mean value is $\sim 70$ km/s, reaching as high as $170$ km/s -- see \textcolor{blue}{Pillepich et al. (2019)}. The maximum rotational velocity is $\sim 350$ km/s, giving $V/\sigma \sim 2-5$ depending on measurement methodology. The redshift one descendant hosts a large stellar bulge surrounded by a thin stellar disk composed of irregular spiral-like features. A stellar clump is visible migrating inwards to eventually coalesce with the bulge. Bombardment by minor mergers at $z < 1$ will convert this galaxy into the spheroidal BCG of a massive group by $z=0$.

\begin{figure*}
\centering
\includegraphics[angle=0,width=6.0in]{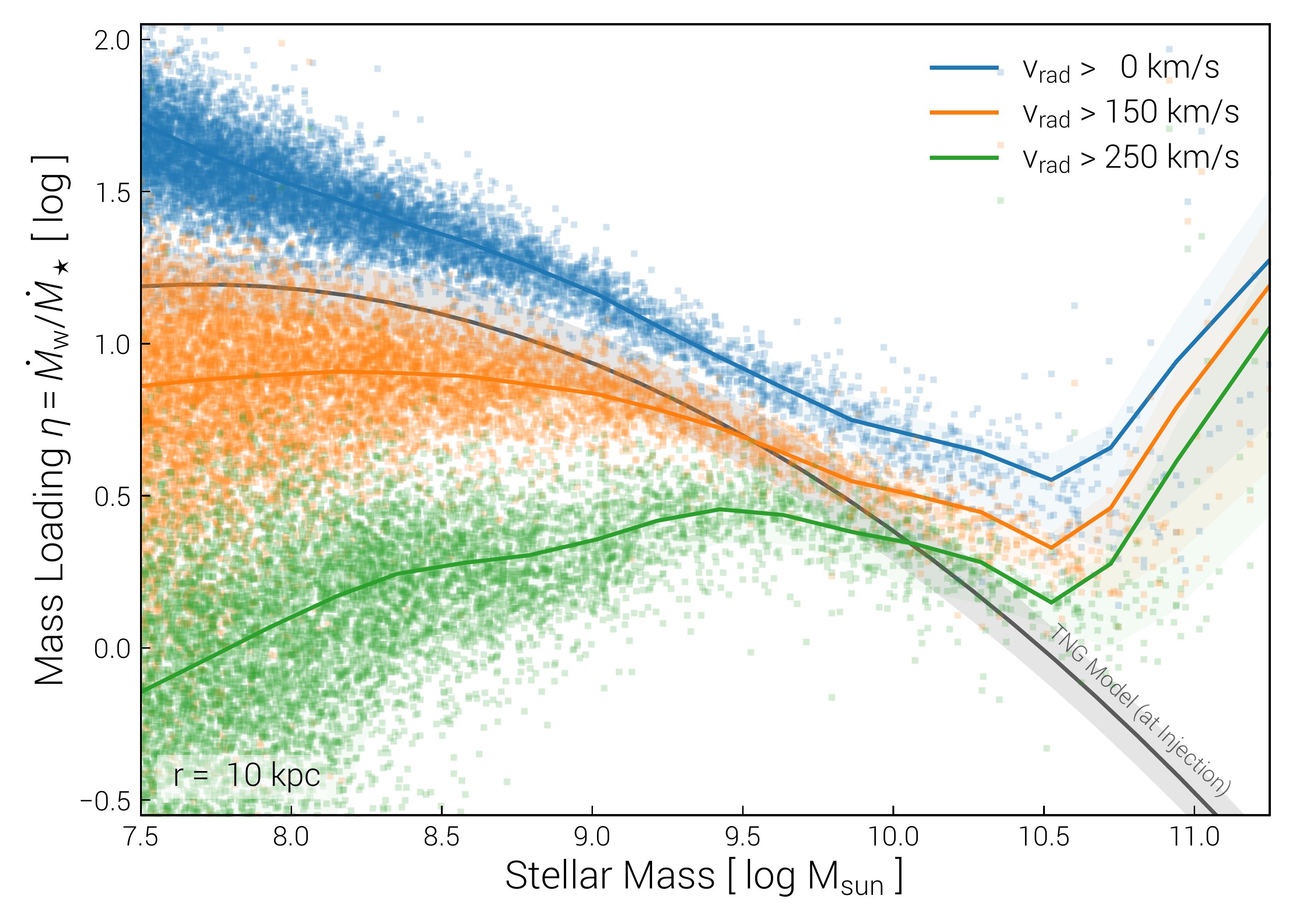}
\includegraphics[angle=0,width=3.4in]{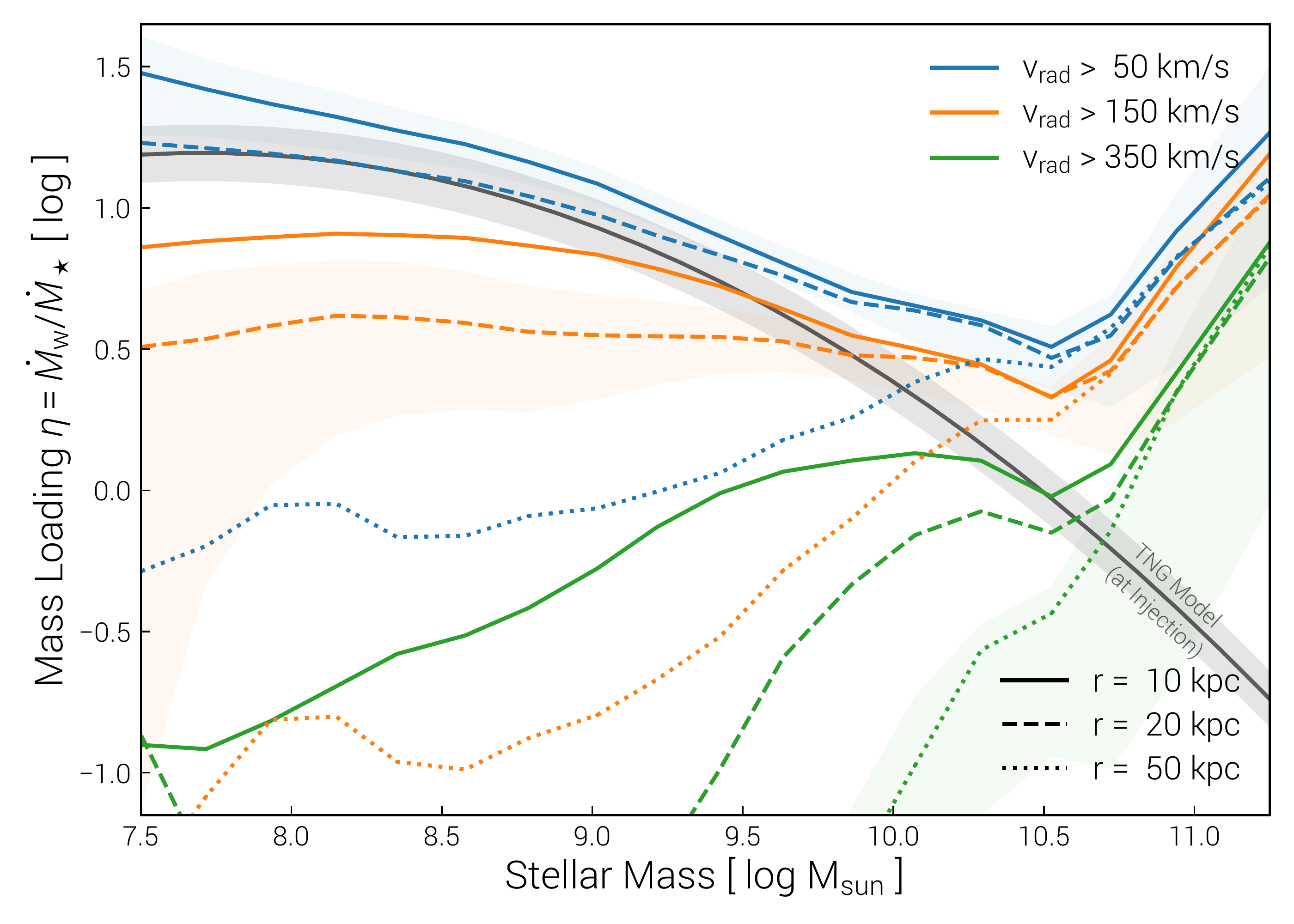}
\includegraphics[angle=0,width=3.4in]{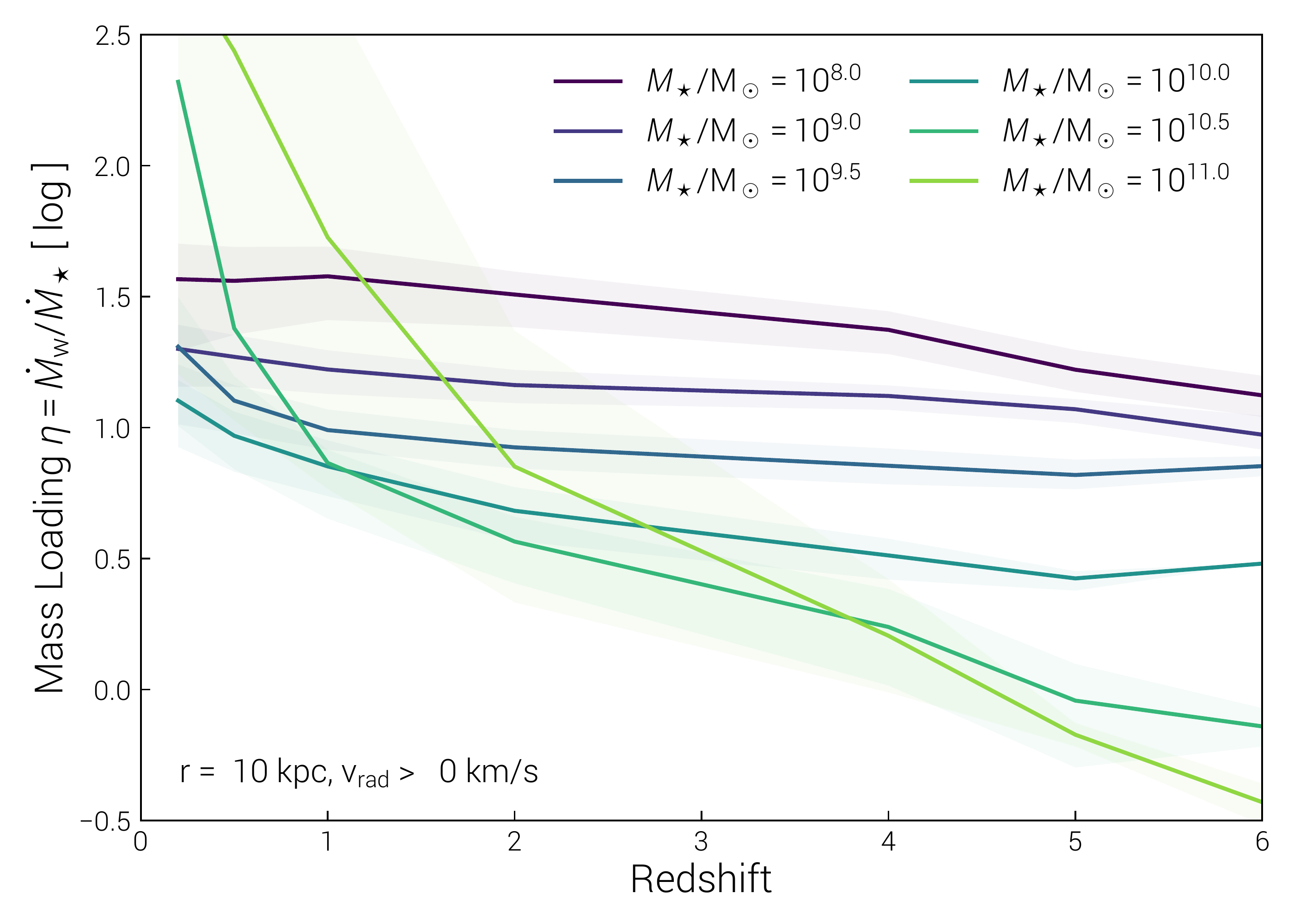}
\caption{ Mass loading factor as a function of stellar mass at $z=2$. \textbf{(Top)} For three different values of the minimum speed \vrad to be classified as an outflow (blue, orange, and green) the mass loading factor \etaM($M_\star$) at a distance of 10 kpc from each galaxy. Lines show median relations, shaded areas 16-84 percentile intervals, and small squares individual systems. Requiring only \vrad$ > 0$\,km/s, \etaM decreases sharply with stellar mass from values as high as 20-50 at $10^8$\msun to 1-5 at $10^{10.5}$\msun. The black line denotes the TNG model input of \etaM at the injection scale, derived as the mean across all star-forming gas cells in each galaxy \protect\citep[as in Fig. 7 of][]{pillepich18a}, which is monotonically decreasing with mass. This is not true of the emergent 10 kpc-scale outflows, where the trend of \etaM reverses its decline and begins to rise rapidly for $M_\star \gtrsim 10^{10.5}$\msun as highly efficient BH feedback begins to drive strong outflows at low-SFR. \textbf{(Bottom left)} The sensitivity of \etaM to both distance from the galaxy (three linestyles) and the minimum \vrad cut to be classified as outflowing (three colors). \textbf{(Bottom right)} Redshift evolution of the mass loading factor \etaM in six different bins of stellar mass (different colors). Although the mass outflow rates decrease towards $z=0$, star formation rates decrease faster, leading to roughly flat $\eta_{\rm M}(z)$ curves for $M_\star \lesssim 10^{10}$ \msun, while the onset of quenching produces rapidly rising \etaM values for massive galaxies towards late times.
 \label{fig_massloading1}}
\end{figure*}

By showing the highly resolved structure of an individual galaxy we emphasize that TNG50 enables the study of the connection between small-scale (i.e. few hundred pc) feedback and large-scale (i.e. few hundred kpc) outflows. Furthermore, we clarify that winds launched by stellar feedback originate from the entirety of the star-forming gaseous disk, on a cell-by-cell basis at the scale of the simulation resolution. As the local star formation properties vary substantially within a galaxy, the resulting SN-driven outflows have non-constant, and non-trivial, distributions of outflow velocity, mass outflow rate, and even temperature (all at launch). That is, galaxies of a given mass or circular velocity do not produce outflows at a single constant velocity or mass loading, as is sometimes mistakenly inferred for large-volume feedback models like TNG. Simultaneously in time and possibly coincident in space, the central black hole will also be injecting either thermal or kinetic energy into its immediate surroundings -- contributing to, or possibly dominating the production of an outflow.

Figure \ref{fig_mosaic} also highlights the statistical power of the galaxy population realized in TNG50 already by $z=2$ (lower half), visualizing the most massive $\sim$ 750 galaxies present in the volume, in the same face-on gas density (left) and face-on stellar light mocks (right) as above. They are sorted from the most massive, in the upper left corner, to the least massive, in the lower right, ranging from \mbox{$8.0 < \log (M_{\star}/\rm{M}_\odot) < 11.7$}. With few exceptions, galaxies at this mass and redshift are gas rich, with rotationally supported gas reservoirs which are typically much more extended than the stellar component. The diversity present even within this sample highlights the importance of the large cosmological volume. The \mbox{$\sim 1.4 \times 10^5$ cMpc$^{3}$} of TNG50 provides a sampling of rare, high-mass halos, together with the statistics needed to make robust statements about a representative galaxy population.


\subsection{Outflow rates and mass loading factors vs. M\texorpdfstring{$_\star$}{*} and z} \label{sec_results_rates}

In Figure \ref{fig_massloading1} we begin to explore the quantitative properties of TNG50 outflows. Here we measure the outflow mass loading \etaM, the ratio of the mass outflow rate and the galaxy star formation rate, as a function of galaxy stellar mass at $z=2$ (top panel). Considering a fixed distance of 10 kpc from the galaxy, we evaluate the mass outflow rates subject to three different minimum \vrad values to be classified as an outflow (blue, orange, and green).

Considering all $v_{\rm rad} > 0$ km/s material in the mass outflow rate (blue), \etaM drops rapidly with increasing stellar mass, from $\eta_{\rm M} \simeq 50$ at $M_\star = 10^{7.5}$\msun to $\eta_{\rm M} \simeq 3-5$ at $M_\star = 10^{10.5}$\msun. Enforcing a more stringent $v_{\rm rad} > 150$ km/s cut as measurably outflowing material (orange) weakens the dependence and flattens the relation with stellar mass considerably -- \etaM decreases by only a factor of two over the same mass range. For even higher velocity thresholds (green), the slope of $\eta_{\rm M} \propto M_\star$ changes sign, because low mass galaxies drive very little mass at such high velocities.

The black line shows the value of $\eta_{\rm M}^{\,i}$ at the injection scale, as prescribed by the TNG model. We derive a single mean value for each galaxy from its ensemble of star-forming gas cells, then present an approximate fit through these mean values as a function of $M_\star$, as well as its scatter (gray band). The model $\eta_{\rm M}^{\,i}$ is nearly flat below $M_\star \lesssim 10^{8.5}$\msun and then decreases monotonically with increasing mass, dropping to unity by $10^{10.5}$\msun and $< 1$ thereafter. However, this behavior is fundamentally different than what is seen in the resolved flows which result at 10 kpc scales. Here the trend of \etaM reverses, reaching a minimum at $M_\star \simeq 10^{10.5}$\msun and before rising rapidly again towards higher masses. This corresponds to the mass scale for the onset of quenching as galaxies begin to transition out of the blue, star-forming population in the TNG model \citep{nelson18a,weinberger18}. Star formation rates are suppressed by the action of efficient BH feedback which simultaneously produces a large mass of outflowing gas through direct ejection of the nuclear (i.e. central) ISM, driving $\eta_{\rm M} = \dot{M}_{\rm out} / \dot{M}_\star$ up. As we move beyond $M_\star \gtrsim 10^{11}$\msun, mass loading factors approach similar values ($> 10$) as in the lowest mass dwarfs. As a result, $\eta_{\rm M}(M_\star)$ has a broken `v' shape: the low-mass behavior regulated by stellar feedback, while the high-mass behavior is set by the energetics and coupling efficiency of BH feedback.

Even where \etaM is established largely by the SN energy budget alone, its value and scaling with $M_\star$ differs from the TNG model input at the injection scale. In the lower left panel we clarify this difference, by showing the dependence on threshold \vrad (different line colors) as well as at different distances away from the galaxy (different line styles). The wind mass loading decreases strongly as a function of distance from 10 kpc outwards at $M_\star < 10^{10.5}$\msun, dropping for low-mass galaxies in the $v_{\rm rad} > 50$ km/s case from $\eta_{\rm M} \simeq 20-50$ at 10 kpc to only $\eta_{\rm M} \simeq 1$ at 50 kpc. Depending on velocity cut and radius, the actual mass outflow rate can exceed that prescribed at injection by the TNG model, presumably due to local entrainment as winds begin to interact in the outskirts of the disk. Declining $\eta_{\rm M}(r)$ is generically expected for a flow which is at least partially fountain, i.e. not entirely escaping \citep[e.g.][]{sarkar14}. In Appendix \ref{sec:appendix} we discuss mass loading trends scaled by the virial velocity of the parent halo.

The radial dependence of \etaM in TNG weakens towards high stellar mass, pointing towards a different launch mechanism. At \mbox{$M_\star \sim 10^{10}$\msun} we see that even for $v_{\rm rad} > 150$ km/s the wind mass loading exceeds that produced by the SF-driven wind model alone. We speculate that the mutual, positive coupling between the two feedback mechanisms is roughly maximal at this mass scale - where black holes are able to contribute energy to drive outflows, but not yet severely impacting galactic star formation and so total available SN energy. As the majority of BHs transition into the low accretion state at $M_\star \gtrsim 10^{10.5}$\msun the radial dependence is entirely eliminated, even for fast $v_{\rm rad} > 350$ km/s outflows. As we discuss later, the vast majority of gas driven by kinetic BH feedback exceeds this velocity to distances greater than $r > 50$ kpc, leaving \etaM invariant to distance in this regime.

In the lower right panel of Figure \ref{fig_massloading1} we explore the redshift dependence of \etaM in different bins of stellar mass, from $\log(M_\star/\rm{M}_\odot) = 10^8 - 10^{11}$. For low mass galaxies with $M_\star \lesssim 10^{10}$\,M$_\odot$ the wind mass loading is roughly constant as a function of redshift. This is because, although the actual outflow rates around such galaxies decrease with redshift at fixed mass, by roughly one order of magnitude from $z=6$ to $z=0.2$, their star formation rates do likewise. On the other hand, galaxies at the high-mass end exhibit a rapidly increasing \etaM towards late times. For example, galaxies with $M_\star \sim 10^{10.5}$\,M$_\odot$ have $\eta_{\rm M} \sim 1$ at $z=6$, increasing to $\eta_{\rm M} \sim 100$ by $z \simeq 0$. In this case mass outflow rates are roughly constant with redshift, while star formation rates begin to decline after some characteristic redshift as a result of the quenching process. At redshift two we see the inversion of \etaM noted above as the crossover of the high-mass curves; at $z \gtrsim 4$ it is clear that \etaM is monotonic with $M_\star$, unlike at low redshift.


\subsection{Outflow velocities: SN versus BH-driven winds} \label{sec_results_vel}

\begin{figure*}
\centering
\includegraphics[angle=0,width=6.2in]{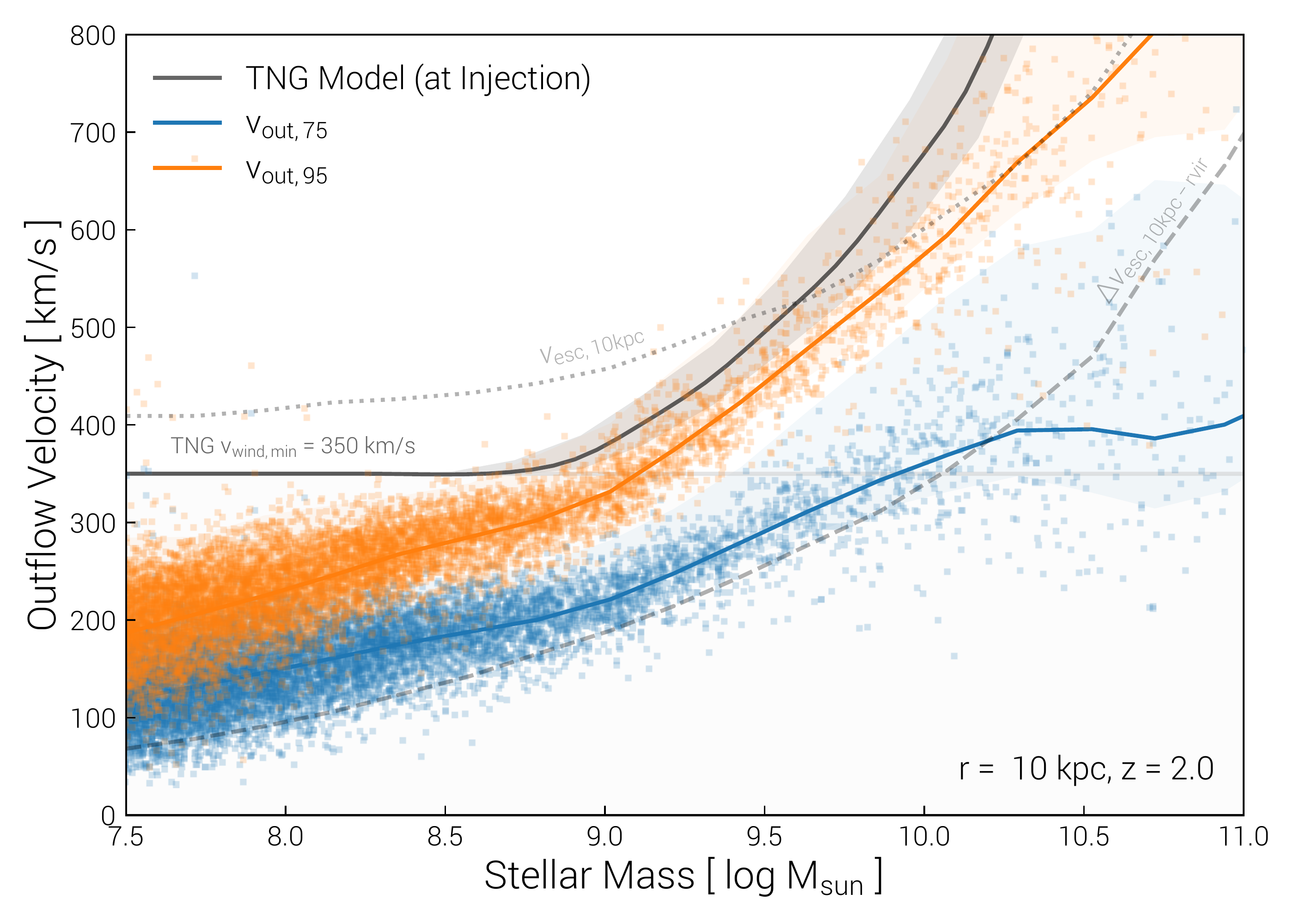}
\includegraphics[angle=0,width=3.4in]{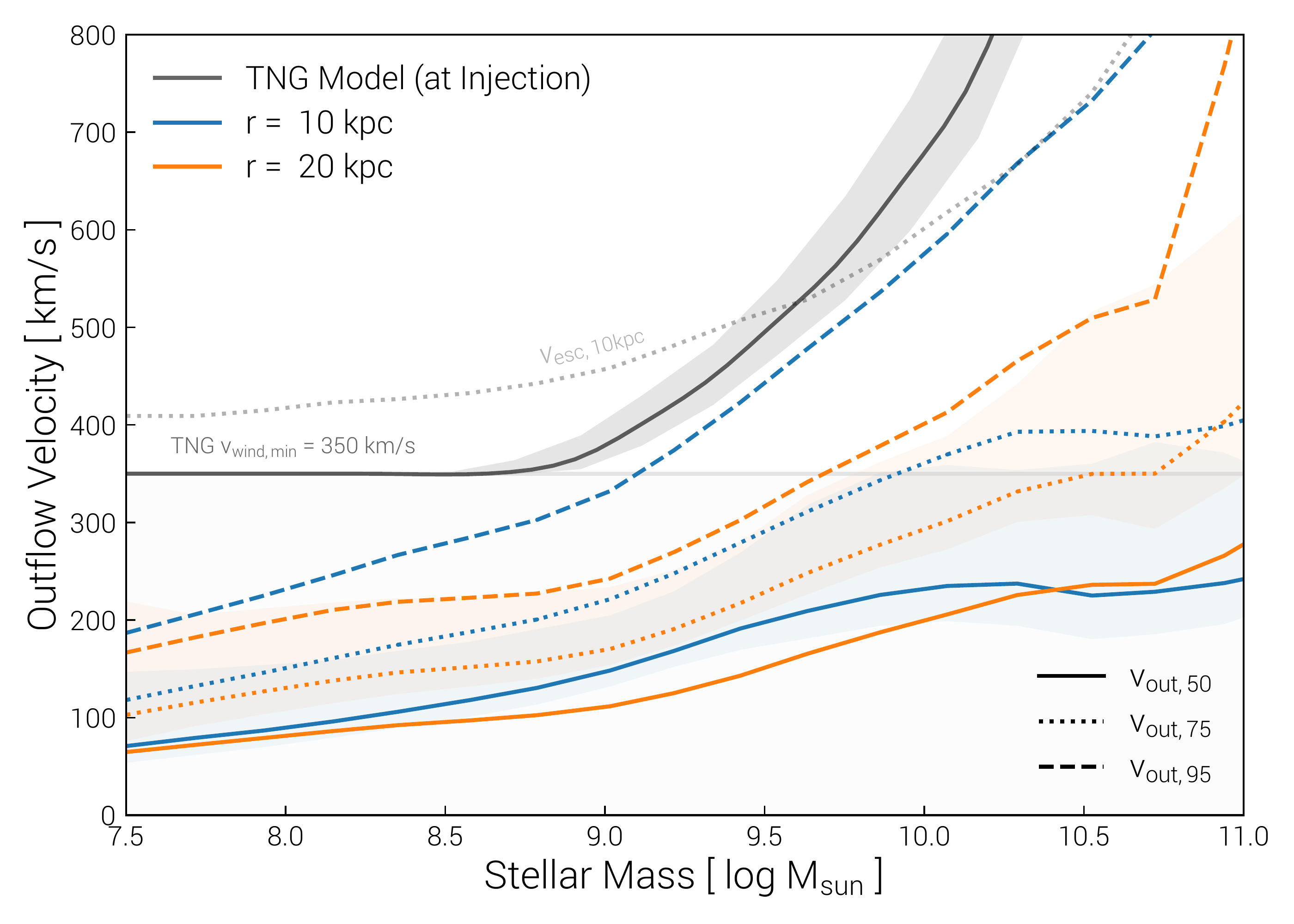}
\includegraphics[angle=0,width=3.4in]{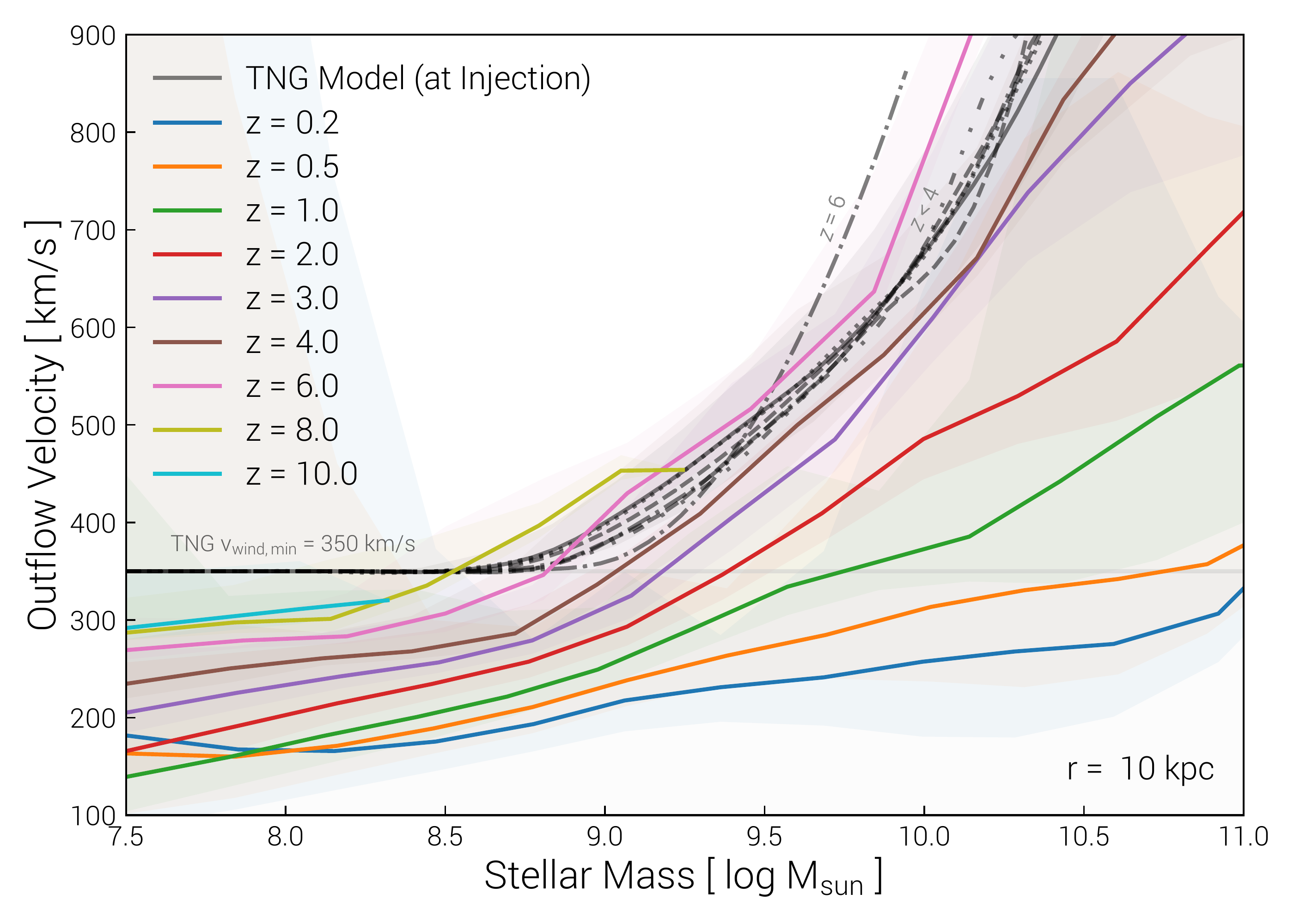}
\caption{ Outflow velocity as a function of stellar mass. Computed for outflowing gas and wind ($v_{\rm rad} > 0$ km/s) within the given radial shell. \textbf{(Top)} The 75$^{\rm th}$ and 95$^{\rm th}$ outflow velocity percentiles, as a function of galaxy stellar mass, for TNG50-1 at $z=2$. The black line indicates an approximate fit to the \textit{injection} velocity for the TNG wind model, derived as the mean launch speed for all star-forming gas cells in each galaxy \protect\citep[similar to the computation for Fig. 6 of][]{pillepich18a}. \textbf{(Bottom left)} Dependence on radius and velocity percentile, for TNG50-1 at $z=2$. Note that the TNG model line is roughly although not strictly an upper limit: $v_{\rm out,99}$ typically corresponds closely, implying that the wind launch speed at injection establishes the envelope of the fastest moving material, although not necessarily the speed of the bulk of the outflowing mass. \textbf{(Bottom right)} Redshift evolution at fixed mass, from $z=0.2$ to $z=10$, for a fixed percentile $v_{\rm out,90}$. Outflows are much faster at earlier times, in both the star formation and black hole driven regimes.
 \label{fig_vout}}
\end{figure*}

Moving to a second fundamental property beyond mass outflow rates, Figure \ref{fig_vout} measures outflow velocities as a function of stellar mass. As before, we select outflowing material at some distance away from the galaxy. We also condense the full spectrum of the outflowing velocity distribution around a galaxy into a single number - mass outflow rate weighted percentiles $v_{\rm out,50}$ to $v_{\rm out,95}$ (as described in Section \ref{sec_methods_outflows}).\footnote{For example, $v_{\rm out,95}$ gives the velocity which only 5\% of the outflow mass flux exceeds, i.e. the velocity envelope of 95\% of the mass flux. In some observational studies, this value might be referred instead to as $v_{\rm out,05}$. As a characterization of the extreme tails of the outflow velocity distribution, it will approach the similar though less well defined $v_{\rm max}$ measure.} In the top panel we show the 75th and 95th percentiles of $v_{\rm out}$ at a distance of 10 kpc at $z=2$. Both increase monotonically with $M_\star$ from $100-200$ km/s at stellar masses of $10^{7.5}$\msun up to $\sim$ 400 km/s ($v_{\rm 75}$) or $\gtrsim$ 800 km/s ($v_{\rm 95}$) for \mbox{$M_\star \gtrsim 10^{10.5}$\msun}. They are typically below the escape velocity $v_{\rm esc} = [-2 \phi(r = \rm{10 kpc})]^{1/2}$ at this distance. Given that the global potential also traces the large scale environment, particularly for smaller galaxies \citep{oppenheimer08}, we also include a second line for $v_{\rm esc}$ (dashed), computed using the potential difference between 10 kpc and $r_{\rm vir}$. This roughly captures, therefore, the energy required to move between these two locations, against the force of gravity. In this case the conclusion would be that $v_{\rm out}$ can commonly exceed the escape velocity, although we caution that in neither case does this actually mean that an outflow can necessarily escape the halo, as the comparison neglects hydrodynamical interactions, drag, and other similar effects. In Appendix \ref{sec:appendix} we discuss velocity trends scaled by the virial velocity of the parent halo.

As before, we include the prescription from the TNG model for the injection (launch) velocity $v_{\rm out}^{\, i}$ of winds generated by stellar feedback as the gray line and shaded band, which represent a fit to the mean value derived for each galaxy. We see that $v_{\rm out}^{\, i}$ represents more an upper envelope to the resulting flow at 10 kpc, rather than its characteristic velocity. We note however that $v_{\rm out,99}$ does exceed the injection velocity even at this distance. The minimum wind velocity of 350 km/s enforced by the TNG wind model leads to the constant $v_{\rm out}^{\, i}$ below $M_\star \lesssim 10^9$\msun. Interestingly, we see that this does not mean that material flowing away from a galaxy at any resolved scale actually inherits such a velocity (nor even its trend with stellar mass). The $v_{\rm out,75}$ velocity percentile is actually below this `minimum' for all $M_\star \lesssim 10^{10}$\msun. We interpret this as an indication of halo drag, either gravitational or hydrodynamical \citep{silich01,dallavecchia08}, an important demonstration of how emergent outflow properties can differ from model inputs due to resolved physical processes.

In the lower left panel of Figure \ref{fig_vout} we explore the dependence of $v_{\rm out}$ on distance (colors) and velocity percentile (linestyles). Outflow velocity notably decreases with distance for all $r > 10$ kpc. At the same time, our velocity percentile `summary statistic' will be influenced by all gas with $v_{\rm rad} > 0$ km/s, i.e. it could also be pulled down by a progressively larger fraction of quasi-equilibrium, slowly moving halo gas relative to an actual, faster outflow component. Outflow velocity scales strongly with the velocity percentile, particularly at the massive end. The distributions of $v_{\rm out}$ clearly have extended tails which are not captured by $v_{\rm 50}$ for instance. At $M_\star = 10^{11}$\msun, the 95th tail reaches $\sim$ 1000 km/s while $v_{\rm out,50}$ is only $\sim 200$ km/s. By measuring the maximal velocities achieved by gas shortly after being kicked by the low-state BH feedback, we determine a corresponding `injection velocity' of $\gtrsim$ 12,000 km/s, applicable to the high-mass end, which shows how rapidly such nuclear-originating flows must decelerate.

The lower right panel of Figure \ref{fig_vout} shows the redshift dependence of outflow speed at fixed stellar mass, where we focus on $v_{\rm out,90}$ and $r = 10$ kpc for simplicity. Winds are faster at higher redshift for a given $M_\star$, both in the SF-driven and BH-driven regimes. In the former, a plausible cause is that, at fixed stellar mass, the DM velocity dispersion will be higher at earlier times. We note an important new aspect of the TNG wind model is the explicit scaling $v_{\rm out}^{\, i} \propto [H_0 / H(z)]^{1/3}$. This dependence produces a co-evolution of wind velocity and virial halo mass, i.e. redshift-independent velocity at a fixed host halo mass \citep{pillepich18a}. After convolution with the evolving stellar mass to halo mass relation, we see that in practice the model injection velocities are only higher at a given $M_\star$ for very high redshift ($z \sim 6$), but settle down to a roughly constant set of the overlapping black curves for all $z < 4$. In practice, this scaling is not relevant for $M_\star \lesssim 10^9$\msun due to the imposition of the minimum wind injection velocity of 350 km/s. However, we see that the resolved wind velocities evolve with redshift even at these low masses. In this regime $v_{\rm out,90} \propto (1+z)^{0.2 - 0.4}$ or $v_{\rm out} \propto H(z)^{0.2 - 0.3}$ depending on $M_\star$. We note that the measured scaling with $H(z)$ has the opposite sign as the model injection scaling, emphasizing that the mapping between model inputs and resolved outputs can be non-trivial. At the high mass end, the redshift scaling is much stronger, $v_{\rm out,90} \propto (1+z)^{0.8 - 1.3}$ for $M_\star \simeq 10^{11}$\msun, indicating that the velocities of BH-driven outflows evolve differently with redshift.

\begin{figure}
\centering
\includegraphics[angle=0,width=3.4in]{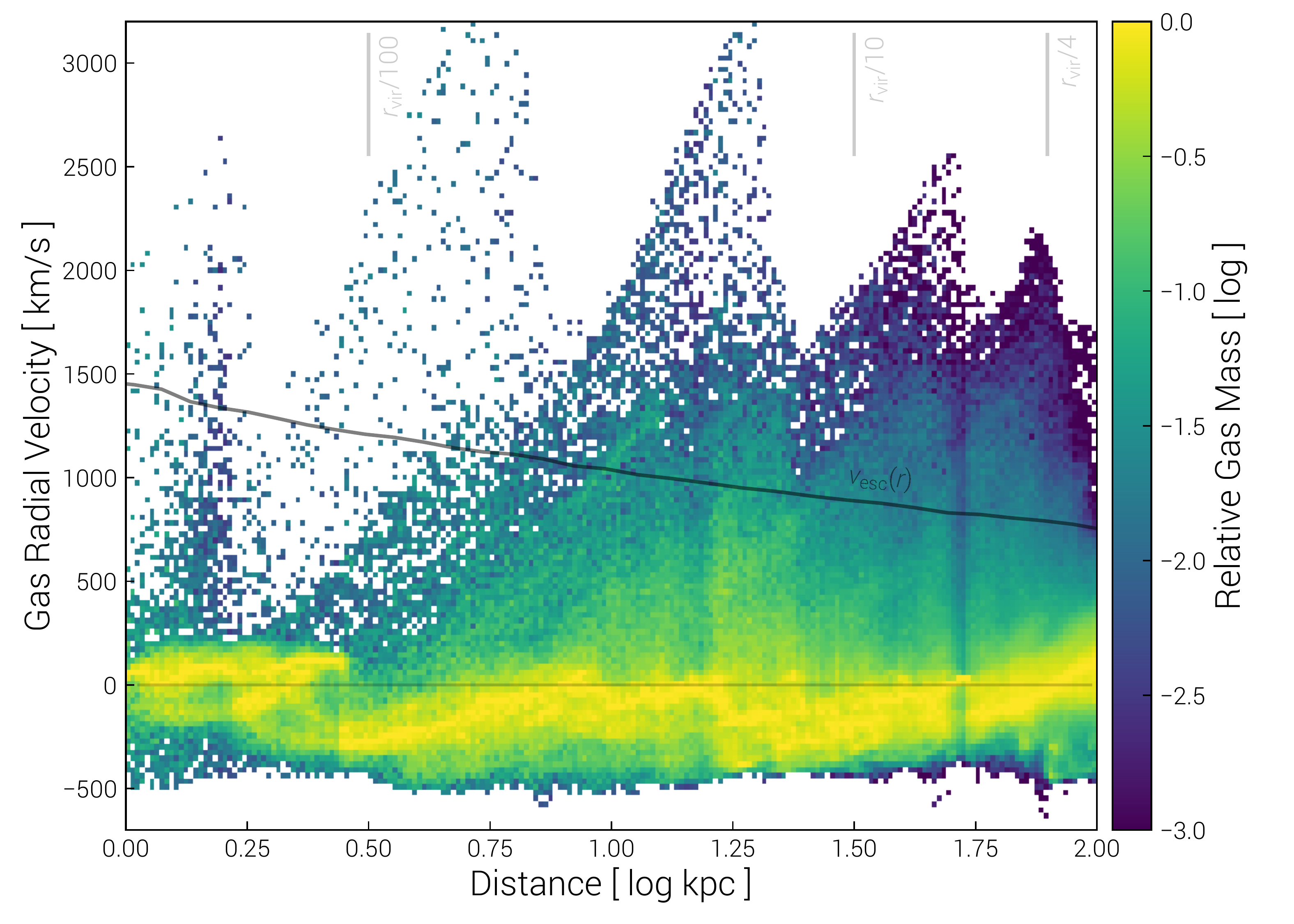}
\caption{ Two dimensional phase space diagram showing high-velocity outflows ($v_{\rm rad} > 0$) in the conditional gas mass distribution on the ($v_{\rm rad}$,\,$r$) plane for one illustrative halo being strongly affected by BH feedback \mbox{$(M_{\rm halo} \simeq 10^{12.8}$\msun, $r_{\rm vir} \simeq 300$ kpc, $v_{\rm circ} \simeq 280$ km/s, $z=0.7$).} The bulk of gas mass is confined to roughly $\pm v_{\rm circ}$ of the halo, with accelerating infall visible as features shifting towards more negative velocities in the halo center. At small distances, distinct branches extend to high velocity, corresponding to individual outflow episodes. These arcs flow outward (in radius) and downward (in velocity) with time, although their upward shape implies an instantaneously increasing radial velocity with distance. 
 \label{fig_phase2d}}
\end{figure}

Having presented the population-level view of outflow velocity, we go on to explore the detailed velocity structure of outflows. Figure \ref{fig_phase2d} shows the phase space diagram of gas mass in the plane of \mbox{$(v_{\rm rad},\rm{radius})$} around a single halo with total mass $\simeq 10^{12.8}$\msun, from just outside the direct region of influence of the black hole (1 kpc) to about a third of the host virial radius (100 kpc). The bulk of mass at almost all radii is sitting just below $v_{\rm rad} \sim 0$ indicating consistent inflow through the inner halo. Outflowing winds are evident as a succession of discrete features at positive radial velocities. These outflows launched by black hole feedback form a sawtooth-like pattern in $(v_{\rm rad},r)$, consisting of a series of triangular components -- each results from a single energy injection event at a small scale off the left-edge of this panel. There are three such flows evident at $r < 20$ kpc, and $\sim$ 5 visible  in the inner halo.

The maximum radial velocity of each outflow component \textit{increases} with distance, i.e $v_{\rm out} \propto r^{\,\alpha} \,(\alpha > 0)$. This shape is a natural consequence if material has a spectrum of launch velocities and the most rapidly outflowing gas reaches, at any given time, the largest distances. The maximal outflow velocity at each radius is established by the upper envelope of the successive peaks, corresponding to the highest velocity tails of each flow. Over time, each individual feature collapses outward, shifting to larger distance and lower velocity. As a result, the envelope of outflowing radial velocities actually \textit{decreases} with distance, i.e $v_{\rm out,max} \propto r^{\,\alpha} \,(\alpha < 0)$. This velocity structure, coming from the superposition of the individual outflow components, is continually reinforced by new injection events at $r \sim 0$ and generically maintains a quasi-steady structure over time. The high velocity tails of the outflowing material lie above the escape velocity curve $v_{\rm esc}(r) = [-2 \phi(r)]^{1/2}$ at all radii (black line).

\begin{figure}
\centering
\includegraphics[angle=0,width=3.4in]{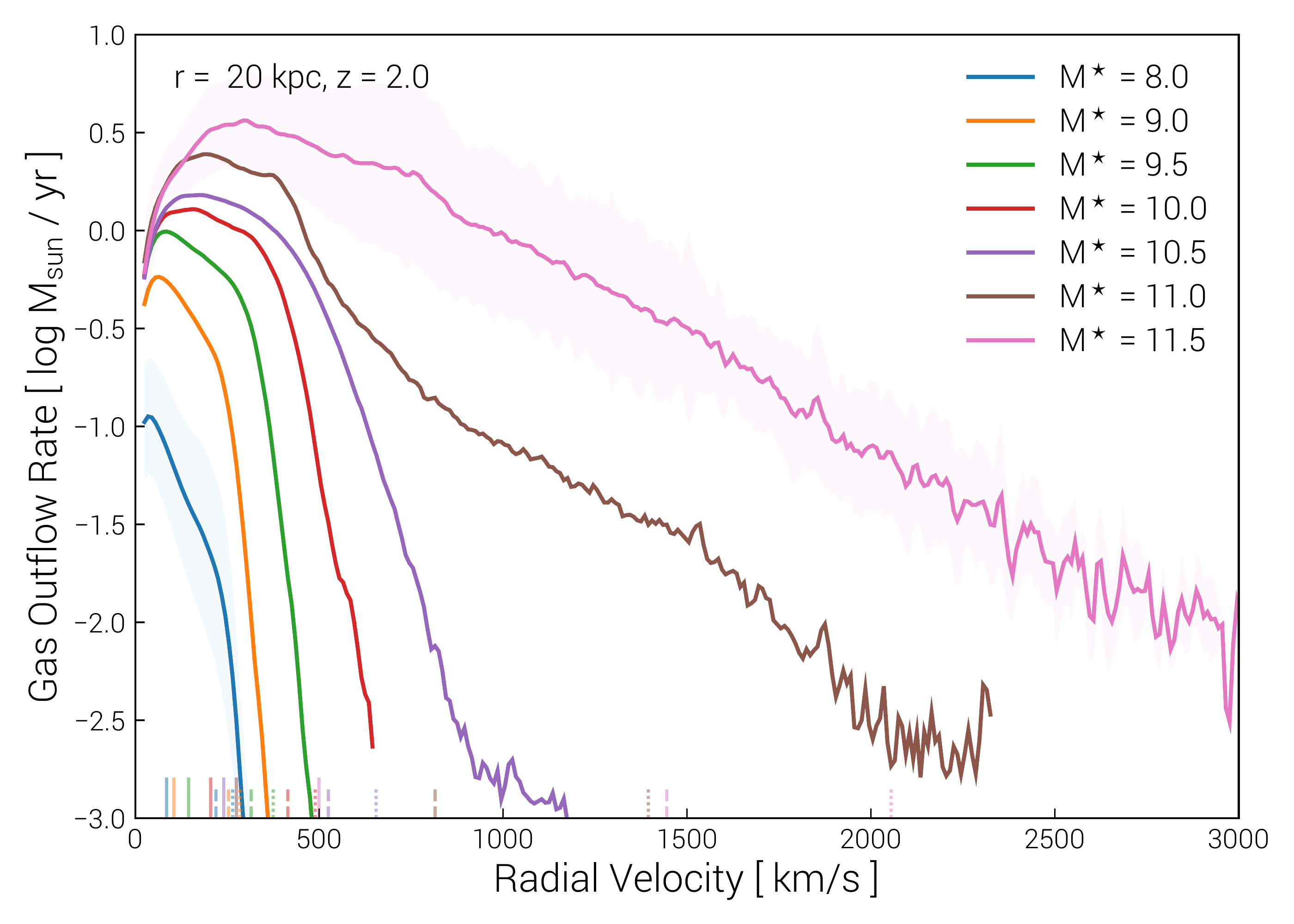}
\caption{ The average distributions of outflow velocity around galaxies in different stellar mass bins at $z=2$. Specifically, we decompose the radial mass outflow rate as a function of radial velocities \vrad, focusing on the flow across $r = 20$ kpc, in units of \msunyr per 10 km/s bin of \vrad. At low mass, the outflow velocity distributions have a similar, centrally symmetric shape originating from SN-driven winds. They peak at velocities $\lesssim$ a few hundred km/s, with a maximal velocity cutoff which scales with $M_\star$. At high masses, these slower outflows are augmented by a long tail to high velocities due to the emergence of BH-driven outflows, which easily reach $\gtrsim$ 3000 km/s. Small solid, dashed, and dotted vertical lines denote $v_{\rm 50}$, $v_{\rm 90}$, and $v_{\rm 95}$ for each mass bin, respectively.
 \label{fig_vrad_dist}}
\end{figure}

Figure \ref{fig_vrad_dist} presents the distribution functions of outflow velocity, decomposing the mass outflow rate $\dot{M}_{\rm out}$ as a function of $v_{\rm rad}$. We stack galaxies in bins of stellar mass, from $8.0 < \log(M_\star/\rm{M}_\odot) < 11.5$ (different line colors), and focus on a distance of $r = 20$ kpc at $z=2$ for simplicity. At lower stellar masses the distributions have a similar shape, being peaked at low velocities \mbox{$\lesssim$ a few} hundred km/s, which increases slowly with $M_\star$. The amount of outflowing gas with higher velocity drops rapidly beyond this central core, such that there is an effective maximum outflow velocity of $\sim$ 300 km/s for $M_\star = 10^8$\msun, $\sim$ 400 kms for $M_\star = 10^9$\msun, and $\sim$ 500 km/s for $M_\star = 10^{10}$\msun. This is the regime of the stellar feedback driven wind model, with the approximate scaling of $v_{\rm out} \propto M_{\rm halo}^{1/3}$. As a result, typical wind velocities scale up with mass, but have bounded maximal values.

However, at stellar masses $M_\star \gtrsim 10^{10.5}$\msun a high-velocity tail begins to emerge. By $10^{11}$\msun this feature is fully developed and results in a roughly linear extension out to high $v_{\rm rad}$ (in this log-linear plane). As a result, we can start to find outflowing mass at $>$ 1000 km/s velocities, where the distributions are then composed of two distinct components: the SF-driven core at low velocity, and the tail at high velocity. In the TNG model, only the kinetic BH feedback mechanism can generate such high velocities, and this mode becomes efficient at $M_\star \gtrsim 10^{10.5}$\msun as black holes transition below the threshold accretion rate. The mass outflow rate decreases more slowly with $v_{\rm rad}$ than in the SF-driven wind case, such that there is no strong cutoff at some maximal velocity. However, the decline is still precipitous: for the highest mass bin considered here at $M_\star = 10^{11.5}$\msun, for $v_{\rm out} \sim 1000 \pm 50$ km/s, the total flux is on average $\dot{M}_{\rm out} \simeq 10 \,\rm{M}_\odot \,\rm{yr}^{-1}$. However, for gas moving with $v_{\rm out} \sim 2000 \pm 50$ we have only $\dot{M}_{\rm out} \simeq 1 \,\rm{M}_\odot \,\rm{yr}^{-1}$, and the flux of gas moving at $\sim$ 3000 km/s is a further factor of ten less. 

As we discuss further in Section \ref{sec_results_vsgal}, it is important to note that these highest velocity BH-driven outflows in the TNG model do not arise from ultra-luminous or `radio-loud' quasars (e.g. with space densities $n \sim 10^{-7}$ Mpc$^{-3}$ or less, and/or $L_{\rm bol} \sim 10^{45}$ erg/s or greater), which are much too rare to arise in the volume of TNG50. Instead, they are generated from the much more common low luminosity, low-accretion rate ($\lambda_{\rm edd} \sim 10^{-2}$ or less) population of BHs for which we posit a radiatively inefficient flow which converts gravitational binding energy into a non-relativistic wind \citep{blandford99,yuan14}. The implication is that -- in the TNG model -- low luminosity, slowly accreting black holes can drive some of the most powerful outflows. Other models produce strong BH-driven outflows by invoking different physical mechanisms and/or numerical implementations \citep[e.g.][]{mccarthy11,henden18}, and differentiating between these scenarios will provide useful theoretical constraints.


\subsection{Multiphase outflows: temperature, density, metallicity} \label{sec_results_multiphase}

\begin{figure}
\centering
\includegraphics[angle=0,width=3.3in]{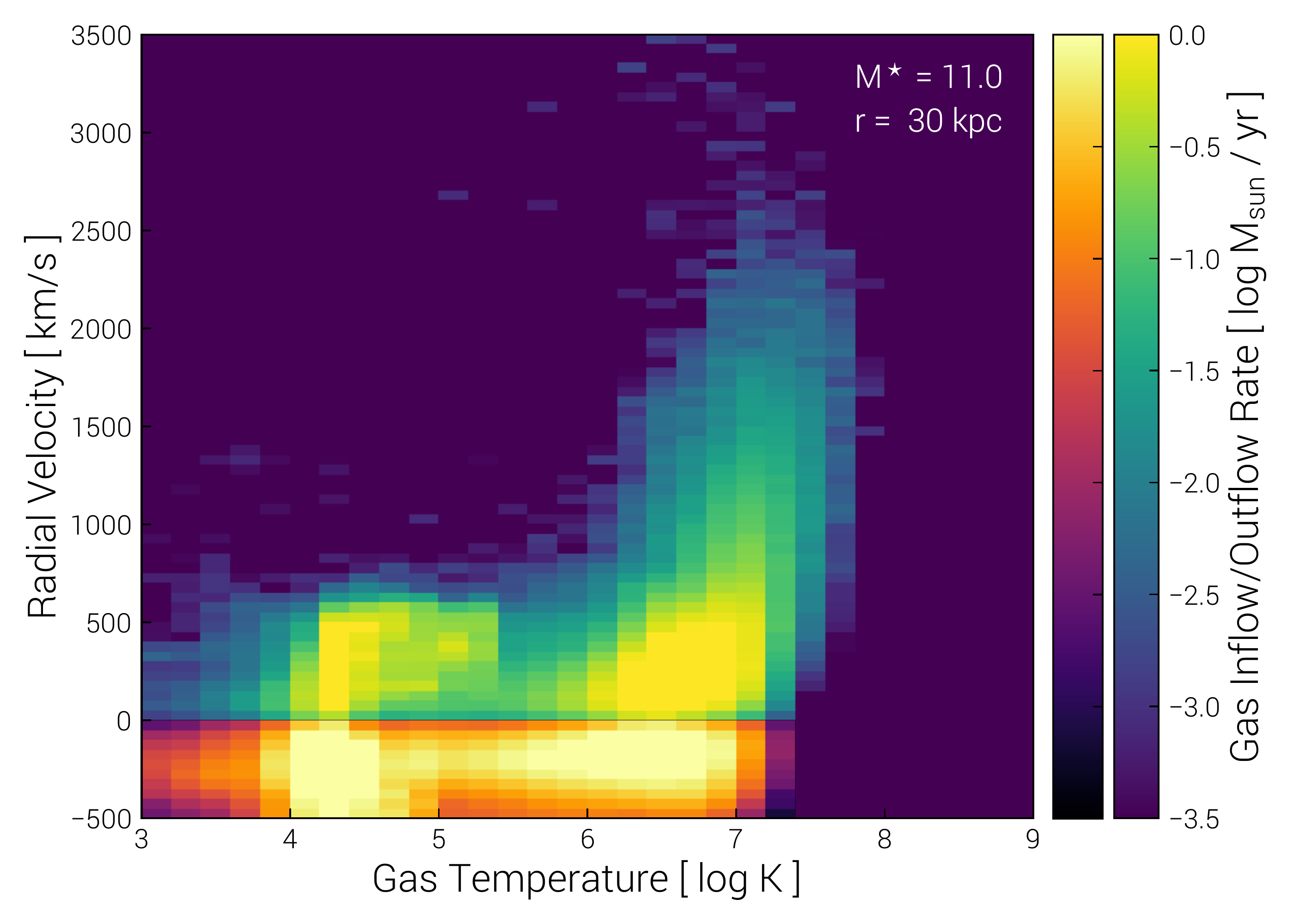}
\caption{ Gas inflow/outflow rates decomposed as a function of gas temperature and outflow velocity at $z=2$. Here we focus on a regime where black holes dominant outflow properties: massive galaxies with $M_\star \simeq 10^{11}$\msun at a distance of $r = 30$ kpc. We find that different phases attain different outflow speeds (or, equivalently, that different outflow velocity components occupy different regions of $\rho-T$ space). In this case, it is only hot gas at temperatures $\gtrsim 10^6$ K which is able to reach $>$1000 km/s, while the cooler component is confined to around $\sim$500 km/s or less.
 \label{fig_outflowrate_temp_vrad}}
\end{figure}

We have so far considered only the kinematics of the outflow as a whole, i.e. phase-agnostic. Observationally there exists a broad correlation between outflow velocity and gas phase temperature, such that neutral phases are slower than ionized components \citep{heckman01,rupke02}, which are themselves slower than a hot wind fluid phase \citep{veilleux05}. 

\begin{figure*}
\centering
\includegraphics[angle=0,width=6.2in]{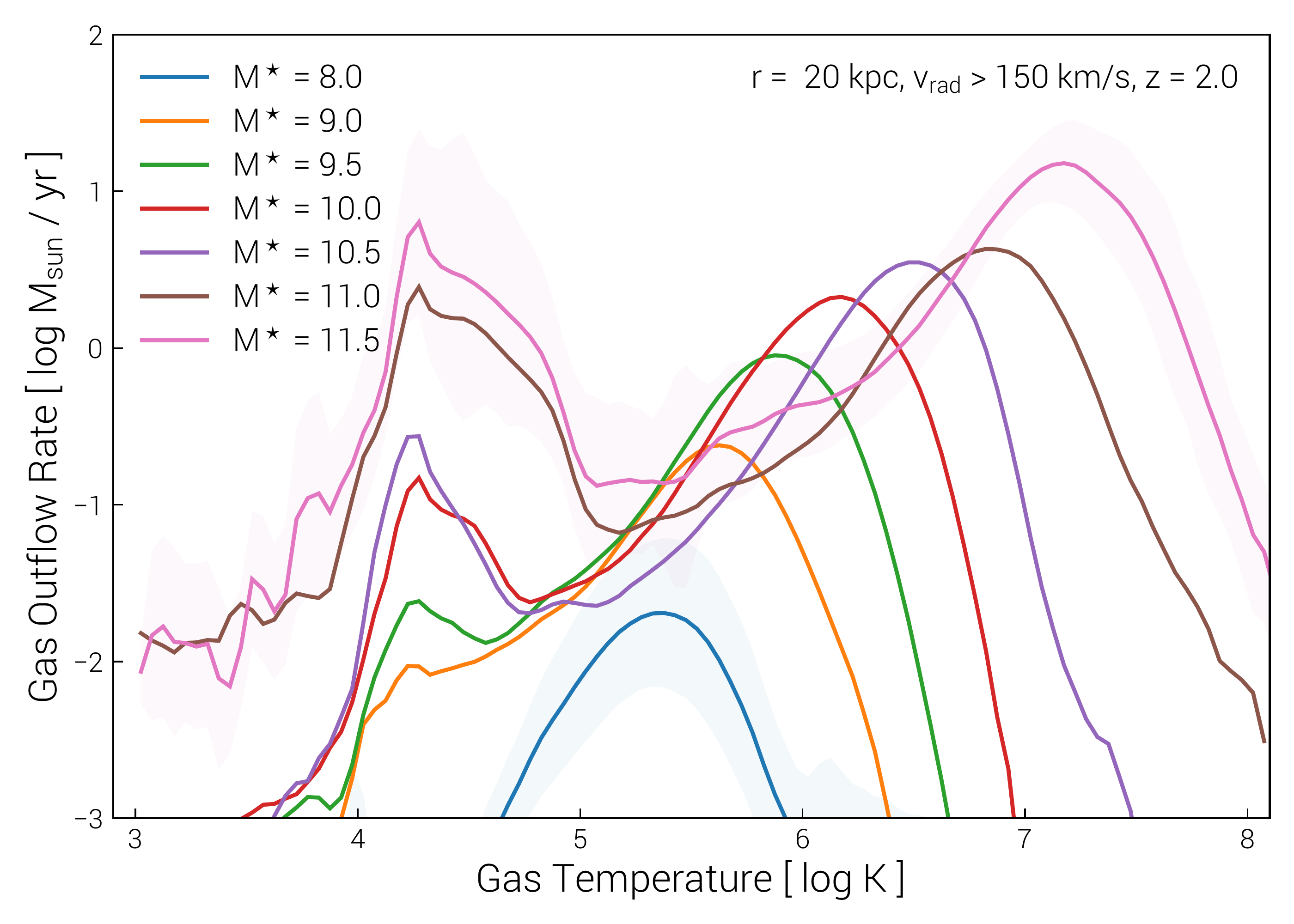}
\includegraphics[angle=0,width=3.4in]{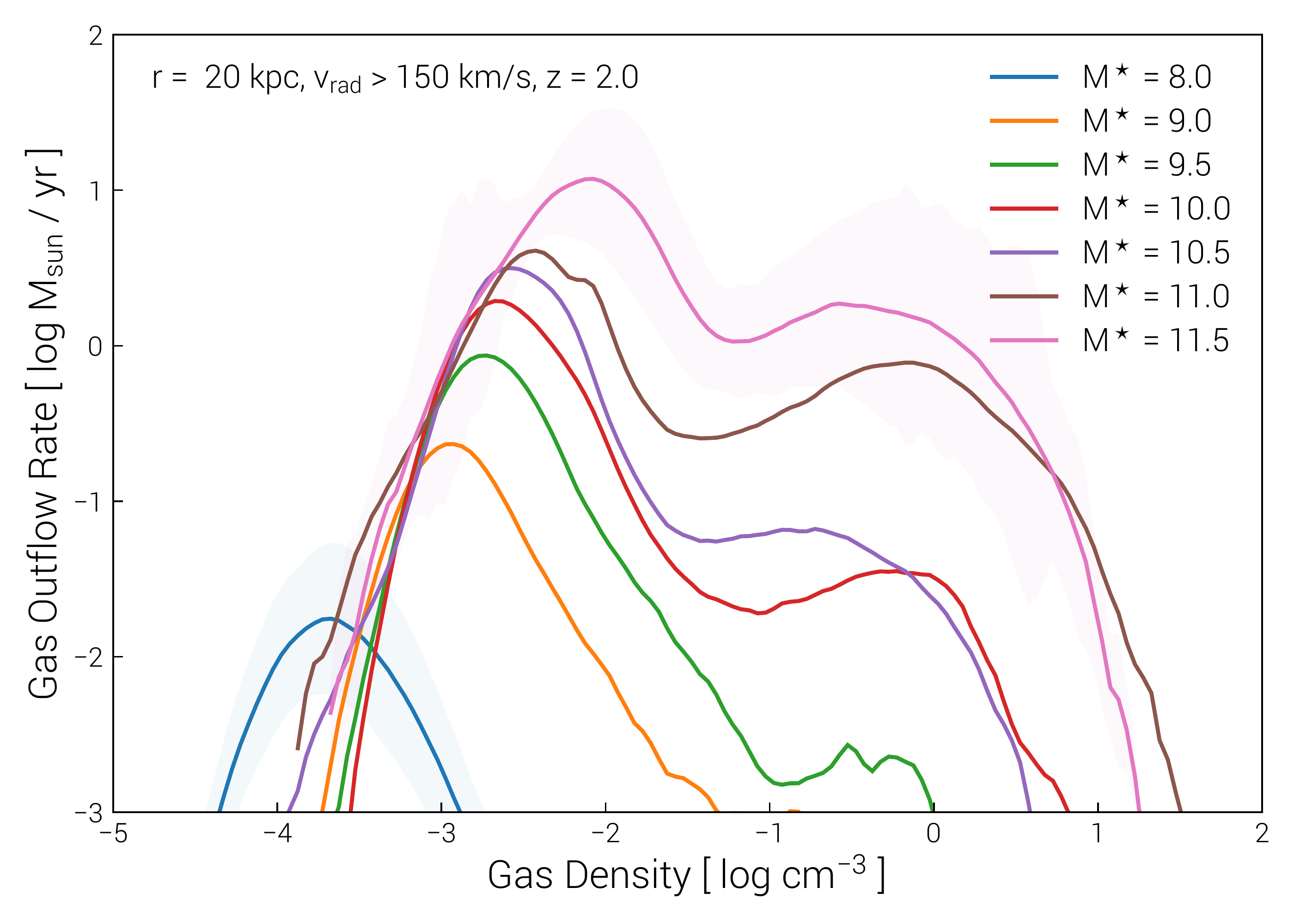}
\includegraphics[angle=0,width=3.4in]{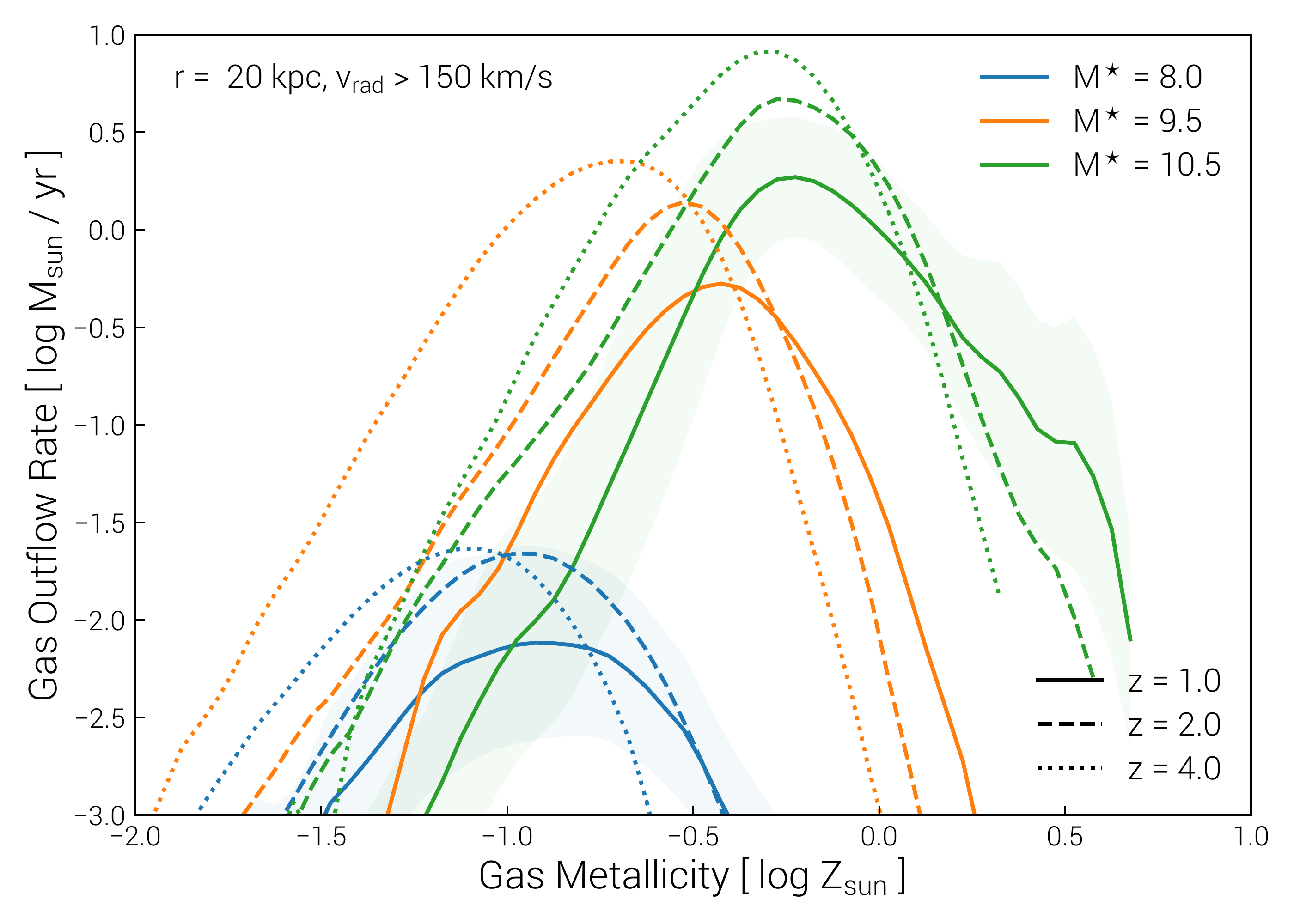}
\caption{ Radial mass outflow rates as a function of gas temperature (top), density (lower left), and metallicity (lower right) at $z=2$. Each panel shows gas in a radial shell centered at $r = 20$ kpc from the galaxy, and with outflow defined by a radial velocity cut of $v_{\rm rad} > 150$ km/s, with flow rates in \msunyr are per 0.05 dex bin. Distributions of $\dot{M}_{\rm out}(T_{\rm gas}; n_{\rm gas}; Z_{\rm gas})$ are given in stacked bins of stellar mass. Demonstration of the multi-phase nature of outflows, and the emergence of the cold, dense component for massive systems. 
 \label{fig_outflowrate_phase}}
\end{figure*}

In Figure \ref{fig_outflowrate_temp_vrad} we show that different outflow phases in TNG have different kinematics. We present the galaxy-stacked 2D histogram of mass outflow rate, as a function of temperature (x-axis) and instantaneous velocity (y-axis). We focus on a mass scale ($M_\star = 10^{11}$\msun) where the action of the central black hole plays an important role in setting the properties of the galactic-scale outflows. For winds which reach an appreciable distance away from the galaxy we find a relative dichotomy of hot/cooler gas phases and rapid/slower outflow velocities. In particular, the only outflowing gas which achieves velocities of $>$ 1000 km/s is hot, with \mbox{$T \gtrsim 10^6$ K}. Within this phase there is a minor correlation between $v_{\rm out}$ and temperature: the hottest gas travels the fastest, with \mbox{$v_{\rm out} \sim 2000$ km/s} outflows having temperatures of order $10^{6.5}$ K.

On the other hand, cooler gas is slower, and so unlikely to propagate as far. The outflow component which would be visible in ionized H$\alpha$ or metal transition lines, between $10^4$ K and $10^5$ K, has typical velocities of a few hundred km/s and a maximal $v_{\rm out}$ of $\sim$ 600 km/s. The phase separation at high velocity implies that the temperature of a high-velocity component can be clearly deduced. The bulk of the outflowing gas, however, coexists at a common, lower velocity interval, preventing one from easily identifying the phase structure of this dominant, low-velocity component. We note that for lower stellar masses $M_\star \sim 10^{9-10}$\msun (not shown), where stellar rather than BH-driven winds are most relevant, this correlation of hotter outflow phases moving faster likewise holds.

Figure \ref{fig_outflowrate_phase} quantifies the multiphase nature of TNG outflows in terms of temperature, density, and metallicity. In the main panel (top) we decompose the mass outflow rate as a function of gas temperature, where each line shows the mean stack within a given stellar mass bin, from $8.0 < \log(M_\star/\rm{M}_\odot) < 11.5$ (different line colors). Two important trends emerge.

First, there exists a warm/hot outflowing component with a well-behaved, roughly gaussian distribution of mass flux weighted log temperature. The peak of this component shifts smoothly upwards with increasing galaxy stellar mass, from $\sim 10^{5.5}$ K at \mbox{$M_\star = 10^8$\msun} to $\sim 10^6$ K by $10^{9.5}$\msun, and all the way up to \mbox{$\sim 10^7$} K at $10^{11}$\msun. The width (in log K) is roughly constant, implying that this constituent of the outflow is set by the temperature structure of the virialized hot halo gas and tracks the increasing host halo virial temperature. Note that bulk (i.e. non-circular) motions in the inner halo gas which exceed the chosen velocity threshold would contribute here, so the hot outflow component could partially reflect increasing amounts of such non-equilibrium kinematics of virialized gas. The peak outflow rate of this component increases by roughly a factor of ten with each decade in stellar mass -- i.e., it is a roughly constant `specific gas outflow rate'. In the integral it dominates, at all mass scales, the total outflowing mass flux.

Next, a second cool outflow component emerges towards higher stellar masses, resulting in a bimodal outflow temperature distribution. This bimodality is clear at all $M_\star > 10^{10}$\msun and becomes stronger towards higher mass, where the peak outflow rates at $T \sim 10^4$ K and $T \sim 10^7$ K come into near equality.\footnote{Some outflowing gas at temperatures even lower than $10^4$ K is also evident. While it would be tempting to interpret this in the context of material which could be neutral or even partially molecular, we remind that the effective cooling floor of TNG is $\simeq$ 10$^4$ K due to the radiative cooling assumptions, which neglect metal fine-structure contributions. As a result, we interpret this gas as representative only of a cool ionized component.} The origin is twofold - the hint of this secondary peak already developing even for galaxies as small as $M_\star = 10^9$\msun, together with the invariance of the peak temperature itself in relation to the shape of our adopted cooling curve \citep{wiersma09} suggests a cooling process \citep[possibly in the outflow itself, e.g.][]{thompson16,schneider18b}. In the future we can use the Monte Carlo tracer particle scheme in TNG \citep{genel13} to directly track the thermal history of outflowing gas and (dis)prove this point. On the other hand, the strong increase of outflowing mass at this temperature above $M_\star > 10^{10.5}$\msun indicates a relation to the onset of efficient black hole feedback. We observe that the kinetic BH mechanism in this regime directly evacuates the central ISM reservoir, leading to centrally suppressed gas densities (and star formation rates, as we discuss later). At stellar masses $\gtrsim 10^{11}$\msun this dichotomy in outflow temperature is therefore a direct consequence of ejective feedback launching formerly dense ISM out into the halo \citep[e.g.][]{scannapieco15}.

Observationally, \cite{chisholm17} recently measured the mass loading of OVI absorbing gas in a single lensed $z \sim 3$ galaxy, concluding that the fastest outflowing material (with $v_{\rm out} > v_{\rm esc}$) has a mass outflow rate larger than in the cooler, photoionized phase. Similarly, from a sample of local starbursts \cite{grimes09} find that OVI-traced gas at $\sim 10^{5.5}$K has a higher mean outflow velocity than seen in lower ionization states. If the expectation presented here is roughly correct, then the most important outflow phase for $M_\star \gtrsim 10^{10}$\msun moves quickly away from available ionized tracers and even past OVI bearing `coronal' gas into a much hotter x-ray traced plasma, where observational constraints are presently challenging.

In the lower left panel of Figure \ref{fig_outflowrate_phase} we similarly decompose the mass outflow rate now as a function of gas density, in the same stacked distributions as a function of stellar mass. A two component structure is also evident: the low density peak is associated to the high temperature gas, with the high density peak corresponding to the low temperature component which prominently emerges only at $M_\star > 10^{10}$\msun. As before, the location of the lower density component tracks the increasing halo gas density with higher host mass. On the other hand, the high density ISM component has a roughly constant maximal outflow rate for densities of \mbox{$n \sim 1$ cm$^{-3}$}, although the most massive galaxies above $M_\star > 10^{11}$\msun can launch outflows to 20 kpc which contain densities of $n > 10$ cm$^{-3}$ and higher. This also indicates a direct ejective origin for this dense material, having originally compressed in the center of the galactic potential. Given our star formation threshold density, some of this ejecta will therefore be star-forming and on the effective equation of state.

The lower right panel of Figure \ref{fig_outflowrate_phase} measures the metallicity of outflowing gas, in three stellar mass bins (colored lines) at $z=1$, $z=2$, and $z=4$ (solid, dashed, dotted). At all redshifts, the outflow metallicity increases with the stellar mass of the galaxy as expected. At redshift two, the dominant metallicity of wind material is $0.1 Z_\odot$ for $M_\star = 10^8$\msun, increasing to $0.3 Z_\odot$ at $M_\star = 10^{9.5}$\msun and reaching $0.7 Z_\odot$ by $M_\star = 10^{10.5}$\msun. The distributions are broad and gas at $\pm 0.5$ dex of these values is always also present at roughly 10\% of the mass outflow rate when compared to the dominant metallicity. At low masses the redshift evolution is roughly independent of stellar mass and fairly weak -- the average $Z_{\rm out}$ increases by $\sim$ 0.2 dex from $z=4$ to $z=2$ and by a further $\sim$ 0.1 dex to $z=1$. At high masses the redshift evolution is even shallower due to the flattening of the gas-phase mass metallicity relation. In both cases the evolution of outflow metallicity is approximately similar to the evolving MZR of TNG \citep{torrey18}, although we defer to future work a detailed comparison of wind metallicity relative to the galaxy ISM metallicity itself \citep[e.g.][]{muratov17}.


\subsection{Angular dependence: emergent bipolarity} \label{sec_results_angle}

\begin{figure}
\centering
\includegraphics[angle=0,width=3.3in]{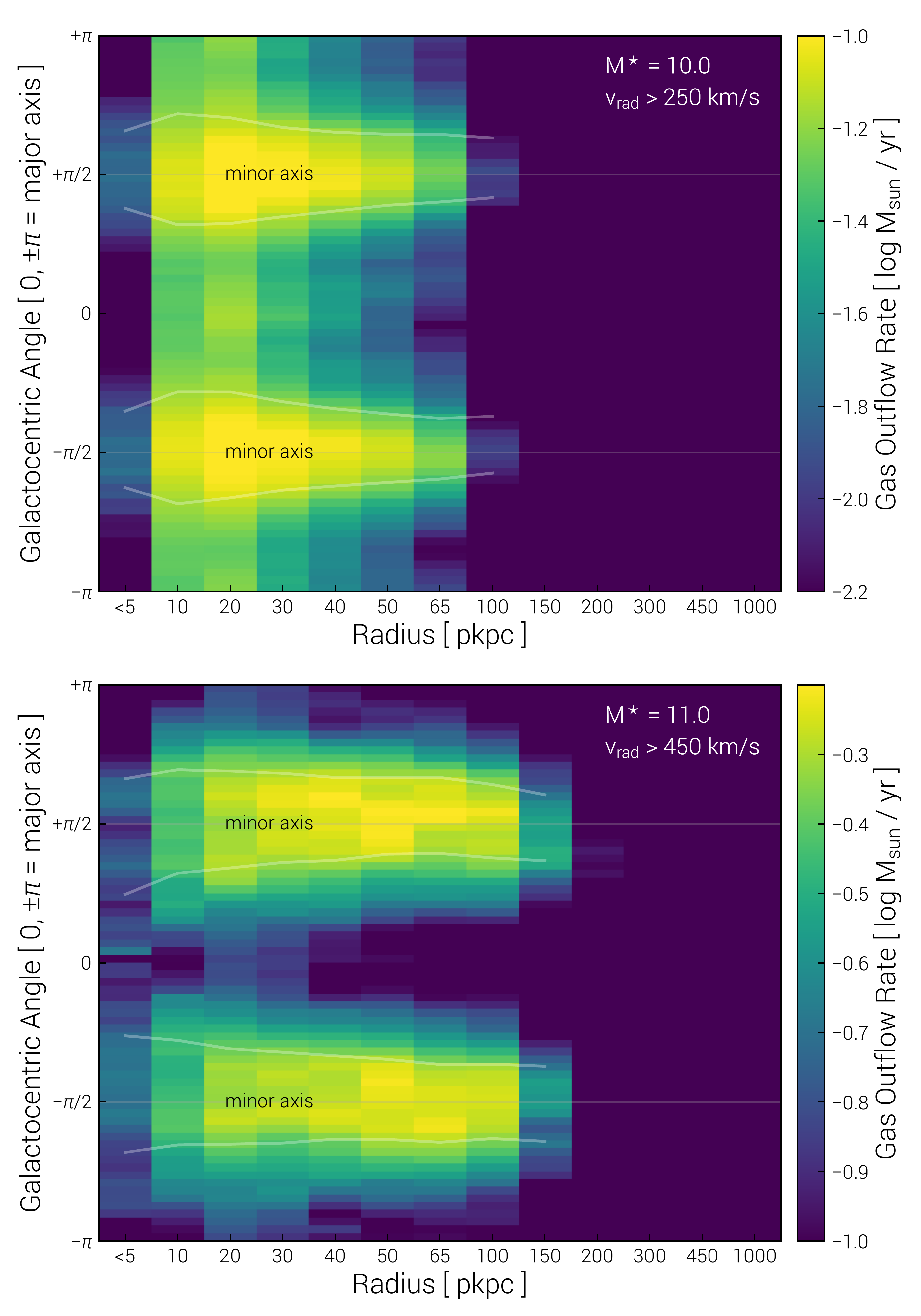}
\caption{ The dependence of mass outflow rate on galactocentric angle at $z=1$, where directions corresponding to the minor axes are $\theta = \pm \pi/2$, marked with horizontal gray lines. Each panel stacks all galaxies in mass bins centered on $M_\star \simeq 10^{10}$\msun (top; $\sim$ 280 systems) or $M_\star \simeq 10^{11}$\msun (bottom; $\sim$ 70 systems). These two mass scales are dominated by stellar and black hole driven outflows, respectively. White lines quantify the opening angle bounding half the total mass outflow at each radius. Despite the isotropy of both feedback models at the energy \textit{injection} scale, a hydrodynamic collimation naturally occurs and results in wide opening angle, bipolar outflows which preferentially escape along the minor axes of galaxies.
 \label{fig_outflowrate_rad_theta}}
\end{figure}

\begin{figure*}
\centering
\includegraphics[angle=0,width=6.8in]{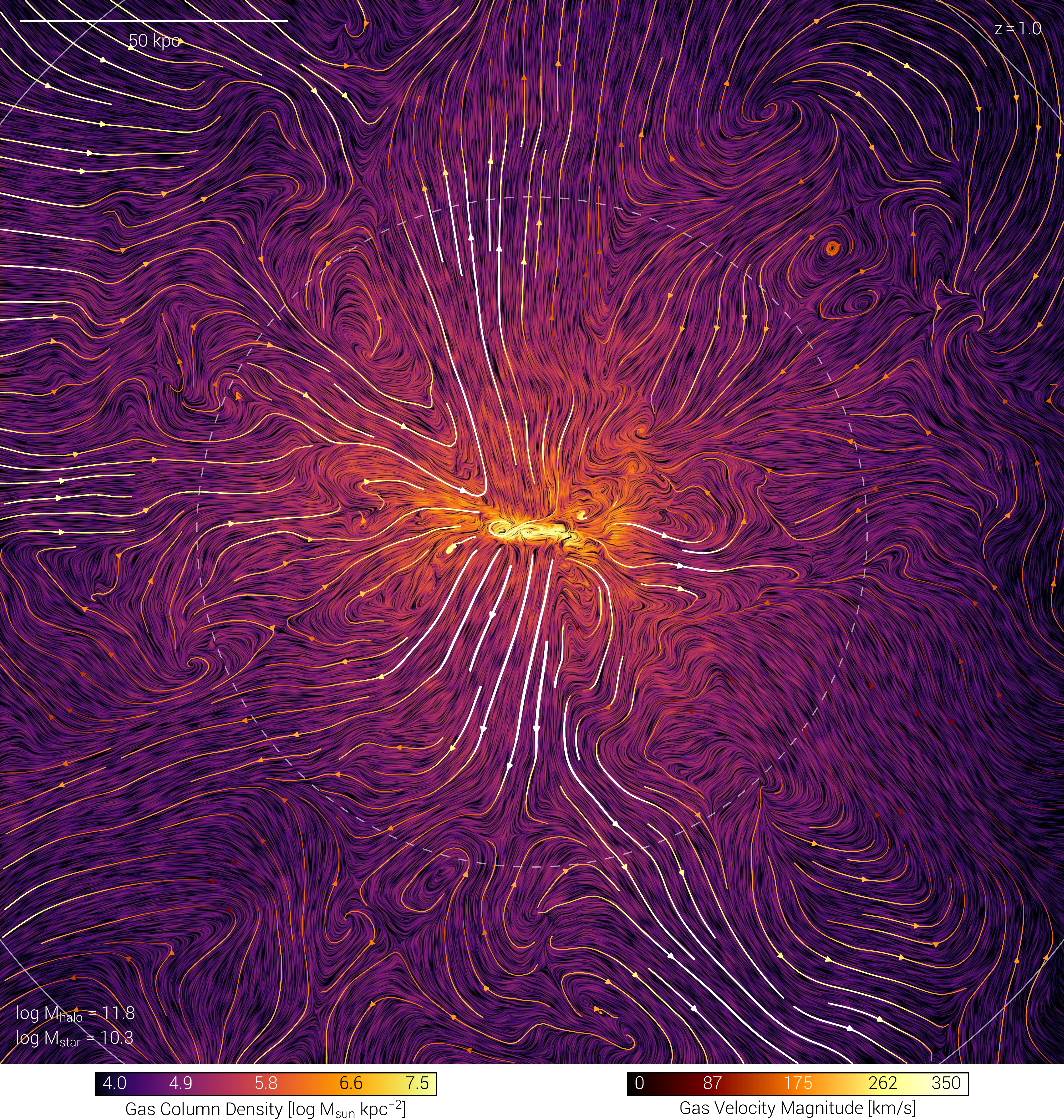}
\caption{ Visualization of the velocity structure of galactic outflows in the lower mass, stellar feedback driven regime. We show a single galaxy, rotated edge-on, with stellar mass $M_\star = 10^{10.3}$\msun and total $M_{\rm halo} = 10^{11.8}$\msun at $z=1$. If observed at higher redshift ($z \sim 3$) such a system would be similar to a Lyman-break galaxy. The field of view is 200 physical kpc, and the background image shows a gas density projection through an equal depth, modulated by the gas velocity field in a 10 kpc thick slice using the line-integral convolution technique \protect\citep[LIC;][]{cabral93}. Color indicates total gas density, while small-scale variation in black shows the local topography of gas motion. Overlaid on top we include streamlines of the same velocity field, to indicate direction and magnitude. Outflows which emerge collimated from the central galaxy traverse $r_{\rm vir}/2$ (dotted circle), establishing a large-scale, circulatory, galactic-fountain type flow largely confined within the virial radius of the halo (solid circle).
 \label{fig_vis_wind}}
\end{figure*}

In Figure \ref{fig_outflowrate_rad_theta} we show the dependency of the mass outflow rate (as the background color) on both galactocentric angle and galactocentric distance at $z=1$. Within each panel the two horizontal lines indicate the direction aligned with the minor axis of the galaxy.\footnote{We rotate each galaxy to an edge-on orientation, such that the longest axis is aligned with $\hat{x}$ (i.e. the $y=0$ line) when viewed in projection in the $x-y$ plane, and the angle $\theta$ for every gas cell is calculated from its $(x,y)$ coordinates in this projection, such that $\theta = \{0, \pm \pi\}$ corresponds to the $\pm x$ (major) axes, respectively, while $\theta = \pm \pi/2$ corresponds to alignment along the $\pm y$ (i.e. minor) axes, respectively.} The top panel stacks relatively low-mass galaxies with $M_\star \simeq 10^{10}$\msun ($\sim$ 280 systems), a mass regime where outflows are generated by stellar feedback. The bottom panel stacks high-mass galaxies with $M_\star \simeq 10^{11}$\msun ($\sim$ 70 systems), where higher-velocity outflows result almost exclusively from BH feedback.

In both cases the conclusion is the same: the mass outflow rate of winds is not directionally isotropic. Rather, strong $\dot{M}$ outflows are preferentially found aligned with the minor axes of galaxies. This is particularly intriguing in the context of TNG, as it represents an emergent quality of galactic-scale outflows. In the TNG model, wind launching and energy injection from stellar as well as black hole feedback is entirely isotropic, in the time-average, multiple event sense. For stellar feedback driven winds, the directionality of the wind-phase mass is entirely random at launch \citep[see][]{pillepich18a}, while BH energy injection in the low accretion state kinetic mode, which drives the outflows seen here at high-mass, is also entirely random with each injection event \cite[see][]{weinberger17}. The preferential propagation of outflows along the minor axis of a galaxy therefore represents a natural, hydrodynamical collimation of the flow. White lines demarcate the `half mass outflow rate angle', quantifying a measure of the opening angle; this is $\theta_{\rm 1/2} \simeq 70^{\circ}$ at $r \sim 10$ kpc and $\theta_{\rm 1/2} \simeq 40^{\circ}-50^{\circ}$ at $r \sim 50$ kpc. 

Figure \ref{fig_outflowrate_rad_theta} quantifies the visual impression provided by Figures \ref{fig_timeevo1} and \ref{fig_timeevo2} for the case of a strong AGN driven outflow. To complement this high mass, high outflow velocity regime, we visualize the velocity structure of a smaller galaxy with $M_\star = 10^{10.3}$\msun ($M_{\rm halo} = 10^{11.8}$\msun) in Figure \ref{fig_vis_wind}. Such a system observed at a higher redshift would be similar to a Lyman-break galaxy. Here we show a gas density projection of the galactic and halo gas, alloyed with its instantaneous velocity field using the line-integral convolution technique. Small-scale transport and swirling, vortical motions are revealed in regions without strong bulk outflow or inflow. Overlaid on top, streamlines indicate flow direction and velocity, with outflows reaching up to $\sim$ 350 km/s as they establish a circulatory, galactic fountain like baryon cycle across $r_{\rm vir}/2$ (dotted circle).\footnote{In addition to this one halo, a comprehensive set of examples, spanning a range of galaxy masses and redshifts, is available on the TNG50 website at \url{www.tng-project.org/explore/gallery/}.}

\begin{figure*}
\centering
\includegraphics[angle=0,width=6.0in]{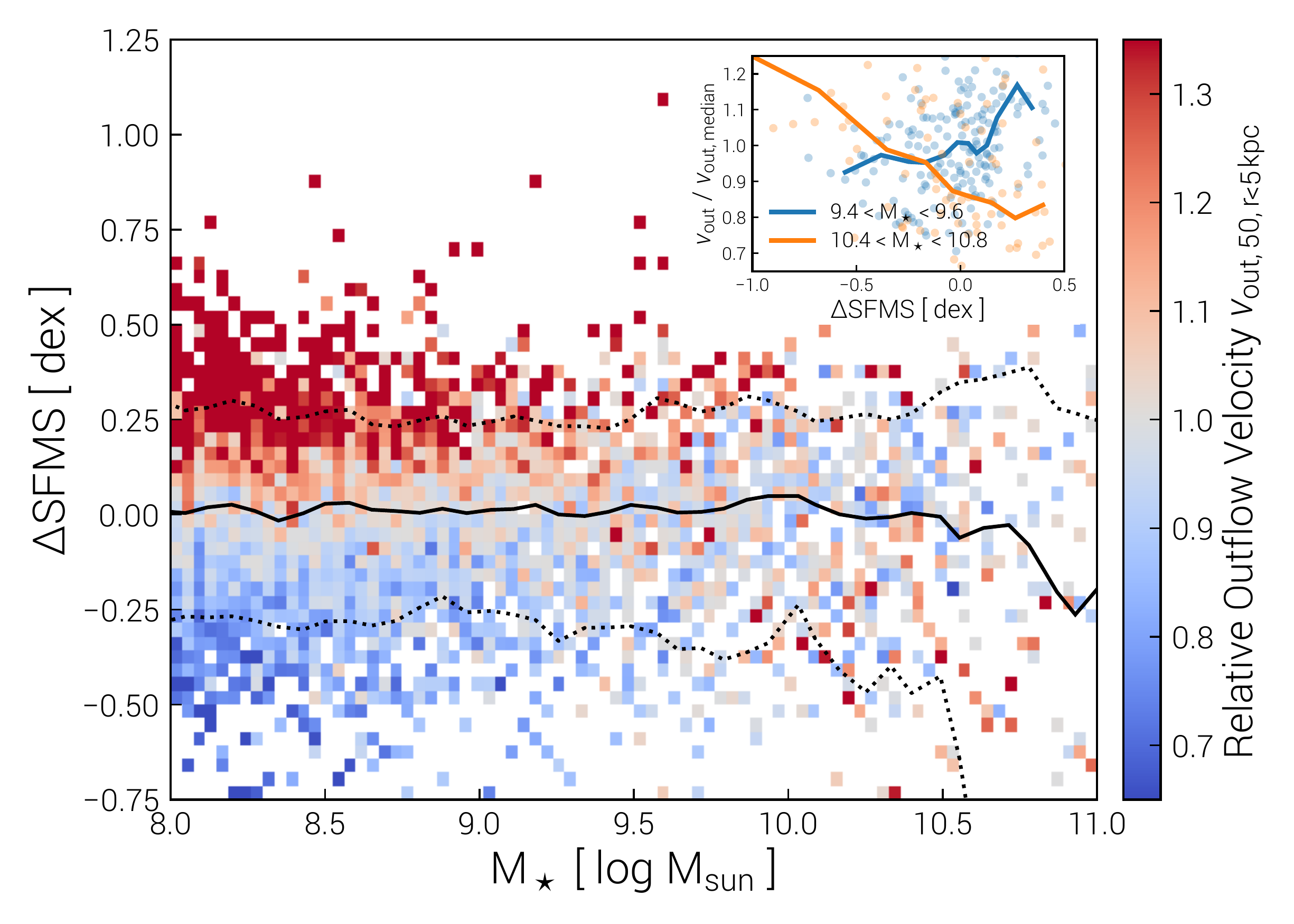}
\caption{ Dependence of outflow velocity $v_{\rm out}$ on offset from the star-forming main sequence $\Delta$SFMS at $z=1$. In the main panel, color encodes the outflow velocity with respect to the median at that stellar mass -- i.e., each column is normalized by its median value, and the dynamic range shown is $\pm$35\%. Across most of the mass range, a strong correlation between $v_{\rm out}$ and $\Delta$SFMS is present, such that outliers above the main sequence drive characteristically faster outflows. This relation inverts at $M_\star \gtrsim 10^{10.5}$\msun, whereby massive galaxies show an anti-correlation -- outliers below the main sequence, i.e. quenched or quenching systems, now drive the fastest outflows, as a result of BH feedback (red points appearing below the solid black line). These two trends are quantified in the inset panel, which shows the trend of relative $v_{\rm out}$ as a function of $\Delta$SFMS for two different stellar mass bins: low-mass star-forming (blue; positive slope) and high-mass (orange; negative slope). The magnitude of the effect is $\sim$30-50\% in velocity across $\sim$1 dex in $\Delta$SFMS, and holds up to $z=6$ at least.
 \label{fig_vout_delta_sfms}}
\end{figure*}

We speculate that this collimation is largely a `path of least resistance' effect, whereby outflows directed into the plane of a disk, comprised of the densest material, experience enhanced resistance. Consequently, we find that a roughly bipolar or biconical outflow is typical behavior. Sculpting of the flow is evident already by $\sim 10$ kpc in the inner halo, and persists out as far as strong outflows are found ($\gtrsim 50$ kpc at least). The inferred opening angle is large, decreasing for stronger outflow thresholds which become more concentrated along the minor axes. We note that the degree of this collimation appears to evolve with redshift (not shown), becoming stronger at late times and emerging particularly from $z=2$ to $z=1$. It is difficult to identify such a clear signal even at $z=2$. The degree of collimation is likely to depend on galaxy morphology, particularly the amount of rotational support of the system.

Observationally, preferential detection of outflow signatures along the minor axes of galaxies is a common conclusion at low-z \citep{heckman00} based on SDSS \citep{chen10,concas17b,bae18} and also towards $z \sim 1$ \citep{martin12,kornei12,bordoloi14c,rubin14}. At the same time, data from even higher redshifts $z \geq 2$ support a lack of collimation at such early times \citep{law12}. Typical evidence along these lines comes from cold gas-phase tracers such as MgII or NaD, or hydrogen, and the association is typically with stellar feedback driven outflows, although asymmetric outflows from low-luminosity black holes may also be common \citep{cheung16,wylezalek17,roy18}. 
For hotter phases, \cite{kacprzak15} find OVI around $z \leq 0.7$ galaxies preferentially along either the major or minor axes (e.g. inflows and outflows, respectively), but not in between; similarly so for colder MgII absorbers at $z \geq 0.3$ \citep{nielsen15}.

In addition to spanning both feedback regimes, we also find no clear temperature dependence of this signal. We therefore expect (i) hotter outflowing gas phases to be similarly collimated, (ii) bipolar geometry outflows to be present across a broad stellar mass range, and (iii) a lack of similar signatures at high redshift ($z>2$; not shown). We speculate that this collimation arises only towards lower redshift due to the rise of ordered rotation and settling of gaseous galactic disks which helps to sculpt the flows.

Theoretically such a natural collimation was anticipated already by \cite{tomisaka88} whose numerical simulations demonstrated galactic-scale superwind emergence preferentially elongated along the direction of the minor axis of a launching disk, caused by propagation in the direction of the maximum pressure gradient of the confining gaseous halo. Early wind models for global simulations sometimes including such hard-coded directionality by hand \citep{spr03}, including in the original Illustris model \citep{vog13}, but this was removed in TNG mainly as an opportunity to simplify an unnecessarily complex model feature \citep{pillepich18a}.


\subsection{Relating outflow and host galaxy properties} \label{sec_results_vsgal}

With the exception of global stellar mass trends, we have not yet explored the dependence of outflows on additional properties of the galaxies from which they are launched. Here we specifically consider star formation activity (SFR) as well as black hole activity ($L_{\rm bol}$, $\lambda_{\rm edd}$). As both types of feedback activity are a strong function of galaxy mass, we naively expect a strong scaling of most outflow properties such as $v_{\rm out}$ with $M_\star$, and move beyond such zeroth order expectations by predominantly investigating trends at fixed mass.

In Figure \ref{fig_vout_delta_sfms} we consider the dependence of outflow velocity on the star formation rate of a galaxy, cast in terms of $\Delta$SFMS, the deviation from the median star-forming main sequence relation. To isolate out the general trend of increasing velocity with $M_\star$, we further normalize $v_{\rm out}$ by its median value in each stellar mass bin. Color then indicates higher (red), lower (blue), or typical (light gray) outflow velocity relative to galaxies at that same mass. As our operating definition of outflow velocity we take $v_{\rm out,50,<5kpc}$, the 50th velocity percentile of outflowing mass near the galaxy. The trends we discuss do not depend strongly on this choice.

First, we find a clear correlation between $v_{\rm out}$ and $\Delta$SFMS for star-forming galaxies, whereby outliers above the main sequence launch faster winds, whereas outliers below the MS drive slower outflows. This statement is true at fixed stellar mass, and holds from $z=1$ up to $z=6$ if not higher.

However, this correlation actually \textit{inverts} at $M_\star \gtrsim 10^{10.5}$\msun. Galaxies which have grown above this mass scale drive faster outflows if they are below the main sequence, i.e. if they are in the process of quenching towards the red population. In the main panel, this is visible as an excess of red pixels for low $\Delta$SFMS at high $M_\star$. We quantify this anti-correlation in the inset panel, which plots relative $v_{\rm out}$ as a function of $\Delta$SFMS for high-mass galaxies (orange) versus low-mass star forming galaxies (blue). The two slopes have opposite sign, and the strength of this relation is such that $v_{\rm out}$ varies by $30-40$\% for outliers offset by $\Delta$SFMS = $\pm 0.5$ dex. 

Observationally, \cite{chisholm15} correlate outflow velocity and SFR in nearby star-forming galaxies, finding that mergers identified via optical morphology drive faster outflows, and that merging systems have the fastest outflows overall. If mergers are preferentially found at $\Delta$SFMS $>$ 0, this may be a similar signal as discussed here. \cite{cicone16} explicitly compare outflow velocity measurements with $\Delta$SFMS (their Fig. 21) for a `normal' SDSS star-forming population, finding a clear positive trend above the main-sequence, with a weak or flat trend below. Towards higher masses, \cite{sato09} use a SF-agnostic sample from EGS to uncover a continuation of detectable NaI outflows into the red population, concluding that outflows outlive the star-formation phase and contribute to quenching galaxies en route to the quiescent population. Our analysis here implies that these `relic' outflows may also be mixed with more recent outflows driven by black holes rather than supernovae.

We note that for star-forming galaxies a large number of properties correlate strongly with $\Delta$SFMS. That is, (a) outliers above the main sequence could be outliers because they drive faster outflows, or (b) they could simply have faster winds because they are already part of the $\Delta$SFMS $>$ 0 population for another reason. In this regime, relative to those systems below the median main-sequence at fixed stellar mass, galaxies have: faster outflows (as noted), lower gas-phase metallicities, higher gas fractions, lower central stellar surface densities, lower stellar metallicities, bluer (g-r) colors, larger stellar sizes (half mass radii), lower outflow mass loadings, less rotationally supported gas disks, higher halo masses, and lower mass BHs (all not shown, but see \textcolor{blue}{Pillepich et al. 2019}). Oscillation and quasi-equilibria about the evolving SFMS leads to an interrelation between many of these galaxy properties. Although the root cause of the trend of $v_{\rm out}$ with $\Delta$SFMS is difficult to determine, and to some degree both (a) and (b) must apply, we note that the injection scaling of $v_{\rm out} \propto \sigma_{\rm DM}$ in the TNG model is sensitive to a spatially localized measurement of the dark matter velocity dispersion, which would naturally respond to merger activity (i.e. where $\sigma_{\rm gas}$ would also be enhanced) in contrast to more isolated, dynamically quiet environments. Likewise, the positive correlation of $M_{\rm halo}$ with $\Delta$SFMS implies larger $\sigma_{\rm DM}$ at fixed $M_\star$ \citep[similar to the finding of][with the EAGLE model]{matthee17}.

\begin{figure}
\centering
\includegraphics[angle=0,width=3.3in]{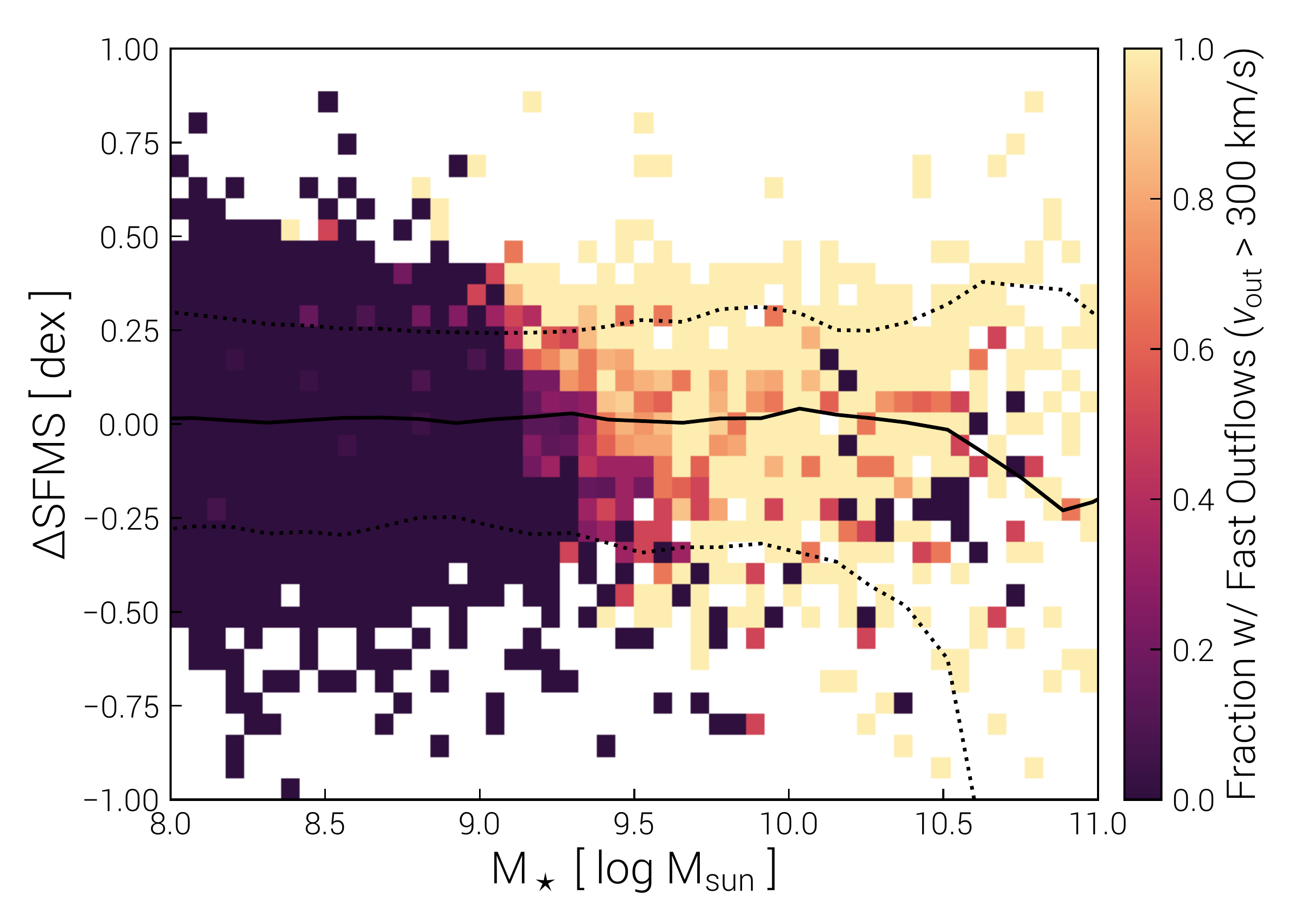}
\caption{ The fraction of galaxies with `fast' outflows, as a function of their position on the $\Delta$SFMS-M* plane at $z=1$. Here we take one possible choice and define $v_{\rm out,90,r=10kpc} > 300$ km/s as the threshold for a fast outflow. The color of each pixel then shows the fraction of galaxies in each bin which satisfy this cut, from zero (dark purple) too one hundred percent (yellow). Galaxies with fast outflows exist only above a diagonal threshold, from the upper left to lower right, in this plane.
 \label{fig_fastout_frac}}
\end{figure}

In order to reincorporate the dominant trend of increasing $v_{\rm out}$ with stellar mass, Figure \ref{fig_fastout_frac} presents a view on the fraction of galaxies with `fast' outflows, with respect to their position on the \mbox{$\Delta$SFMS$-M_\star$} plane. We do this by defining $v_{\rm out,90,10kpc} > 300$ km/s as a somewhat arbitrary though useful threshold. For example, if observational detectability of an outflow depended mainly on the existence of a component traveling above 200 km/s (i.e. above an instrumental limit or intrinsic level), our choice would map to an outflow detection fraction. The color of each pixel shows the fraction of galaxies in that bin which satisfy the velocity cut, from none (dark purple) to 100\% (yellow). As a common feature, we find a diagonal boundary across the $\Delta$SFMS$-M_\star$ plane, with fast outflow galaxies residing towards the upper right. This can be understood as the superposition of the $v_{\rm out}-\Delta$SFMS secondary correlation on top of the dominant $v_{\rm out}-M_\star$ relation. A higher threshold velocity moves this boundary to the right, with roughly constant slope.

This trend can be compared in spirit to Fig. 17 of \cite{rubin14} at $z \sim 0.5$ based on MgII absorption modeling, or Fig. 3 of \cite{forsterschreiber18b} at $z=0.6 - 2.7$ based on broad emission line decomposition. Both hint towards a tilted trend in $\Delta$SFMS at fixed $M_\star$, in the sense that galaxies above the main sequence have a larger detected outflow fraction, while galaxies approaching $10^{11}$\msun and higher, even those below the extension of the main-sequence, do likewise. Similarly, Fig. 18 of \cite{cicone16} based on OIII emission from ionized outflows at $z \lesssim 0.7$ also highlights higher outflow velocities at positive $\Delta$SFMS. We emphasize such similarities are largely speculative and hindered by the current inability to make quantitative comparisons to these observables, as discussed below in Section \ref{subsec_discussion_comparisons}. 

Given our current inability to do a full forward-modeling, and the uncertainties inherent with inversion in the opposite direction starting from the observational quantities, our ability to make a robust comparison between the simulations and observations is presently limited. However, we would like to give a sense of the overlap between the outflow properties expected from the TNG model and available observational datasets. We therefore provide a series of suggestive plots, which are intended to be \textit{qualitative only} in nature, and from which \textit{no direct statements can be made} as to the level of (dis)agreement between the simulations and observations. Further modeling efforts from the theory side are needed before we can contrast against, or provide interpretation of, the wealth of high-quality observational data on outflows \citep[see][]{rupke18}.

\begin{figure*}
\centering
\includegraphics[angle=0,width=3.4in]{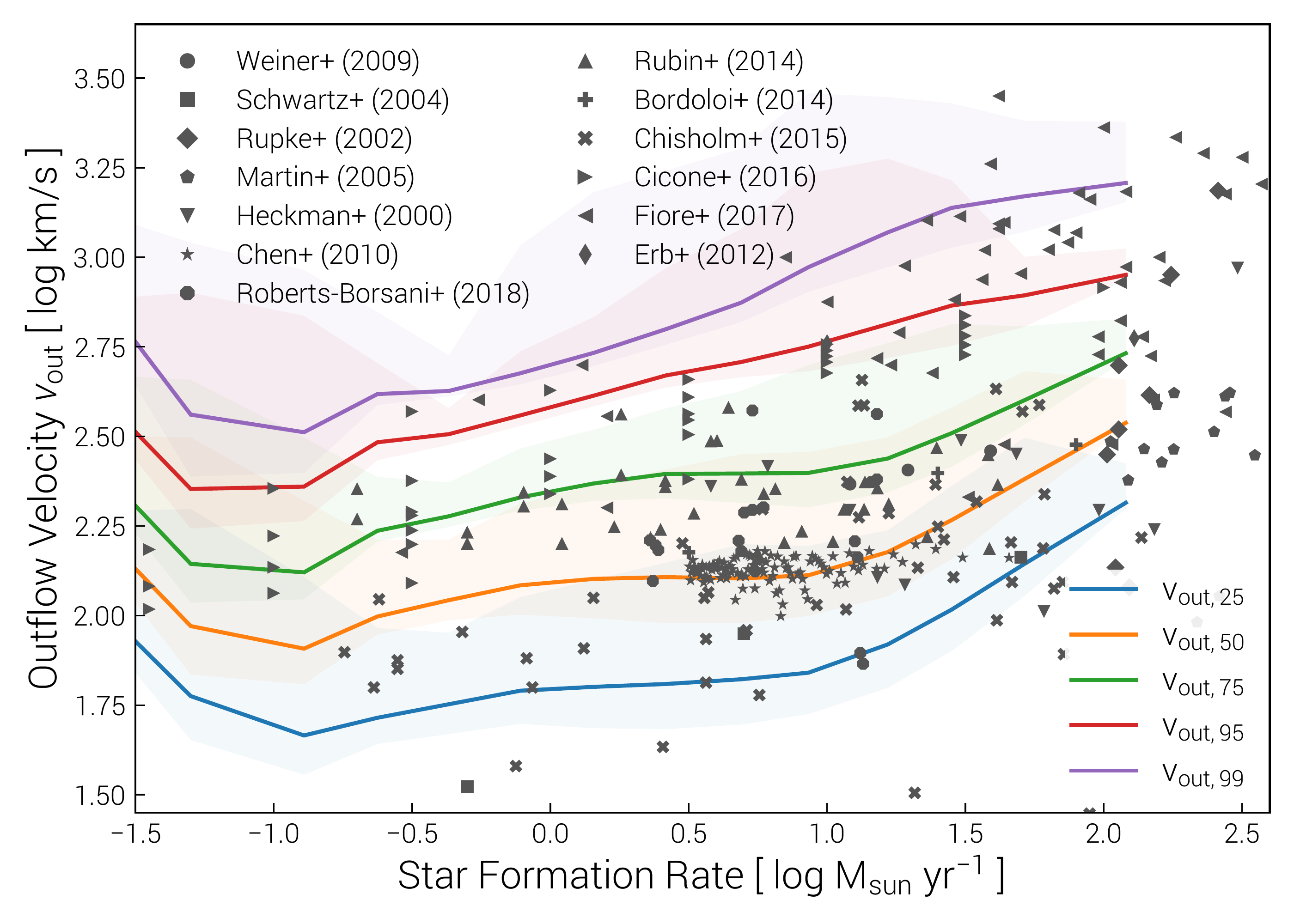}
\includegraphics[angle=0,width=3.4in]{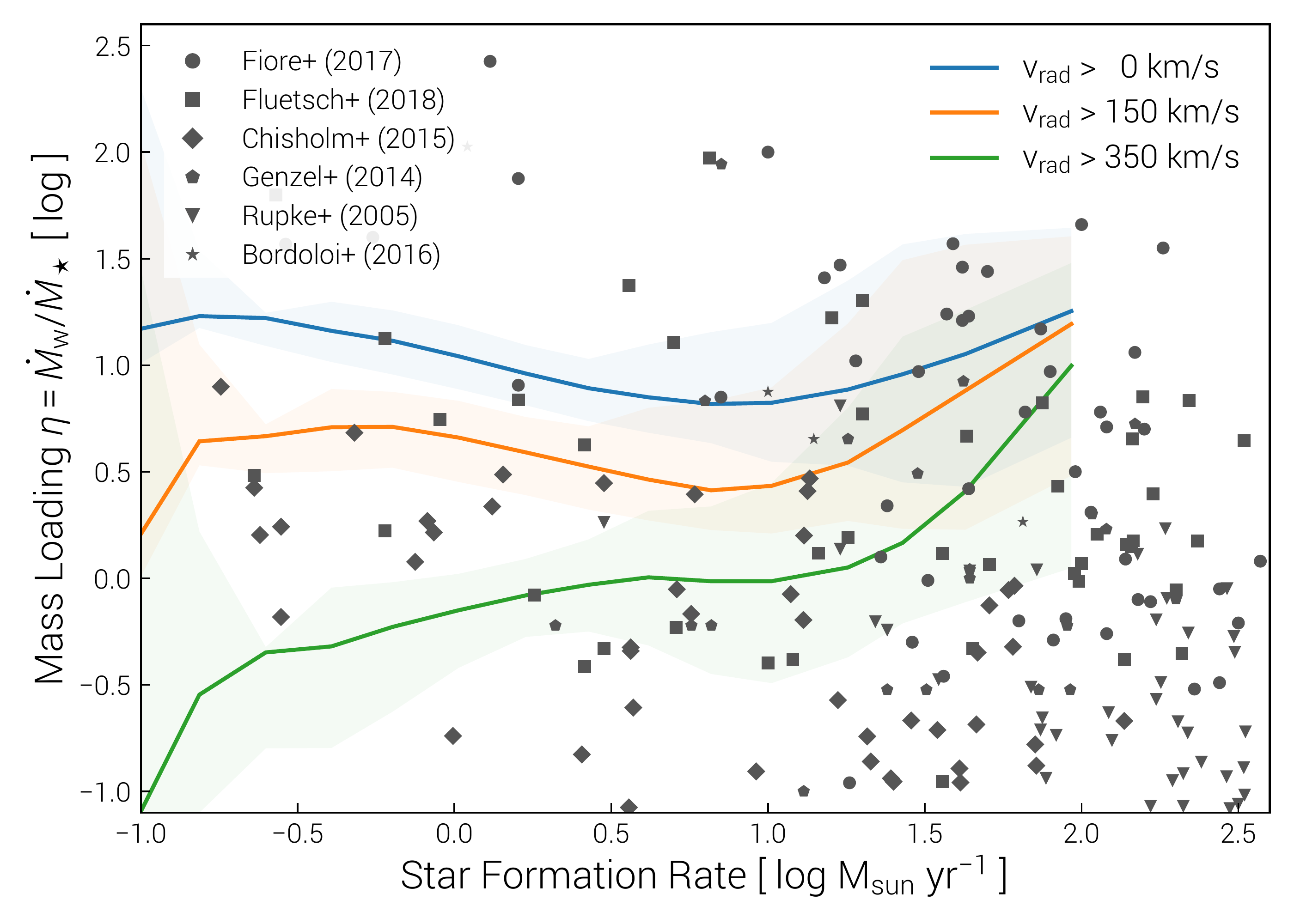}
\includegraphics[angle=0,width=3.4in]{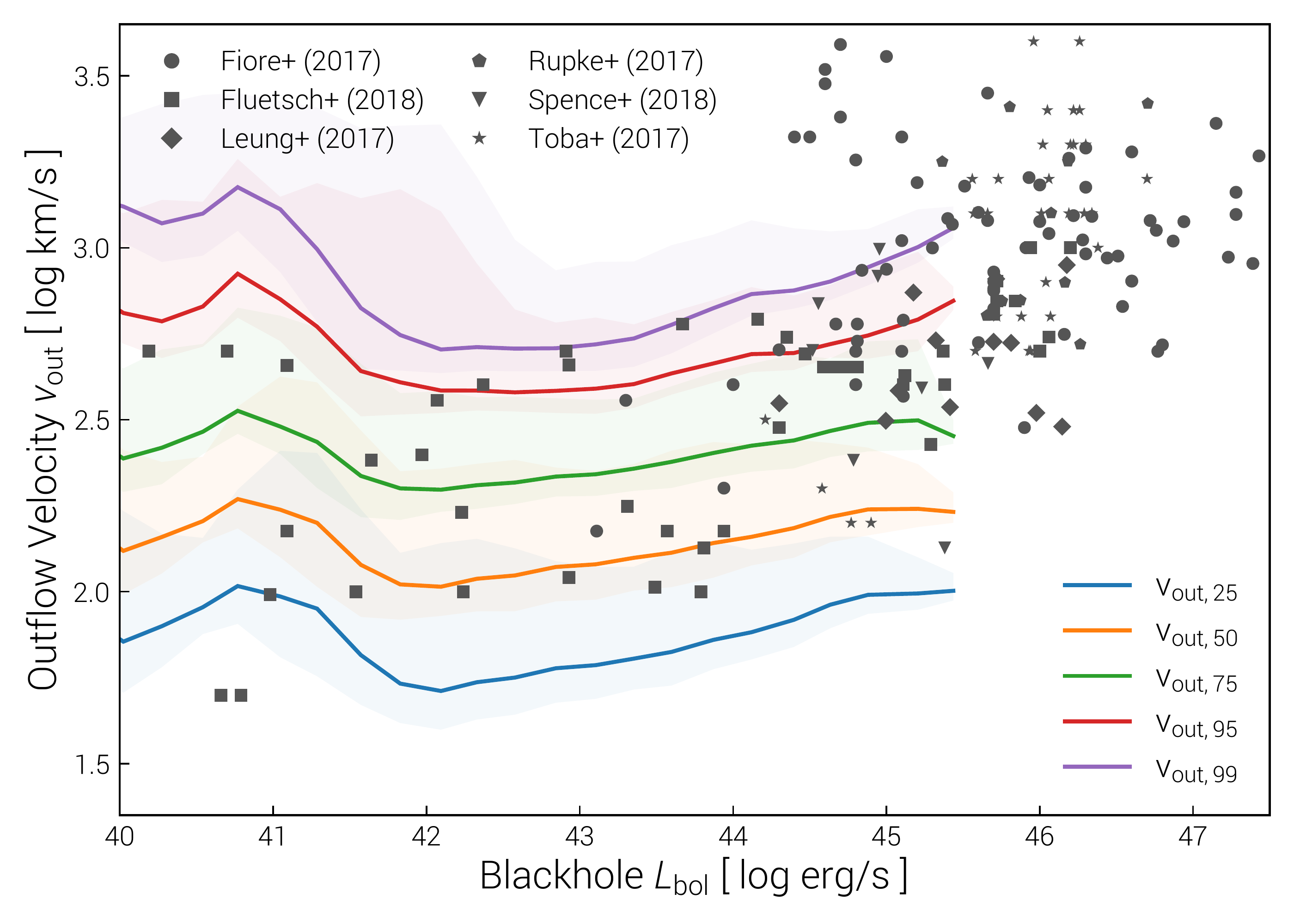}
\includegraphics[angle=0,width=3.4in]{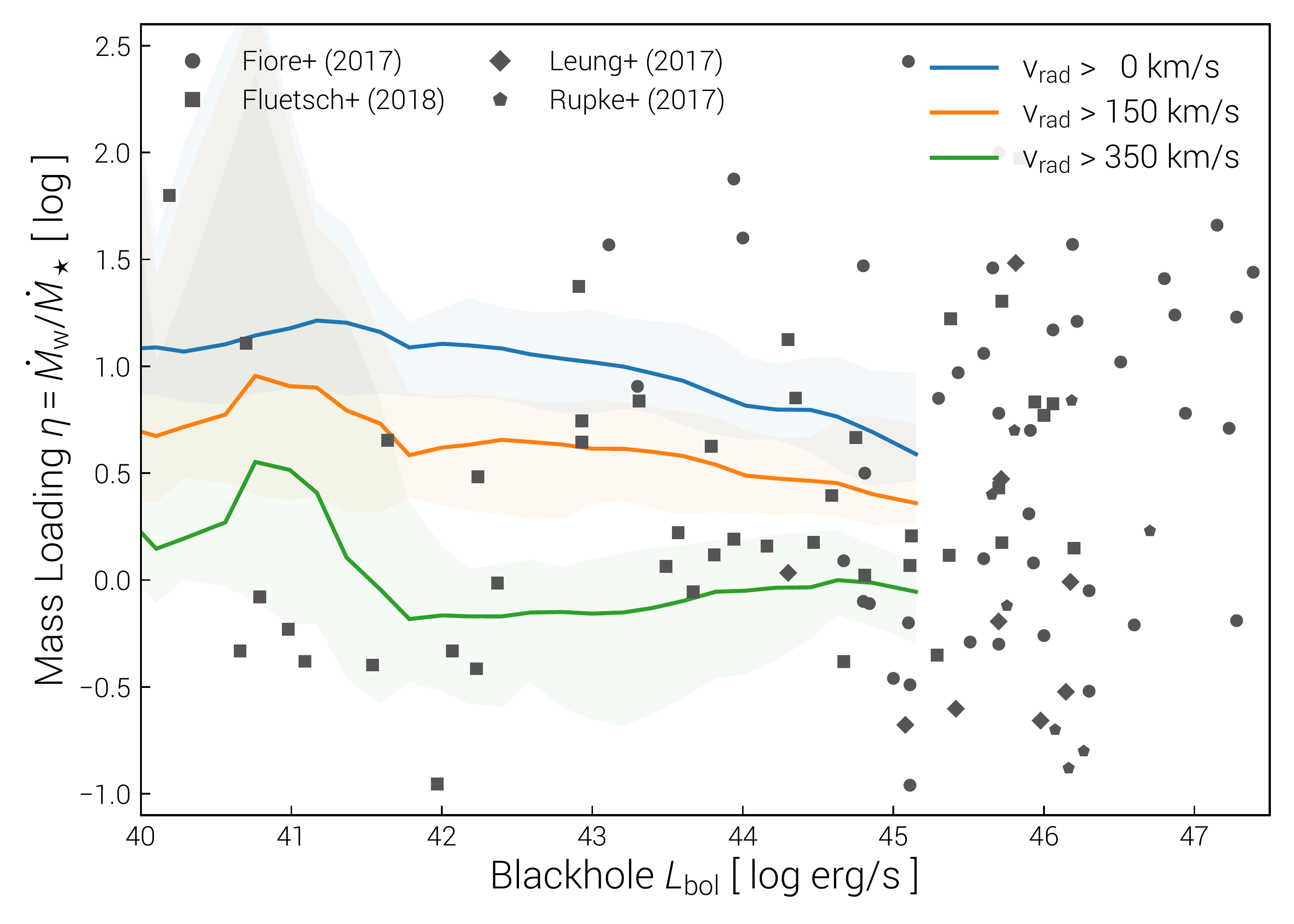}
\includegraphics[angle=0,width=3.4in]{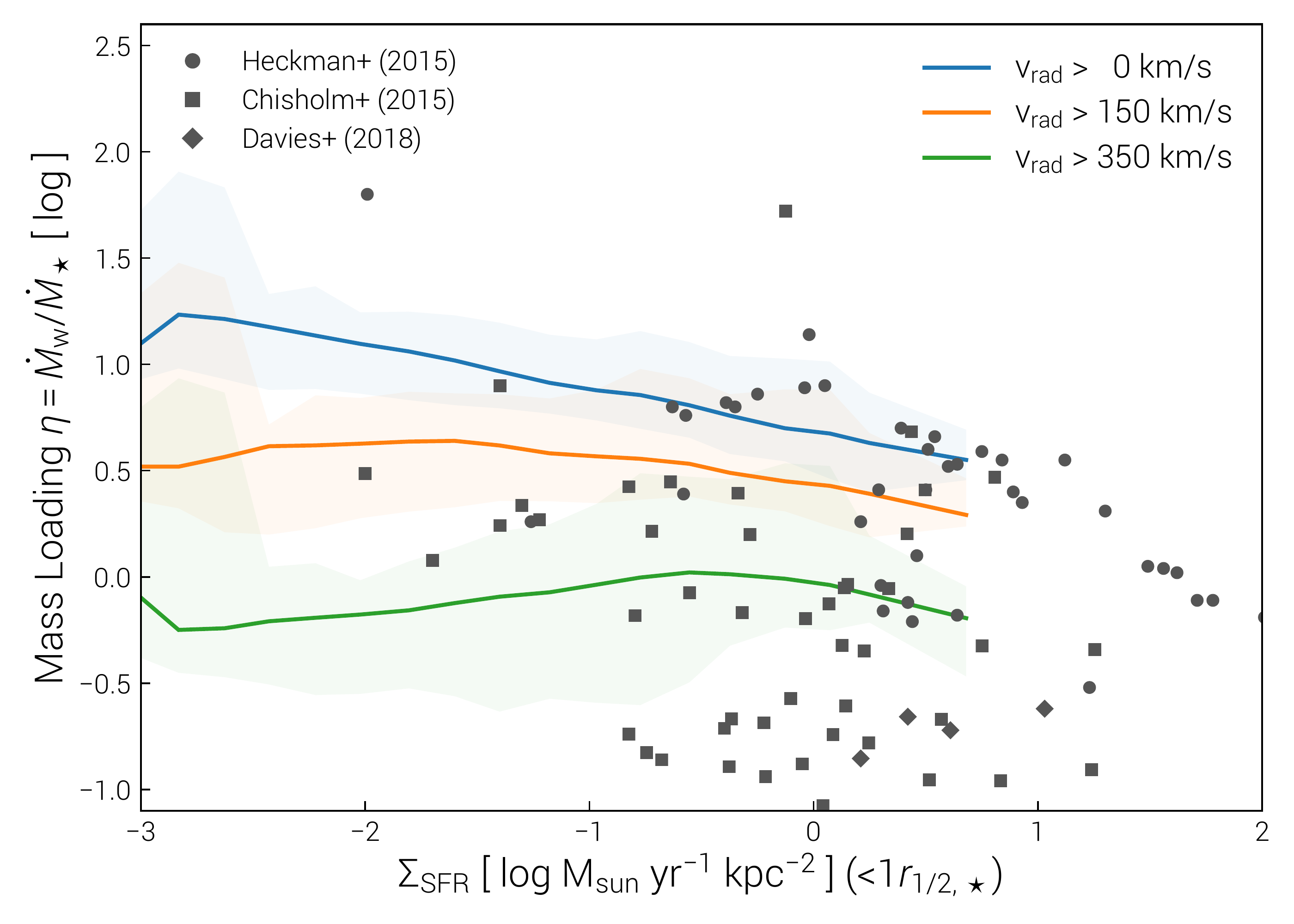}
\includegraphics[angle=0,width=3.4in]{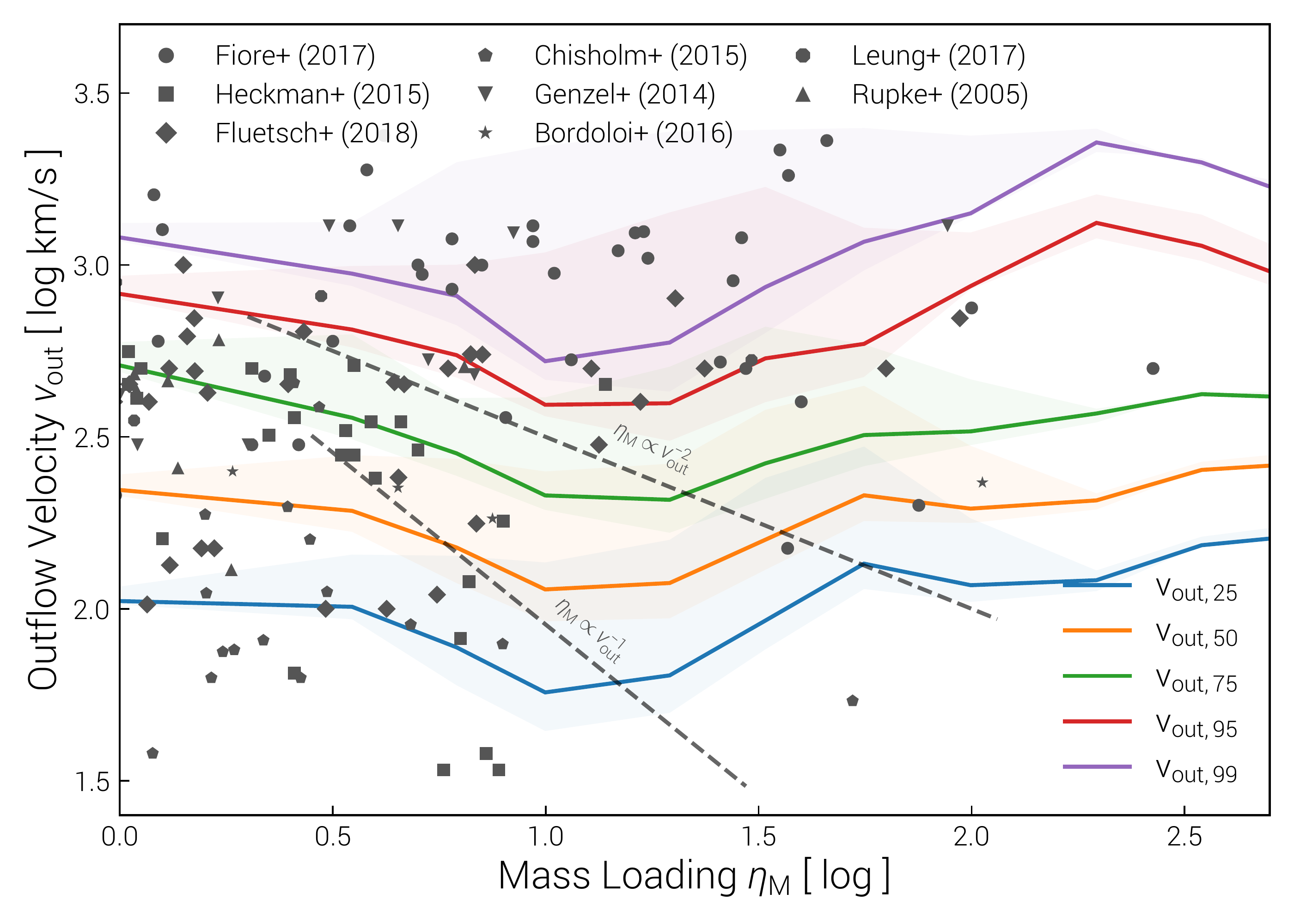}
\caption{ Scalings of outflow velocity and mass loading factor with galaxy or black hole properties, in comparison to many literature results and compilations. The upper two panels show $v_{\rm out}$ and \etaM as a function of star formation rate. The middle two panels show $v_{\rm out}$ and \etaM as a function of BH bolometric luminosity. The bottom two panels show mass loading as a function of the surface density of star formation $\Sigma_{\rm SFR}$ as well as $v_{\rm out}$ versus \etaM. All simulation lines are shown at $z=1$ and adopt a consistent minimum stellar mass of $M_\star \geq 10^{9}$\msun, but otherwise include all galaxies as representative of a mass complete volume-limited sample. Expanded percentile bands at the [5,95] levels (i.e. $2\sigma$) are shown to highlight the existence of outliers and tails of the distributions. To emphasize the definitional range, each $v_{\rm out}$ panel includes five velocity percentile ranges ($v_{25}$ to $v_{99}$), and each \etaM panel includes three velocity cut criteria (0 - 350 km/s). Observations are taken from \protect\cite{heckman00,rupke02,schwartz04,rupke05c,martin05,weiner09,chen10,erb12,rubin14,genzel14,bordoloi14b,chisholm15,heckman15,bordoloi16,cicone16,leung17,fiore17,rupke17,toba17,fluetsch18,robertsborsani18,spence18,davies18}, and cover a large diversity of galaxy sample selection functions and biases, operational definitions, measurement techniques, `star formation driven' as well as `AGN driven' outflows, gas phase/ion, and redshift -- no homogenization has been undertaken. We highlight the occupation of existing observational data in these six planes, contrasting against intrinsic trends from the simulation, and emphasize that \textit{no direct comparison is intended or can be made between the observations and simulations}. Many observations include or specifically target rare galaxy types, such as ULIRGS or ultra-luminous AGN, which are not sampled in the TNG50 volume (e.g. SFR $\geq 100$ \msun yr$^{-1}$, or $L_{\rm bol} \geq 10^{45}$ erg/s). See text for discussion.
 \label{fig_outflows_vs_obs}}
\end{figure*}

In Figure \ref{fig_outflows_vs_obs} we show a number of correlations between outflow properties and central galaxy or supermassive black hole properties. In these various spaces we add a diverse set of observations, with citations indicated in the caption. These span all redshifts, at least as high as $z \sim 4$, and include many local universe data points from $z \sim 0$. They also span a range of sample selections -- in particular, a significant number specifically target some of the most energetic nearby ULIRGs or the most luminous observable starbursts. All detectable gas phases are represented, including molecular, neutral, and ionized tracers. Importantly, we combine datasets where outflows are believed to arise from stellar feedback as well as BH feedback -- their separation being often difficult.

In addition, outflow properties including velocity and mass loading factors are measured in a diversity of ways and with a number of definitions across the observational works, and no homogenization has been undertaken. For example, outflow velocities may be derived from line center offsets from systemic, line wings below some percentile of the continuum, or based in some way on line width, with or without inclination corrections. Error bars are excluded in every case for clarity.

All simulation results are restricted to a consistent minimum stellar mass of $M_\star \geq 10^{9}$\msun, but otherwise include galaxies of all stellar masses as representative of a mass complete volume-limited sample, and are shown at $z=1$ (colored lines). To highlight the definitional range, we include five velocity percentile ranges ($v_{25}$ to $v_{99}$) for $v_{\rm out}$, or three velocity cut criteria (0 - 350 km/s) for \etaM, as appropriate. Measurements are made at $r = 10$ kpc; varying galactocentric distance would produce an additional fan of lines in each panel. Observational data points are shown as black symbols.

\begin{itemize}

\item The upper left panel shows the relationship between $v_{\rm out}$ and star formation rate of the central galaxy. The stellar mass of galaxies in this plane increases monotonically from left to right, while outliers which scatter towards high $v_{\rm out}$ at a given SFR are massive, low-sSFR, and black hole outflow dominated. Although outflow velocity increases with SFR, the slope is fairly shallow: from \mbox{$0.1 \,\rm{M}_\odot \,\rm{yr}^{-1}$} to \mbox{$100 \,\rm{M}_\odot \,\rm{yr}^{-1}$} the outflow velocity increases by a factor of $\sim 3-5$. Observational datasets which span only a small dynamic range in SFR are unlikely to robustly identify a trend and would conclude a flat $v_{\rm out}$ \citep[as appreciated in e.g.][]{chen10}. We highlight the results of \cite{cicone16}, where we have taken the LoSVD $v_{\rm 0.1}$ (OIII) values based on a consistent analysis of a large, low-z sample which spans nearly our entire SFR range and shows an encouraging similar scaling as $v_{\rm out,95}$ for instance. We note that due to the finite volume of TNG50, there is only a partial sampling of high $\Delta$SFMS outliers and very few galaxies with star formation rates above 100 $\rm{M}_\odot \,\rm{yr}^{-1}$. 

\item With the upper right panel we show the dependence on the mass loading factor \etaM with the SFR of the galaxy. As above, the stellar mass of galaxies in this plane increases monotonically from left to right, while outliers which scatter towards high \etaM at a given SFR are quenching (or quenched) and black hole outflow dominated. Note that many observations present \etaM of a specific gas phase which is only a sub-component of the total. Many of these \citep[e.g.][]{chisholm15} imply to zeroth order a trend of decreasing \etaM with SFR, but as we have shown in Section \ref{sec_results_multiphase} higher SFR (and so $M_\star$) galaxies progressively shift the majority of their outflowing mass into a difficult to observe hot phase. Without a phase-specific modeling of \etaM we cannot robustly compare to trends in SFR, other than to comment that a large fraction of observed points, which are predominantly at $1.0 < \log(\rm{SFR} / \rm{M}_\odot \,yr^{-1}) < 2.5$ have lower inferred mass outflow rates than the phase-total rates of TNG.

\item In the center left panel we show the dependence of outflow velocity on the bolometric luminosity of the black hole. As with extreme starbursts, the volume of TNG50 is also too small to host the brightest luminosity AGN. There are essentially no black holes with $L_{\rm bol} > 10^{45}$ erg/s, while the majority of observations of BH-driven outflows focus on this regime of bright quasars which host $1000$ km/s or greater winds, which are offset from our simulated sample. When observations of `normal' galaxies with black holes exist, inferred values for $v_{\rm out}$ cover the range of velocity percentiles from TNG \citep[][where $v_{\rm out}$ = FWMH$_{\rm broad}$/2 + $|v_{\rm broad} - v_{\rm narrow}|$]{fluetsch18}. On this plane, stellar mass increases with $L_{\rm bol}$ from left to right, but also with $v_{\rm out}$ from bottom to top. In particular, the tails visible at low black hole luminosities are the most active BH-driven outflows in the simulation. 

\item The center right panel shows \etaM as a function of $L_{\rm bol}$ of the central black hole. Depending on velocity threshold, the trend can be slightly decreasing, flat, or increasing with $L_{\rm bol}$, but is weak in any case. In comparing to observed trends, the previous caveats about phase specific mass outflow rate tracers apply. In TNG, the scatter in mass loading increases strongly below $L_{\rm bol} < 10^{42}$ erg/s, where low and high mass galaxies are simultaneously present, with the latter driving high \etaM outflows. 

\item The lower left panel highlights the relation between \etaM and $\Sigma_{\rm SFR}$, an integrated star formation rate surface density, computed within one times the stellar half mass radius (roughly similar to $R_{\rm e}$). As before, depending on velocity threshold, the trend of mass loading with galaxy-integral $\Sigma_{\rm SFR}$ can be negative, flat, or even possibly positive. At high-z and based on kpc-scale correlations, \cite{davies18} find an increasing trend, suggestive of a small-scale relationship which warrants an in-depth comparison.

\item Finally, the lower right panel shows the relationship between outflow velocity and mass loading factor. The usual scalings of $\eta_{\rm M} \propto v_{\rm out}^{-1}$ ('momentum' driven) and $\eta_{\rm M} \propto v_{\rm out}^{-2}$ ('energy' driven) are indicated by the dashed gray lines. The latter is prescribed by the TNG model at injection \citep{pillepich18a}, and we roughly recover this expectation between \etaM of a few and a few tens. In this range, stellar mass increases leftwards, from $M_\star \sim 10^8$\msun at \etaM of a few tens to $M_\star \sim 10^{10.5}$\msun at \etaM $\sim$ a few. High mass, quenching or quenched systems populate the top of this plane, distributed across all values of \etaM as high outliers to the median lines. For example, at $v_{\rm out} \sim 1000$ km/s and $\eta_{\rm M} \sim 50$. Many of the observational data points in this regime, for instance of the \cite{fiore17} compilation, are thought to be black hole driven outflows. The upturn of outflow velocity at $\eta_{\rm M} \gtrsim 10$ occurs when low-mass galaxies disappear and $M_\star > 10^{10.5}$\msun systems set the median. The monotonic scaling behavior of $v_{\rm out}(\eta_{\rm M})$ breaks down when low-SFR galaxies begin to dominate at a given mass loading. 

\end{itemize}

\begin{figure}
\centering
\includegraphics[angle=0,width=3.2in]{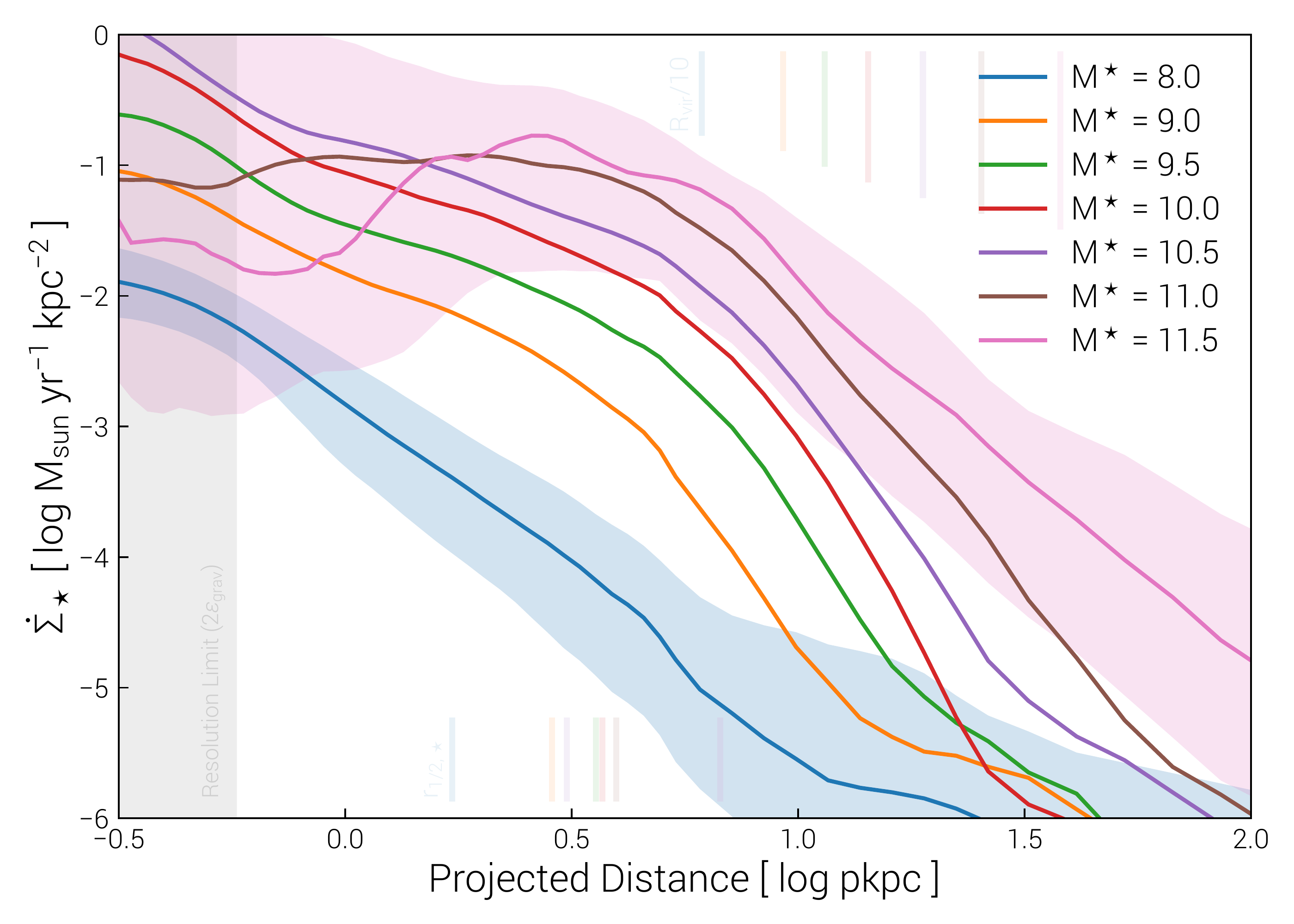}
\caption{ Median radial profiles of star formation rate surface density, in random projections, at $z=1$. Stacked bins of stellar mass, where the half mass radii of the stars ($r_{1/2,\star}$) is given by the short vertical colored lines along the bottom edge of the panel. Similarly, 10\% of the virial radius is shown by the short colored lines along the top edge of the panel.
 \label{fig_radprofiles_sfr}}
\end{figure}

Outflow properties are clearly related to galaxy stellar mass, star formation rate, and black hole activity, as well as the distribution of star formation activity. To emphasize the co-existence of stellar driven outflows across the mass range where BH activity also becomes relevant, Figure \ref{fig_radprofiles_sfr} presents stacked radial profiles of star formation rate surface density $\Sigma_{\rm SFR}(r)$ at $z=1$ for galaxies of different stellar masses, from $10^8$ to $10^{11.5}$\msun.

Overall, the star formation scales up with increasing $M_\star$ at all radii. Except for the two highest mass bins, $\Sigma_{\rm SFR}$ profiles are always monotonically decreasing away from the galactic center. They have a characteristic break at $\sim 1-2 \, r_{\rm 1/2,\star}$, outside of which profiles steepen noticeably. We draw attention to the behavior within the galaxy body itself at high-mass, where $\Sigma_{\rm SFR}$ flattens at $r < 10$ kpc for $M_\star \gtrsim 10^{11}$\msun. For the most massive systems, the profile even inverts towards the center. This central inversion is also present in the total gas density profile as well as the gas-phase metallicity profiles (not shown), indicative of dilution of the remaining central gas content by metal-poor inflows.

The drop in star formation (and so gas mass) in the central ISM is another example of the ejective character of the low-accretion mode BH feedback in TNG, which starts to operate at $M_\star \gtrsim 10^{10.5}$\msun. We previously saw evidence for this in Figure \ref{fig_outflowrate_phase} as the culprit for the cold phase of the bimodal temperature distribution of outflowing gas. 
Even though such galaxies are either in the progress of quenching for the first time, or maintaining their quenched state, Figure \ref{fig_radprofiles_sfr} shows that there can be non-negligible, residual star formation, particularly at characteristic larger radii.


\section{Discussion} \label{sec_discussion}

\subsection{The dichotomy of SN versus BH-driven outflows and the transition to quiescence}

As any star-forming gas is eligible to drive a wind-phase in TNG, galaxies do not by construction require a threshold in $\Sigma_{\rm SFR}$ such as the canonical \mbox{0.1\msun yr$^{-1}$ kpc$^{-2}$} suggested in \cite{heckman02} as a requirement to launch a galactic wind \citep[but see][]{rubin14}. In our case, outflowing mass flux will correlate with the spatial locations of star formation, where the most supernova energy is available. The suppression of central SF at high $M_\star$ implies that stellar feedback driven winds will then become less nuclear and more extended. At the same time, the co-existence of SF-driven and BH-driven outflows also becomes inevitable at $M_\star \gtrsim 10^{10.5}$\msun \citep[for example, as seen in NGC 3079;][]{cecil01}.

In general, energy inputs from stellar and black hole feedback are essentially always both present, except in the case where the halo is not yet massive enough to host a BH seed. That is, one cannot absolutely separate out the energetic contribution from these two sources to the production of an outflow, neither in simulations nor in observations. If part or all of a measured $\dot{M}_{\rm out}$ is generated due to energy injection from black holes rather then supernovae, then the denominator in our expression for $\eta_{\rm M}^{\rm SN} = \dot{M}_{\rm out} / \dot{M}_\star$  (Eqn. \ref{eqn_eta_M_SN}) neglects some if not most of the relevant energetics, as implied by the `SN' superscript.

It would be tempting to write \mbox{$\eta_{\rm M}^{\rm tot} = \dot{M}_{\rm out} / (\dot{M}_\star + \dot{M}_{\rm BH})$} or \mbox{$\eta_{\rm E}^{\rm tot} = \dot{E}_{\rm out} / (\dot{E}_{\rm SN} + \dot{E}_{\rm BH})$} and so measure the outflow properties relative to the total available energetics. However, SN versus BH feedback operate at different (spatial) scales, and have different effective (galactic or halo-scale) coupling efficiencies. For instance, although $\dot{E}_{\rm BH}$ might dominate $\dot{E}_{\rm SN}$ for some main-sequence galaxies with rapidly accreting black holes, the resulting thermal AGN feedback may have little dynamical impact and so not actually be the cause of any outflows. Conversely, even a quenching or almost entirely quiescent galaxy may have $\dot{E}_{\rm SN} \gg \dot{E}_{\rm BH}$; nonetheless, the relatively small amount of energy available from the low-accretion state BH feedback could in practice be solely responsible for a high mass flux, high velocity outflow, possibly due to the spatial (i.e. nuclear) or temporal (i.e. rapid) coherence of that energy injection, or due to the physical/numerical coupling mechanism. 

These differences leave signatures in properties of contemporaneous outflows. In TNG, the process of reddening and migration away from the blue, star-forming population is tied closely to a threshold in black hole and so stellar mass \citep{nelson18a}. As a result, the galactic-scale winds which arise from galaxies in different locations of the SFR$-M_\star$ plane (and with different BH properties) directly constrain underlying assumptions of the feedback models. Emblematically, Figure \ref{fig_vout_delta_sfms} reveals the correlation of $v_{\rm out}$ with $\Delta$SFMS as well as its inversion into an anti-correlation above a particular mass scale. Feedback also leaves signatures in the residual star formation of the galaxy itself -- Figure \ref{fig_radprofiles_sfr} demonstrates that the evolutionary process of quenching proceeds `inside-out' in TNG, with star-formation truncated in the center of the galaxy first. IFU survey data at low redshift will unambiguously measure the spatially resolved progression of SFR through the transition to quiescence \citep{belfiore17a}, providing stringent and complementary constraints to the galactic outflows themselves.

Observationally it has been noted that, at least in the local universe, `normal' AGN-host galaxies do not show a clear difference in outflow prevalence or properties when compared to non-AGN control samples \citep{sarzi16,robertsborsani18}. This may come down to the timescale issue, given that the BH need not be `on' and injecting energy at the moment of observation, although a recent (i.e. Myr-scale) outflow could still be present \citep{forsterschreiber14,gabor14b}. As shown in Figures \ref{fig_timeevo1} and \ref{fig_timeevo2}, Myr-timescale variability at the energy injection scale can produce a coherent, large-scale BH-driven outflow which lasts for at least several hundred million years. In Figure \ref{fig_outflows_vs_obs} we noted only weak dependencies of both \etaM and $v_{\rm out}$ on instantaneous black hole luminosity, likely due in part to this issue.

At the same time, it is evident that the highest velocity BH-driven outflows in the TNG50 simulation do not arise from high luminosity quasars (e.g. with space densities \mbox{$n \lesssim 10^{-7}$ Mpc$^{-3}$}), which are not present in the small volume. Instead, they are generated from low-accretion rate BHs for which the model posits a non-relativistic wind mechanism produced by a radiatively inefficient flow \citep{blandford99}. In TNG black holes which are slowly accreting and have low bolometric luminosities can drive some of the most powerful outflows.

Observational evidence has emerged in this direction, probing low accretion rate black hole populations possibly underrepresented in previous surveys \citep{cheung16,penny18}. Observational constraints on the outflow properties as well as the mass distribution of low luminosity AGN, particularly as a function of location in the SFR$-M_\star$ plane and so in relation to the quenching state of the galaxy, will provide an important assessment of the TNG black hole model in the future.

\subsection{Towards quantitative comparisons with observations} \label{subsec_discussion_comparisons}

Outflows are usually observed through the Doppler shift of specific gas tracers, either through blueshifted features in absorption \citep[e.g.][]{steidel02}, or through broad line components in emission \citep[e.g.][]{heckman90}. In both cases several complexities arise which make the inference of intrinsic physical values difficult.

First, the outflow signature itself must be separated from the host. Studying kinematics using absorption spectra when the galaxy stellar continuum is the background source is complicated by existence of strong stellar absorption features in the same lines; \cite{chen10} find for instance that $\sim$ 80\% of Na D in stacked SDSS spectra arises from stellar atmospheres -- careful subtraction is required and the signal of interest represents a potentially small residual. Handling emission which partially fills in absorption signatures from the ISM or resonantly scattered emission similarly requires careful modeling to back out the intrinsic absorption \citep[e.g. for FeII and MgII;][]{zhu15}. Outflows derived from emission lines also face difficulties, particularly in the decomposition of narrow and broad-line components, inclination corrections, the treatment of the complex underlying ISM kinematics, and the diverse excitation mechanisms for nebular lines \citep{newman12}. 
Significant assumptions are still required to convert observed line properties into e.g. mass outflow rates or mass loading factors; in particular, on the geometry, metal content, and ionization state of the outflowing gas. The uncertainties therein are significant \citep{murray07,chisholm16}, and ignoring or improperly accounting for galaxy-to-galaxy variation in assumed metallicity, ionization corrections, and outflow radii can introduce a factor of ten uncertainty in $\dot{M}_{\rm out}$, possibly hiding (or producing misleading) trends \citep{chisholm17}. For a mass conserving outflow with density varying only as a function of distance from the galaxy, the expression for mass outflow rate is

\begin{equation}
\dot{M}_{\rm out} = \Omega \,r^2 \,\mu(r) \,n_{\rm H}(r) \,v_{\rm out}(r)
\end{equation}

\noindent where $n_{\rm H}(r)$ is the hydrogen number density, $v_{\rm out}(r)$ is the dependence of the outflow velocity on distance, $\Omega = 4\pi$ for a unity covering factor, and $\mu(r) / m_{\rm p}$ the mean molecular weight. In terms of a column density $N_{\rm H}$ observed along the line of sight, the mass flow rate across a shell at radius $r$ is

\begin{equation}
\dot{M}_{\rm out}(r) = \Omega \,r \,\mu(r) \,N_{\rm H} \,v_{\rm out}(r).
\end{equation}

Alternatively, given a total, uniform mass $M_{\rm gas}$ at a constant outflow speed $v_{\rm out}$ within an outer radius of $r_{\rm out}$, dimensional analysis yields an even simpler expression for total outflow rate of

\begin{equation}
\dot{M}_{\rm out} = M_{\rm gas} \,v_{\rm out} \,r_{\rm out}^{-1}.
\end{equation}

In all cases, observations must assume values for $\Omega$ and $r_{\rm out}$, which are largely or entirely unconstrained \citep{harrison18}. The functional form of $v_{\rm out}(r)$ is also typically unknown and often assumed to be constant, taking some derivable characteristic velocity. Finally, whatever phase of the gas is observed represents only a fraction of the total outflow \citep{cicone18}, and for metal ion lines this can be a truly minuscule component. Inversion to the total hydrogen mass is therefore crucial, and yet also error prone due to the large correction factors.
These complexities, which together accumulate into large uncertainties, motivates a comparison in the other direction. That is, forward modeling of the simulation outcome into the space of direct observables. 

From the simulation side, there are several difficulties in modeling the actual gas phases. First, many important observational tracers are very cold, dense phases at $T \lesssim 10^4$ K, including neutral or molecular gas. This is an unresolved regime in many current cosmological models, due primarily to resolution limitations. Second, a standard ionization post-processing treatment, as commonly applied in studies in the low-density CGM for instance, is insufficient. This is because local radiation sources are important so close to the galaxy, from stars as well as possibly a central AGN, which are present in addition to the uniform, meta-galactic background. A treatment here would require an expensive, essentially radiative transfer calculation to derive the total incident radiation on each gas parcel, as well as assumptions on free parameters such as escape fractions from young stars. In addition, the ionization calculation is complicated by non-equilibrium effects due to rapid temporal evolution, as arise in shocks. This is a difficult problem even for specialized photoionization codes \citep[e.g. MAPPINGS;][]{sutherland18}. It is also unclear how to efficiently capture the needed short time-scale information from a cosmological simulation.

Overcoming these challenges would imply a level of modeling detail which would also enable direct predictions of synthetic nebular emission lines from the ISM itself \citep{gutkin16,byler17}, as well as extended and extraplanar ionized gas reservoirs \citep[e.g.][]{jones17}, in the context of cosmological galaxy simulations \citep{hirschmann17}. This is the inevitable direction of future modeling efforts and will be an important step to bridge the gap between hydrodynamical simulations and many of the most accessible and important observational probes of galaxies. For outflows, we will then be able to directly calculate phase-specific predictions -- for example, $\eta_{\rm M, MgII}$ or $v_{\rm out,H\alpha}$. This will facilitate direct quantitative comparisons with observables, and allow us to assess the `tip of the iceberg' effect and the importance of gas phases missing in current observations of galactic outflows.


\section{Conclusions} \label{sec_conclusions}

In this work and together with the companion paper \textcolor{blue}{Pillepich et al. (2019)} we have presented the new TNG50 simulation, the third and final volume of the IllustrisTNG project. TNG50 has been designed to overcome the resolution limitations inherent in cosmological simulations by sampling a large, statistically representative volume at unprecedented numerical resolution. Leveraging this new resource, we describe a first exploration of the properties of galactic outflows -- driven by both supernovae and black hole feedback at $z>1$ -- in the cosmological setting and with respect to the galaxies from which they arise. By resolving the internal structure of individual galaxies, TNG50 enables us to study the connection between small-scale (i.e. few hundred pc) feedback and large-scale (i.e. few hundred kpc) outflows. The diversity of galaxies and outflow properties realized in the fully representative galaxy sample of TNG50 highlights the importance of the large cosmological volume. We summarize our principal results:

\begin{itemize}

\item The new TNG50 simulation occupies a unique combination of large volume and high resolution, with a 50 Mpc box sampled by $2160^3$ gas cells. This provides a baryon mass resolution of $8 \times 10^4$\msun and an average spatial resolution of star-forming ISM gas of $\sim 100-200$ parsecs. This resolution approaches or exceeds that of modern `zoom' simulations of individual galaxies, while the volume contains $\sim$ 20,000 resolved galaxies with $M_\star \ga 10^7$\msun. It offers an unparalleled view into the small-scale structure of galaxies and the co-evolution of dark matter, gas, stars, and supermassive black holes from dwarfs to massive ellipticals. (\S\ref{sec_sims} and \S\ref{sec_results_tng50})

\item By measuring mass outflow rates around galaxies we quantify the mass loading factor $\eta_{\rm M} = \dot{M}_{\rm out} / \dot{M}_\star$ as a function of stellar mass, redshift, distance, and velocity. Whereas the TNG model input value for \etaM at the injection scale of stellar feedback monotonically decreases with $M_\star$, this is not the case for the emergent outflows at 10 kpc. Instead, the mass loading \textit{inverts} and rises rapidly for $M_\star \gtrsim 10^{10.5}$\msun. As a result, $\eta_{\rm M}(M_\star)$ has a broken `v' shape: the low-mass behavior is regulated by stellar feedback, while the high-mass behavior is set by the energetics and coupling efficiency of BH feedback, as it begins to drive strong outflows away from quenching, low-SFR systems. Strong outflows reach larger distances towards low redshift, and \etaM decreases monotonically with increasing galactocentric distance for $r > 10$ kpc. (\S\ref{sec_results_rates})

\item We extract the velocity of outflows as a function of $M_\star$, redshift, and distance, characterizing heterogeneous $v_{\rm out}$ distributions around individual galaxies through mass flux weighted velocity percentiles such as $v_{\rm out,95}$. In this case, median outflow velocities increase from $\sim$ 200 km/s at $M_\star = 10^{7.5}$\msun to $\sim$ 1000 km/s at $M_\star = 10^{11}$\msun. At fixed stellar mass, outflows are faster at higher redshift, by roughly a factor of two from $z=1$ to $z=6$. Despite a `minimum' launch velocity of 350 km/s imposed by the TNG model, low-mass galaxies host much slower outflows due to resolved halo drag. At the Milky Way mass scale, despite high wind velocities of $\gtrsim$ 800 km/s prescribed by the model, the bulk of outflowing material escapes the galaxy at $\sim$ 150 km/s. The provided scalings $v_{\rm out} \propto M_\star^\alpha$ depend on redshift and velocity percentile, and $v_{\rm out}$ always declines with galactocentric distance. At the high-mass end, BH feedback produces high-velocity outflows at speeds exceeding $\sim 3000$ km/s to 20 kpc and beyond into the inner halo. (\S\ref{sec_results_vel})

\item Outflows from TNG galaxies are multiphase, and different phases have different kinematics. In the temperature distribution of outflowing gas, the dominant component is a `hot' phase whose peak temperature increases with the virial temperature of the host halo gas. At $M_\star \gtrsim 10^{10.5}$\msun, however, a second cool outflow component emerges, resulting in a bimodal outflow temperature distribution. This colder gas is also denser, and is a direct consequence of ejective BH-driven feedback launching formerly nuclear ISM material out into the halo. The highest outflow velocities are achieved by the hottest outflow phases. (\S\ref{sec_results_multiphase})

\item Despite the directional isotropy of all energy inputs from both SN and BH feedback mechanisms, outflows are found preferentially aligned with the minor axes of galaxies. These bipolar outflows are therefore an emergent feature of our simulations, resulting from hydrodynamical collimation. Strong angular anisotropy is evident already by $\sim$ 10 kpc, and persists out as far as strong outflows are present ($\gtrsim$ 50 kpc at least). It exists at both low and high mass scales, where SN and BH feedback, respectively, dominate the production of galactic-scale outflows. Collimation increases with time, being ubiquitous by $z=1$ but difficult to convincingly demonstrate even at $z=2$. We speculate that this redshift trend is associated with the epoch of disk settling, where the rise of orderly rotating gas structures helps to sculpt outflow propagation. (\S\ref{sec_results_angle})

\item We demonstrate a correlation of $v_{\rm out}$ with $\Delta$SFMS, such that star-forming galaxies above the main sequence drive faster winds than otherwise. This relationship inverts for \mbox{$M_\star \gtrsim 10^{10.5}$\msun}, where quenching galaxies far below the main sequence launch the fastest outflows as a result of BH feedback. (\S\ref{sec_results_vsgal})

\item We measure the trends of outflow properties ($v_{\rm out}$ and \etaM) with galactic star formation rate, black hole bolometric luminosity, and the spatial distribution and concentration of star formation activity. Where available we compare against a large, heterogeneous mix of observations spanning all possible galaxy types and selections, redshifts, gas phases, and outflow tracers. While data occupies similar parameter ranges as our intrinsic predictions from TNG, a robust comparison with observations of outflow properties requires additional modeling. In TNG, low luminosity, slowly accreting black holes can drive some of the most powerful outflows. (\S\ref{sec_results_vsgal})

\end{itemize}

\noindent In the future, similar measurements of outflow properties can be made in other hydrodynamical cosmological simulations, facilitating meaningful comparisons of the feedback physics implemented in different models. This will be particularly powerful because the gas dynamics of resolved outflows and halo-scale flows in general are followed faithfully without subgrid prescriptions. It remains to be seen if other models can produce similarly realistic galaxy populations with radically different outflow properties, or vice versa.

In this work we have shown that inputs and parameters of the feedback model at the \textit{injection} scale differ from the emergent properties of galactic-scale outflows. Despite the relative simplicity of the physical assumptions invoked on the smallest scales, the resulting outflow properties are diverse and complex. That is, model parameterizations do not translate directly into observable outflow signatures. The rich phenomenology of outflows encodes the effective functioning of feedback physics and provides a unique way to study the impact of the baryon cycle on galaxy formation.


\section*{Acknowledgements}
DN would like to thank Kate Rubin, Gwen Rudie, and Alice Shapley for insightful discussions and suggestions, as well as the anonymous referee for a constructive report. SG, through the Flatiron Institute, is supported by the Simons Foundation. The primary TNG simulations were realized with compute time granted by the Gauss Centre for Supercomputing (GCS): TNG50 under GCS Large-Scale Project GCS-DWAR (2016; PIs Nelson/Pillepich), and TNG100 and TNG300 under GCS-ILLU (2014; PI Springel) on the GCS share of the supercomputer Hazel Hen at the High Performance Computing Center Stuttgart (HLRS). GCS is the alliance of the three national supercomputing centres HLRS (Universit{\"a}t Stuttgart), JSC (Forschungszentrum J{\"u}lich), and LRZ (Bayerische Akademie der Wissenschaften), funded by the German Federal Ministry of Education and Research (BMBF) and the German State Ministries for Research of Baden-W{\"u}rttemberg (MWK), Bayern (StMWFK) and Nordrhein-Westfalen (MIWF). Additional simulations were carried out on the Draco and Cobra supercomputers at the Max Planck Computing and Data Facility (MPCDF).

\bibliographystyle{mnras}
\bibliography{refs}


\appendix

\section{Halo-normalized velocities} \label{sec:appendix}

\begin{figure}
\centering
\includegraphics[angle=0,width=3.4in]{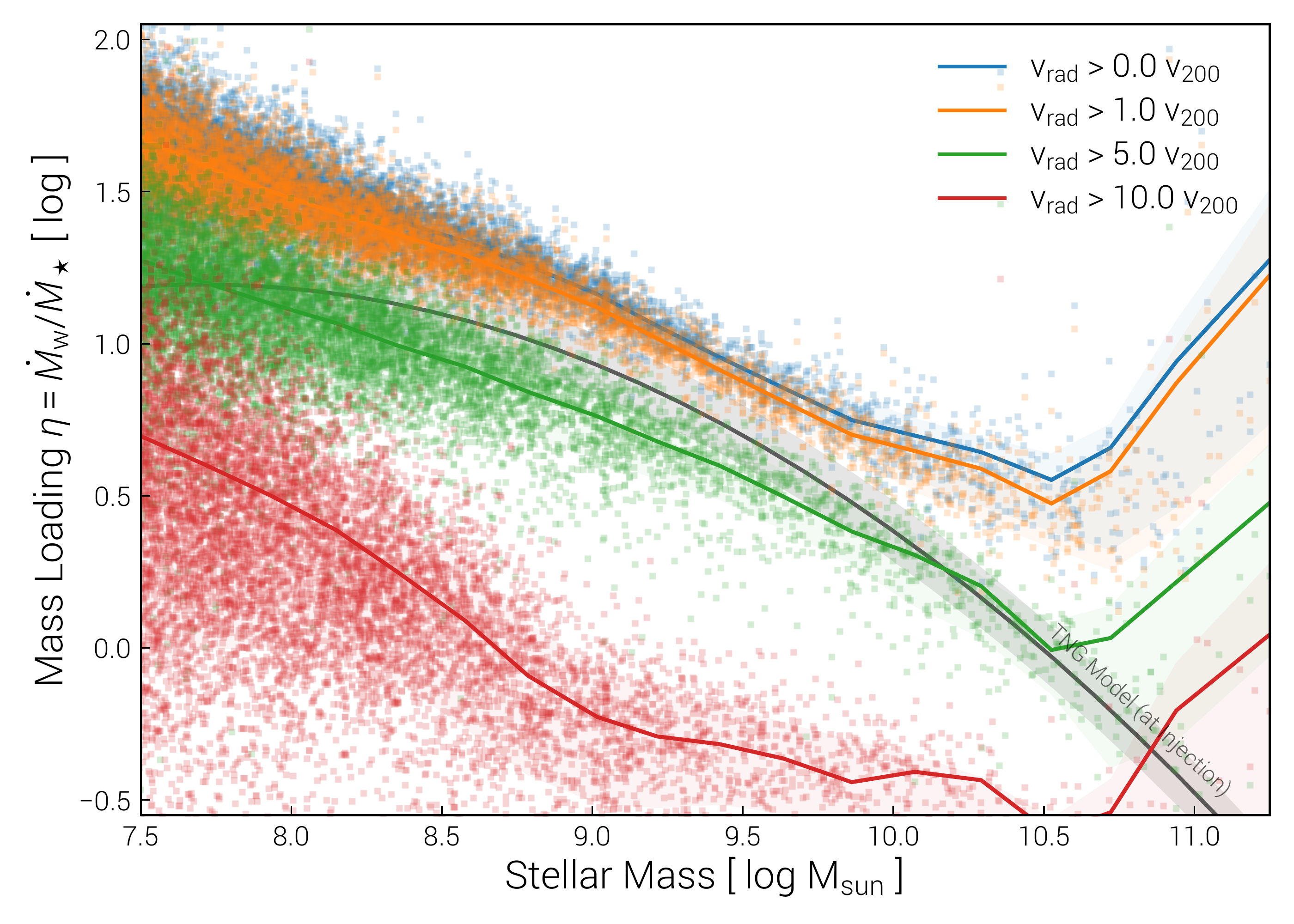}
\includegraphics[angle=0,width=3.4in]{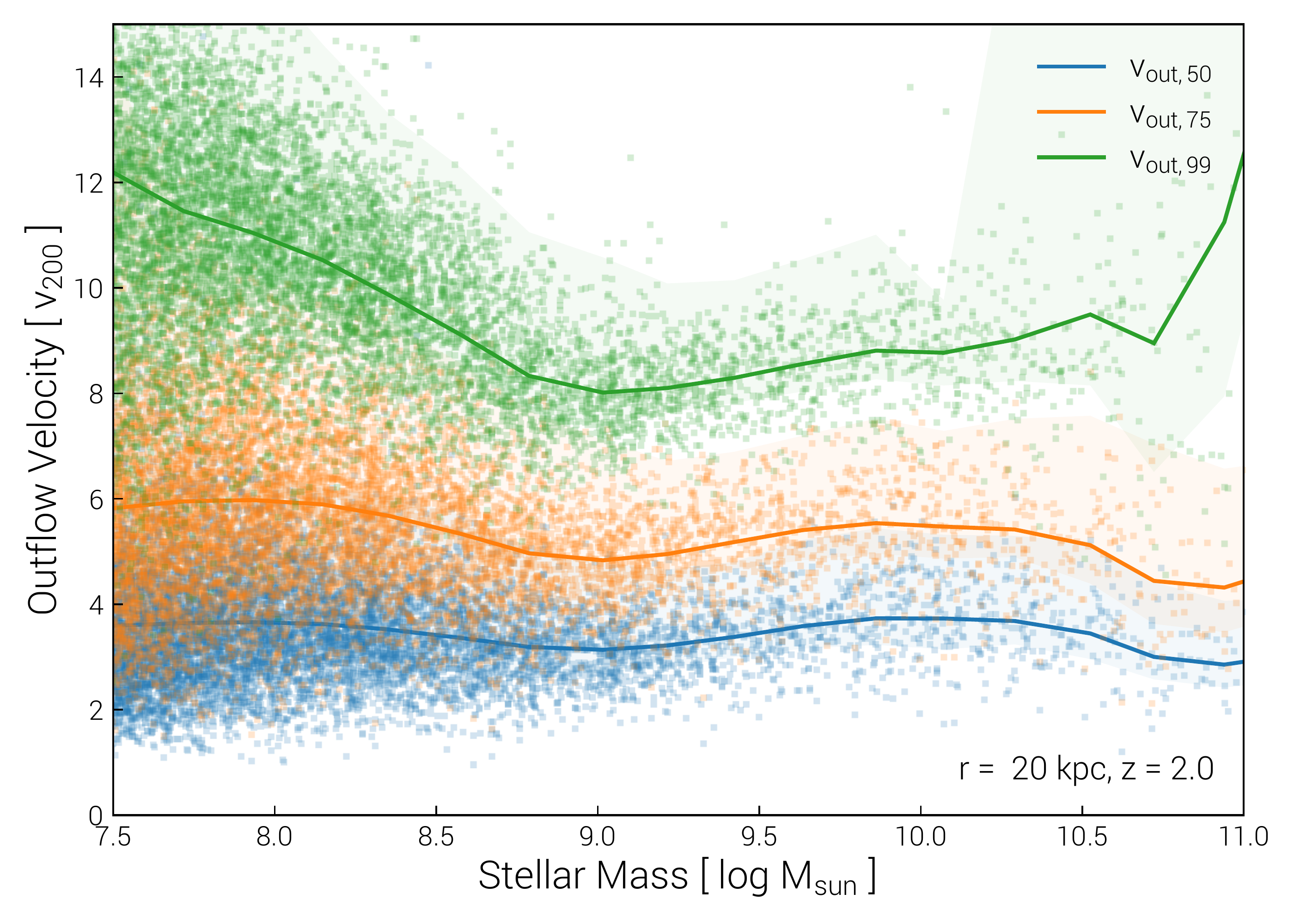}
\caption{ \textbf{(Top)} As in the main panel of Figure \ref{fig_massloading1}, the trend of \etaM as a function of $M_\star$ at $z=2$. However, instead of exploring several constant outflow velocity thresholds, we recast these in terms of the increasing halo virial velocity $v_{\rm 200}$. \textbf{(Bottom)} As in the main panel of Figure \ref{fig_vout}, the trend of $v_{\rm out}$ as a function of $M_\star$ at $z=2$ and at a distance of 20 kpc.
 \label{fig_appendix}}
\end{figure}

In this Appendix we revisit two primary results, namely the trends of mass loading and outflow velocity as a function of stellar mass. For \etaM$(M_\star)$ we change the $v_{\rm rad}$ threshold from a constant value in km/s to a threshold which scales with the halo mass. We likewise present the trends of $v_{\rm out} (M_\star)$ with a halo-based normalization. In both cases, we take a halo virial velocity defined as $v_{\rm 200}$, using the spherical overdensity measurements for each halo. For reference, at $z=2$, $v_{\rm 200}$ increases from $\sim 25$ km/s at $M_\star = 10^8 \,\rm{M}_\odot$ to $\sim 100$ km/s at $M_\star = 10^{11} \,\rm{M}_\odot$, reaching $\sim 150-200$ km/s only for the most massive $M_\star \gtrsim 10^{11.5} \,\rm{M}_\odot$ galaxies.

The top panel of Figure \ref{fig_appendix} shows \etaM as a function of $M_\star$ for four different thresholds: $v_{\rm rad} > \{0,1,5,10\} v_{\rm 200}$. As with a constant threshold, increasingly restrictive choices lead to lower measured mass outflow rates and hence mass loading values. For reasonable choices of velocity threshold, the results of Figure \ref{fig_massloading1} at the high-mass end are qualitatively and quantitatively similar. \etaM still shows a minimum at $M_\star \sim 10^{10.5} \,\rm{M}_\odot$, rapidly increasing above this mass due to strong BH-driven winds.

The main difference arises at the low-mass end, where the use of a halo-normalized velocity threshold allows slow outflows driven by relatively shallow potential halos to still contribute. As a result, no threshold value causes a turnover and subsequent decline of \etaM towards the smallest stellar masses, as was the case for the $v_{\rm rad} > 150$ km/s and $v_{\rm rad} > 250$ km/s choices of Figure \ref{fig_massloading1}. Based on this zeroth order comparison with halo circular velocity, it is clear that low mass galaxies continue to drive outflows which are `fast', relatively speaking.

In the bottom panel of Figure \ref{fig_appendix} we show the trend of outflow velocity as a function of stellar mass, again at $z=2$ and for a fixed distance of 20 kpc from the galaxy. In contrast to Figure \ref{fig_vout}, velocities are in each case normalized to $v_{\rm 200}$ of the parent halo. Three different percentiles are contrasted: $v_{\rm out,50}$, $v_{\rm out,75}$, and $v_{\rm out,95}$. As expected, measurements further into the tails of the distribution reveal gas flowing at larger velocities relative to the halo velocity. The trend with $M_\star$ in the first two cases is roughly constant, and the typical outflow velocities are between \mbox{$\sim 2-5 v_{\rm 200}$} and \mbox{$\sim 4-7 v_{\rm 200}$}, respectively. The high velocity tail starts to reveal different behavior, increasing relative to $v_{\rm 200}$ at both the low mass and high mass ends. The former can be understood in terms of the minimum injection velocity of 350 km/s for the TNG wind model, and the latter in terms of the rise of fast BH-driven outflows. In both cases the fastest moving components of the outflow can exceed \mbox{$10 v_{\rm 200}$} at these distances.

\end{document}